\documentclass[review]{elsarticle}

\usepackage{lineno,hyperref}
\modulolinenumbers[5]

\journal{Computer \& Fluids}

\usepackage{amsmath,amssymb}
\usepackage{CJK}
\usepackage{graphicx}
\usepackage{floatrow}
\usepackage{subfigure}
\usepackage{caption}
\usepackage{color}
\usepackage{bm}
\usepackage{float}
\begin{document}
\begin{frontmatter}

\title{A numerical study of the side-wall effects on turbulent bands in channel flow at transitional Reynolds numbers}

\author{Haoyang Wu}
\author{Baofang Song\corref{mycorrespondingauthor}}
\cortext[mycorrespondingauthor]{Corresponding author}
\ead{baofang_song@tju.edu.cn}
\address{Center for Applied Mathematics, Tianjin University, Tianjin 300072, China}




\begin{abstract}
We investigated the side-wall effects on turbulent bands in channel flow at transitional Reynolds numbers by direct numerical simulations using the open source spectral-element code Nektar++. {\color{black}{The width-to-height aspect ratio of 50:1 is considered for this study.}} Our study shows that turbulent bands can survive the collision with the side wall above bulk Reynolds number of $Re\simeq 1000$ but decay below $Re\simeq 975$, i.e. the critical Reynolds number should be approximately between the two Reynolds numbers. We also discussed about the underlying mechanism for the decay of the band at low Reynolds numbers and potential effects of larger spanwise channel widths than that considered in our study. The results are informative for experimental studies of channel flow turbulence at transitional Reynolds numbers.
\end{abstract}


\end{frontmatter}


\section{Introduction}
In sufficiently large channels, turbulence forms localized banded structures at transitional Reynolds numbers, i.e. the so-called turbulent stripes or bands \cite{Tsukahara2005, TsukaharaKawaguchi2014, Tuckerman2014, Xiong2015, Tao2018, Kanazawa2018, Paranjape2019, Shimizu2019, Paranjape2020, Xiao2020, Tuckerman2020, Liu2020, Duguet2020}. Aside from a tilt angle about the streamwise direction (the direction of the driven pressure gradient or mass flux), a fully localized turbulent band also exhibits an active downstream end and a more diffusive upstream end \cite{Xiong2015, Tao2018, Kanazawa2018, Paranjape2019, Shimizu2019}. Interestingly, recent studies showed that the active downstream end plays an important role in sustaining the entire band, and it was even proposed that a turbulent band is driven by this end at low Reynolds numbers \cite{Kanazawa2018, Paranjape2019, Shimizu2019, Xiao2020}. Indeed, studies showed that turbulence is continually generated at the downstream end and is responsible for the growth of turbulent bands at low Reynolds numbers \cite{Paranjape2019, Shimizu2019, Xiao2020, Song2020, Liu2020}. The downstream end propagates in both streamwise and spanwise directions, and the propagation speed is considerably different from the advection speed of the turbulence in the bulk part of the band (i.e. the part far from the two ends). It was lately proposed that this speed difference quantitatively determines the tilt angle of turbulent bands \cite{Xiao2020b}. Particularly, the downstream end has a non-vanishing spanwise propagation speed, which is approximately 0.1 in unit of the centerline velocity of the parabolic laminar flow and the direction of the spanwise speed is correlated with the tilt angle of the band with respect to the streamwise direction, i.e. bands with opposite tilt angles would have opposite directions of the spanwise propagation of the downstream end \cite{Paranjape2019,Shimizu2019,Xiao2020}. Sustained turbulent bands were reported at as low as $Re\simeq 660$ (based on half channel height and the centerline velocity of the parabolic basic flow) \cite{Paranjape2019,Tao2018,Kanazawa2018}. This banded characteristic was shown to be gradually lost as the Reynolds number increases to above $Re\simeq 950$, at which turbulent bands start to broaden in the streamwise direction and more spatially extended turbulence starts to form \cite{Xiao2020b}. Besides, frequent splitting (bands nucleating bands with the same orientation), branching (bands nucleating bands with opposite tilt directions) and band-band interactions start to occur (splitting and branching can also occur at lower Reynolds numbers but are much rarer) \cite{Paranjape2019,Shimizu2019}. Nevertheless, the turbulence-generating downstream end is a robust feature.
 
Plane Poiseuille flow (in infinite or periodic channels) serves as an ideal model for studying channel flow turbulence without side-wall effects, as most theoretical studies prefer. In a realistic channel with side walls like in experiments, as the downstream end has a spanwise propagation speed, a turbulent band must collide with the side wall, if it does not decay or collide with other bands earlier. Particularly, side walls will certainly affect the turbulence sufficiently close to the side walls. Therefore, the side-wall effects must be considered especially for experiments in channels with not-very-large spanwise widths.

Interactions between turbulent bands and the resulting flow pattern in plane Poiseuille flow have been studied by \cite{Paranjape2019, Shimizu2019, Song2020, Duguet2020}. 
However, the interaction between bands and side wall has not attracted much attention so far because most studies considered plane Poiseuille flow or turbulence far from the side wall. {\color{black}{Although the aspect-ratio effects on the transient nature of localized turbulence in channel flow has been studied in \cite{Takeishi2015}, the authors only considered a width-to-height aspect ratio up to 9, which is too small to accommodate a fully localized turbulent band that is sustained by an active turbulence-generating downstream end. Therefore, the collision between the downstream end and the side wall could not be studied in their setup.}} Given that the spanwise propagation speed of turbulent bands is small (approximately 0.1), it may take some time for a band to reach the side wall if the band was generated far from the side wall. Therefore, the side-wall effects may not be crucial for studies that only concern the short-time behaviors of turbulent bands. However, side-wall effect should be taken care of in studies that concern long-time behaviors. For example, some experimental studies \cite{Sano2016} tried to establish the connection between the transition to turbulence in channel flow and directed percolation phase transition by measuring the turbulence fraction as the order parameter. The turbulence fraction should be measured at a statistical equilibrium state, to which it usually takes a long time to reach. The side-wall effects may significantly affect the turbulence fraction measurement if the domain size is not large enough. 

To our knowledge, the only work that explicitly discussed about this problem is \cite{Paranjape2019}. The authors observed in experiments that a turbulent band immediately starts to decay when its downstream end collides with the channel side walls at sufficiently low Reynolds numbers, whereas the side walls may even perturb the flow such that trigger turbulence above $Re\simeq 1150$ in their setup (note that laminar flow may be kept to higher Reynolds numbers in different setups \cite{Liu2020}). {\color{black}{Ref. \cite{Takeishi2015} also observed that side walls deplete vorticies close to them and result in spanwise localized turbulent patch if the aspect ratio is above approximately 4.}} However, a critical Reynolds number above which a turbulent band can survive after colliding with the side wall was not mentioned, and the mechanism underlying the decay was not discussed either. The objective of the present study is to determine such a critical Reynolds number and discuss about the possible mechanism for the decay at low Reynolds numbers. The results would be informative for experimental studies of channel flow transition.

\section{Geometry and Method}
\subsection{Geometry and mesh}\label{sec:geometry}
\begin{figure}[h]
  \centering
  \includegraphics[scale=0.68]{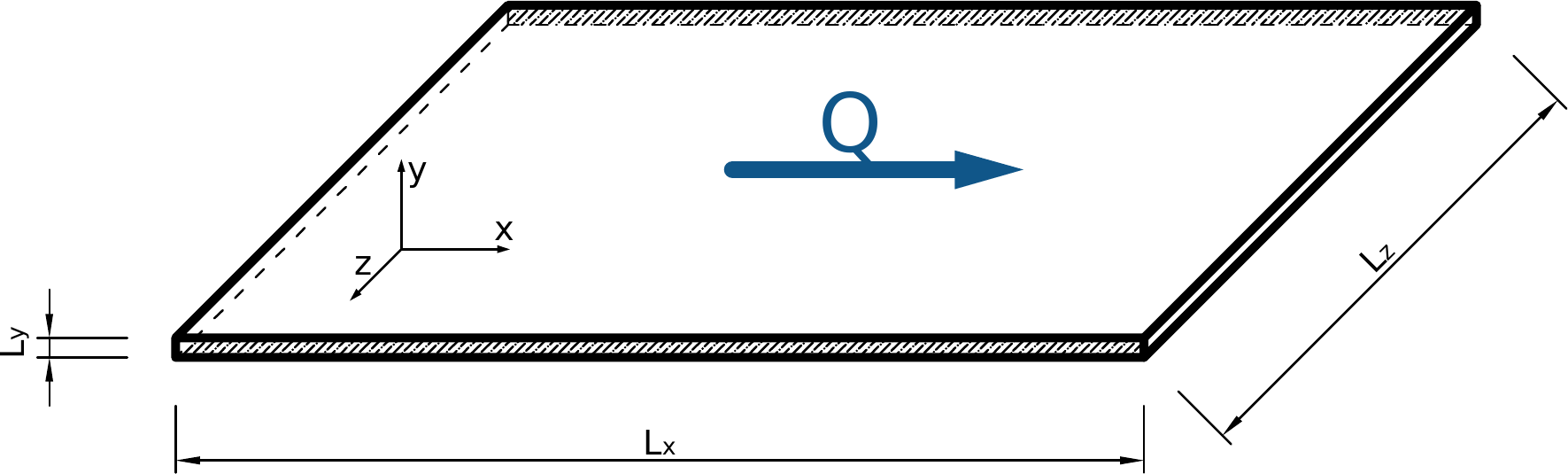}
  \caption{The geometry of the computational domain. The streamwise, spanwise and wall normal directions are denoted as $x$, $z$ and $y$ directions, respectively. The length, width and height of the channel are $L_x$, $L_z$ and $L_y$, respectively. Two side walls are placed at the two ends in the spanwise direction, i.e. at $z=\pm L_z/2$, see the shaded region. } \label{fig:model}
\end{figure}

Figure \ref{fig:model} illustrates the geometry of the flow in this paper. The dimensions of the computational domain are selected as $L_x=L_z=100$ and $L_y=2h=2$, i.e. the half-channel-height $h$ is selected as the length unit. The flow is driven by a constant mass flux $Q$ that equals to the mass flux of the Poiseuille flow in the same flow domain  (without channel side walls). The Reynolds number is therefore defined as $Re=\frac{3U_bh}{2\nu}$ where $\nu$ is the kinematic viscosity of the fluid and $U_b$ is the bulk speed of the flow (i.e. the averaged streamwise velocity in the $z-y$ cross-section). Velocity is normalized by $3U_b/2$ and {\color{black}{time by $\dfrac{2h}{3U_b}$}}.

{\color{black}{It was pointed out that if the domain size with periodic boundary conditions is not sufficiently large, turbulent bands will decay due to self-interaction \cite{Tao2018}. The streamwise size of the channel is selected based on \cite{Tao2018} such that turbulent bands can form without significant self-interaction at the considered Reynolds numbers, while keeping the computation cost as low as possible. Ref. \cite{Takeishi2015} used a close streamwise length of 120. The spanwise channel size is chosen such that turbulent bands can be accommodated in the channel. The aspect ratio of 50 is much larger than that considered in DNS by \cite{Takeishi2015} (up to 9) and comparable with the experimental study of  \cite{Yimprasert2021} where the ratio was 80. Refs \cite{Paranjape2019, Sano2016} used significantly larger widths in experimental studies, but they focused on the flow far from the side walls. In our study, the side-wall effect is meant to be investigated in the near side-wall region, therefore a much wider channel may not he necessary.}}

\begin{figure}[h]
\centering
\includegraphics[width=0.5\textwidth]{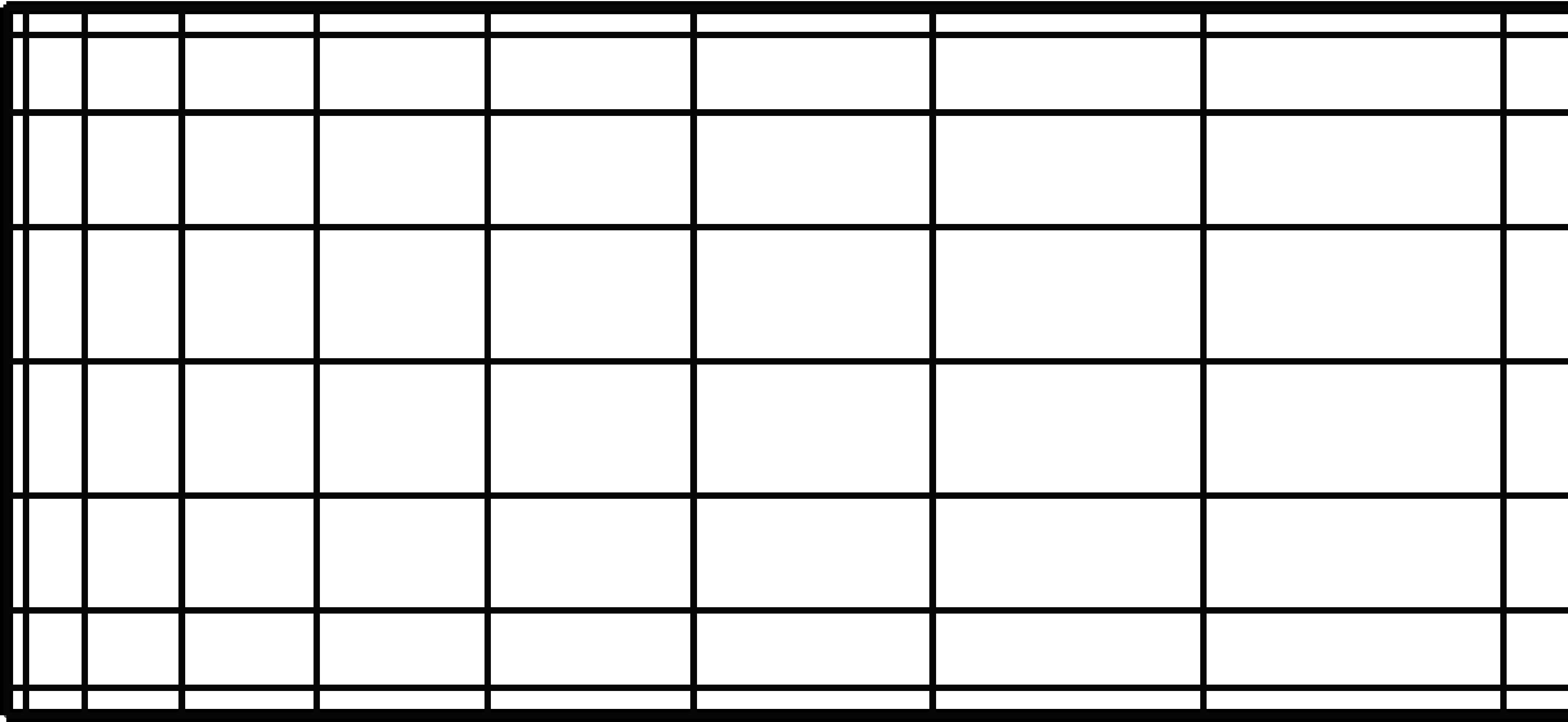}
\caption{The mesh near the channel side wall in the $z-y$ cross-section.}\label{fig:wallmesh}
\end{figure}

As we intend to use periodic boundary condition in the streamwise direction, we employ the quasi-3D formulation in Nektar++, i.e. spectral element discretization in the $z-y$ cross-section and Fourier spectral discretization in the periodic $x$ direction. By exploiting the highly efficient Fast Fourier Transform in $x$ direction, this formulation is much more efficient and less memory-demanding than a fully spectral element formulation. We use quadrilateral elements in the $z-y$ cross-section.
In order to resolve the flow close to the channel walls, in $z$ and $y$ directions, the size of the element decreases towards the wall. Specifically, the coordinates of the vertices of the elements ($z_n$ and $y_m$, where $0\le n\le N$ and $0\le m \le M$, and $N$ and $M$ denote the total number of elements in $z$ and $y$ directions, respectively) are defined by the following mapping:
\begin{equation}\label{grid}
\begin{aligned}
z_n=L_z\frac{\arcsin(-\beta_z \cos(\pi t_n))}{\arcsin(\beta_z)}, \quad\quad y_m=L_y\frac{\arcsin(-\beta_y \cos(\pi t_m))}{\arcsin(\beta_y)},
\end{aligned}
\end{equation}
where $t_n$ and $t_m$ are points uniformly distributed in [-1,1] and $\beta_z$ and $\beta_y$ are mapping parameters for the mesh in $z$ and $y$ directions, respectively. In this paper, we set $\beta_y=0.25$ and and $\beta_z=0.8$. Figure \ref{fig:wallmesh} shows the mesh near one side wall with $80$ elements in the $z$ direction ($N=80$) and $8$ elements in the $y$ direction ($M=8$). The mesh in this work is generated using the open-source package Gmsh \cite{2008Gmsh}.

\subsection{Methods}\label{sec:methods}
The governing equations of the flow are the non-dimensional Navier-Stokes equations

\begin{equation}\label{equ:NS}
\begin{aligned}
\frac{\partial \boldsymbol{u}}{\partial t} + \boldsymbol{u}\cdot \nabla\boldsymbol{u} &= -\nabla p + \frac{1}{Re} \nabla^2 \boldsymbol{u} + \boldsymbol{F}, \\
\nabla\cdot \boldsymbol{u} &= 0,
\end{aligned}
\end{equation}
where $p$ denotes pressure, $\bm F$ denotes the external force and $\boldsymbol{u} = (u, v, w)$ denotes velocity with $u$, $v$ and $w$ denoting the velocity components in $x$, $y$ and $z$ directions in Cartesian coordinates, respectively.

No-slip boundary conditions are imposed at channel walls, i.e.
\begin{equation}\label{equ:BC}
\begin{array}{ll}
\boldsymbol{u}|_{z=\pm 50} &= 0, \\
\boldsymbol{u}|_{y=\pm 1} &= 0,
\end{array}
\end{equation}
and periodic boundary condition is imposed in the streamwise direction.
That is to say, in the present work, we don't consider the finite-length effects in the laboratory experiments but focus on the side-wall effect.

Using the spectral/hp element method of Nektar++ \cite{nektar} in the $z-y$ plane and Fourier spectral method in $x$ direction, the velocity $\boldsymbol{u}$ and pressure $p$ are approximated as the expansion
\begin{equation}\label{equ:expansion}
\begin{aligned}
 A(x,y,z,t) = \sum_{k=-K}^K\sum_{p,q=0}^P\hat{A}_{pqk}(t)\phi_{pq}(z,y)\exp(i\alpha kx),
\end{aligned}
\end{equation}
where $\hat{A}_{pqk}$ denotes the coefficient of the mode $(p,q,k)$, $\phi_{pq}(z,y)$ denotes the polynomial basis, $\alpha$ is the wavenumber of the fundamental wave and determines the streamwise length of the computational domain. We set $\alpha=2\pi/100$ such that $L_x=100$. To achieve high numerical accuracy that is needed for studying the transition problem, we choose Legendre polynomials up to order 9 ($P=9$) as the basis $\phi$ based on Gauss-Lobatto-Legendre (GLL) sub-element grid points in $z$ and $y$ directions (which corresponds to 8 sub-element grids in each direction). In the streamwise direction, we use 768 Fourier modes (i.e. $K$ = 384) for the streamwise length of $L_x=100$. With these parameters, the spatial resolution in this paper is comparable with those used in the literature \cite{Tuckerman2014,Tao2018, Xiao2020b}. {\color{black}{See detailed description about the grid size and resolution test in \ref{sec:appendix}.}}

The high-order splitting method \citep{KARNIADAKIS1991414} (i.e. the velocity correction scheme in Nektar++) is used to solve the discretized incompressible system, and the $3^{rd}$ order semi-implicit IMEXOrder3 scheme of Nektar++ \cite{nektar} is used for the time-stepping. For solving the pressure Poisson equation, the high-order pressure boundary condition at solid walls of \cite{KARNIADAKIS1991414} is employed. A time-step size of $\Delta t=0.015$ is used for all the simulations. 

\subsection{The forcing term}

Below $Re\simeq 800$, it is rather difficult to generate a turbulent band in either numerical simulations \cite{Tao2018} and experiments \cite{Paranjape2019, Yimprasert2021} because special perturbations are needed \cite{Tao2018}, especially when one wants to control the position and orientation (tilt angle about the streamwise direction) of the band. Song \& Xiao \cite{Song2020} proposed an effective perturbation method for numerical simulations at low Reynolds numbers, which enables us to control the position and orientation of the band precisely. This method is ideal for our current study because we can control the collision between the turbulence and the side wall with this method.
The method is based on the study of \cite{Xiao2020} who proposed that the inflectional instability associated with the local mean flow at the downstream end of a turbulent band is responsible for the turbulence generation and the growth of the band. The core idea of this method is to mimic this instability by adding a localized (in the $x-z$ plane) body force to induce a demanded linear instability. The body force can be designed using a target velocity profile with similar instability properties. Briefly, a force $\boldsymbol{f}=\bm f(y)$ (homogeneous in $x$ and $z$ directions) is obtained by solving the equation
\begin{equation}\label{force}
\begin{aligned}
\boldsymbol{f} + \frac{1}{Re}\nabla^{2}\boldsymbol{U} = 0 ,
\end{aligned}
\end{equation}
where $\bm U = \bm U(y)$ is the target velocity profile that will be induced by the force $\bm f$ in a steady flow. Ref \cite{Song2020} presented a polynomial fit of the actual local mean flow profile at the downstream end of a band at $Re=750$ measured by \cite{Xiao2020}, and we used the same profile. For the ease of analysis and discussion later in section \ref{sec:discussion}, here we repeat the formula of $\bm U$
\begin{eqnarray}
\label{equ:target_profile}
U_x&=&-0.2478y^8 + 0.5390y^6 -0.2768y^4-0.1250y^2+0.1106,\\
U_y&=&0, \\
\label{equ:target_profile_2}
U_z&=&-0.2469y^8 + 0.7262y^6 -0.8448y^4+0.3765y^2-0.0110.
\end{eqnarray}
Then the body force term $\bm F$ in equation (\ref{equ:NS}) is constructed by properly scaling and localizing $\bm f$ \cite{Song2020}. Besides, the force $\bm F$ will be moved with a speed of 0.85 (in unit of 3$U_b$/2) in the streamwise direction and a speed of 0.1 in the spanwise direction (or $-0.1$ depending on the orientation of the band), because these speeds are approximately the natural propagation speeds of the downstream end of turbulent bands in the considered Reynolds number regime \cite{Xiao2020b}. The readers are referred to \cite{Paranjape2019, Xiao2020, Song2020, Xiao2020b, Liu2020} for more details about the speed measurement and to \cite{Xiao2020,Xiao2020c} for possible mechanisms that determine the propagation speed of the downstream end. One can set an initial position $(x_0, z_0)$ and a moving direction of the force $\bm F$ (i.e. the intended direction of the spanwise speed of the band to be generated), and the force should be switched off if a sufficiently long band has been generated and the band will be self-sustained above $Re\simeq 660$ according to \cite{Kanazawa2018, Tao2018, Paranjape2019}. This external force term $\bm F$ is added to the Naiver-Stokes solver using the User Defined Function interface provided by Nektar++.

\section{Results}
\subsection{The basic flow}
Firstly, we computed the basic flow in the channel without introducing perturbations and the external forcing $\bm F$. We started the simulation from a homogeneous parabolic flow and imposed the target volume flux. The basic flow experiences some transient adjustment especially near the two side walls given the no-slip boundary condition. The simulation was stopped when the flow had sufficiently developed and nearly reached the steady state.

At sufficiently high Reynolds numbers, one can expect secondary flows at the corners of the channel which may disturb the flow and even trigger turbulence. In experiments, Ref \cite{Paranjape2019} observed that the basic flow can be kept laminar up to $Re\simeq 1150$ whereas the side wall may trigger turbulence at higher Reynolds numbers. In numerical simulations where disturbances can be kept low, the laminar flow may be kept at higher Reynolds numbers, but in the present study we only considered $Re\le 1050$. 

\begin{figure}[H]
\centerline
	\centering
	\subfigcapskip = -4pt
	\begin{minipage}[h]{0.75\linewidth}
	\centering
	\subfigure []{\includegraphics[width=0.9\textwidth]{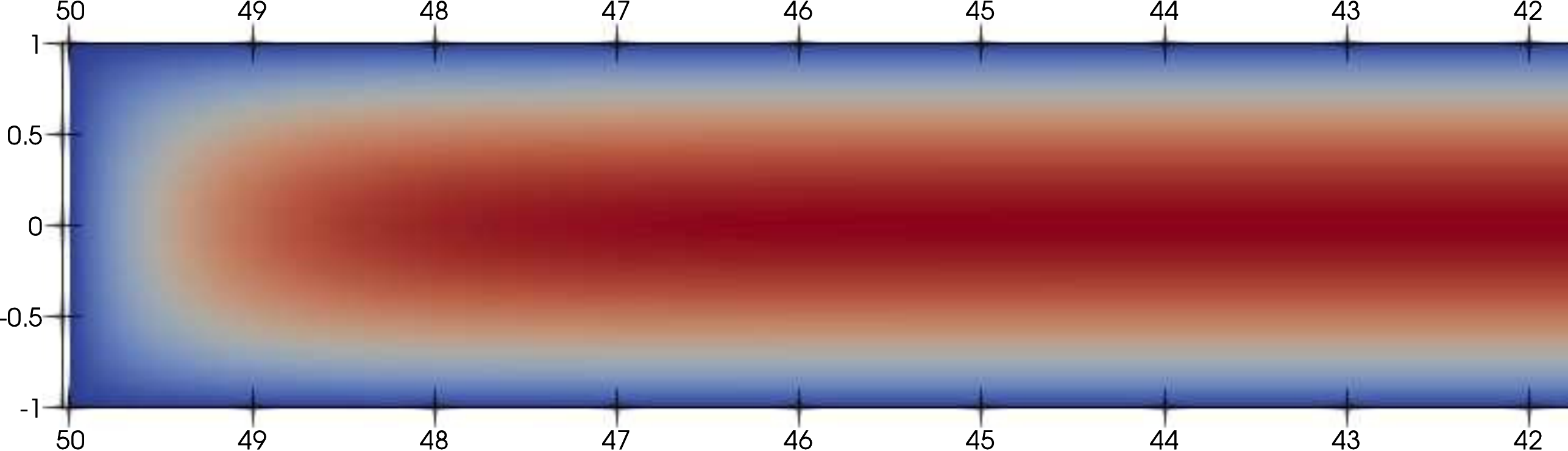}}
	\end{minipage}
	\quad
	\begin{minipage}[h]{0.11\linewidth}
	\subfigure {\includegraphics[width=0.9\textwidth]{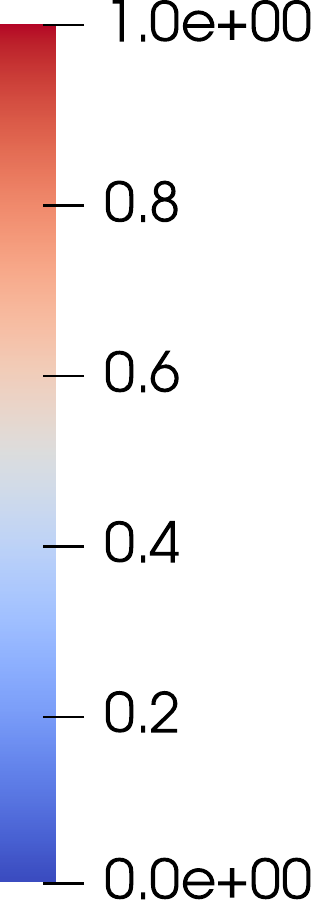}}
	\end{minipage}
	\quad
	\begin{minipage}[h]{0.45\linewidth}
	\subfigure {\includegraphics[width=1.1\textwidth]{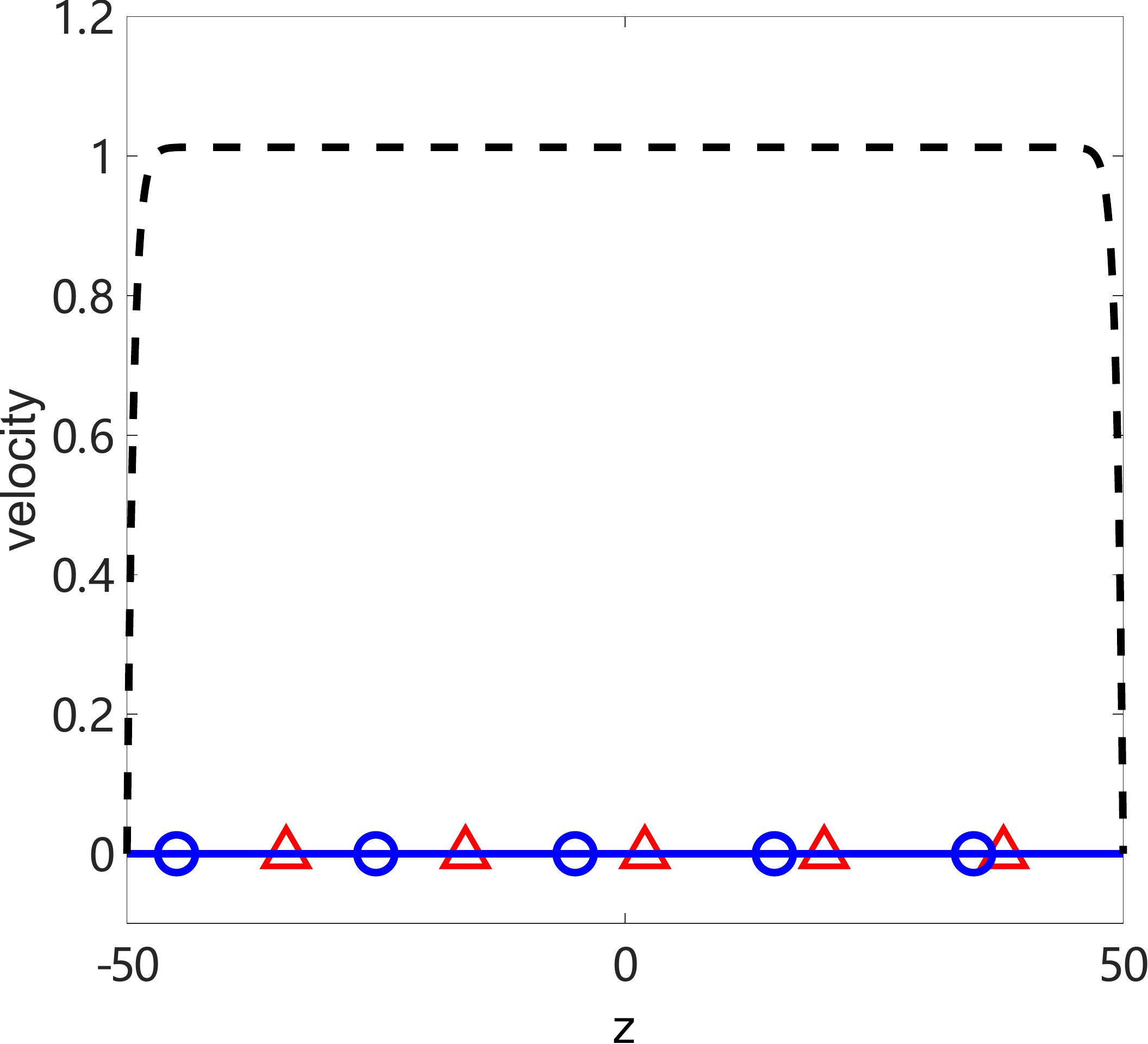}}
	\centerline {(b)}
	\end{minipage}
	\quad
	\quad
	\begin{minipage}[h]{0.45\linewidth}
	\centering
	\subfigure {\includegraphics[width=1.1\textwidth]{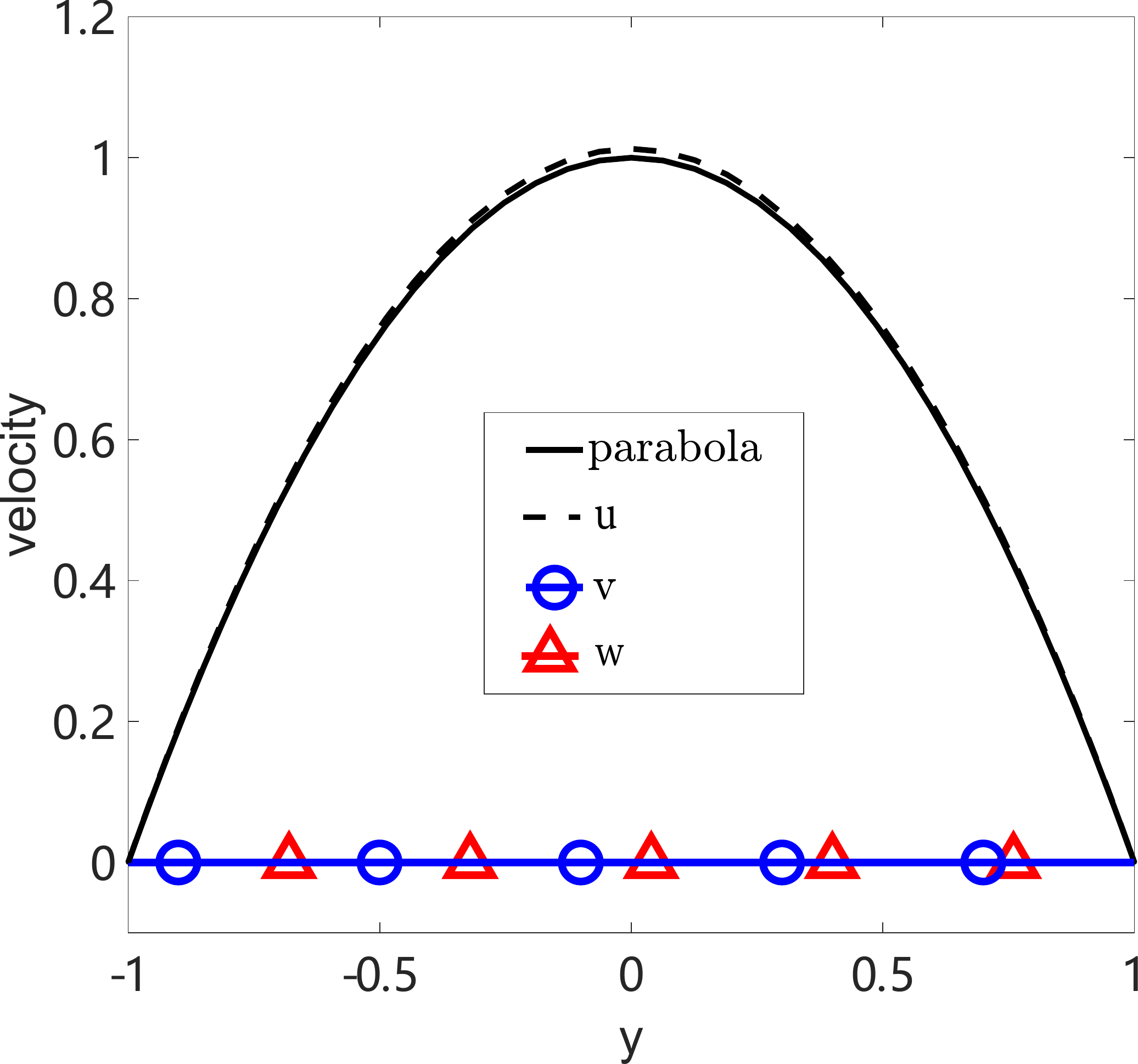}}
	\vspace{2pt}
	\centerline {(c)}
	\end{minipage}
	\begin{minipage}[h]{0.45\linewidth}
	\subfigure {\includegraphics[width=1.1\textwidth]{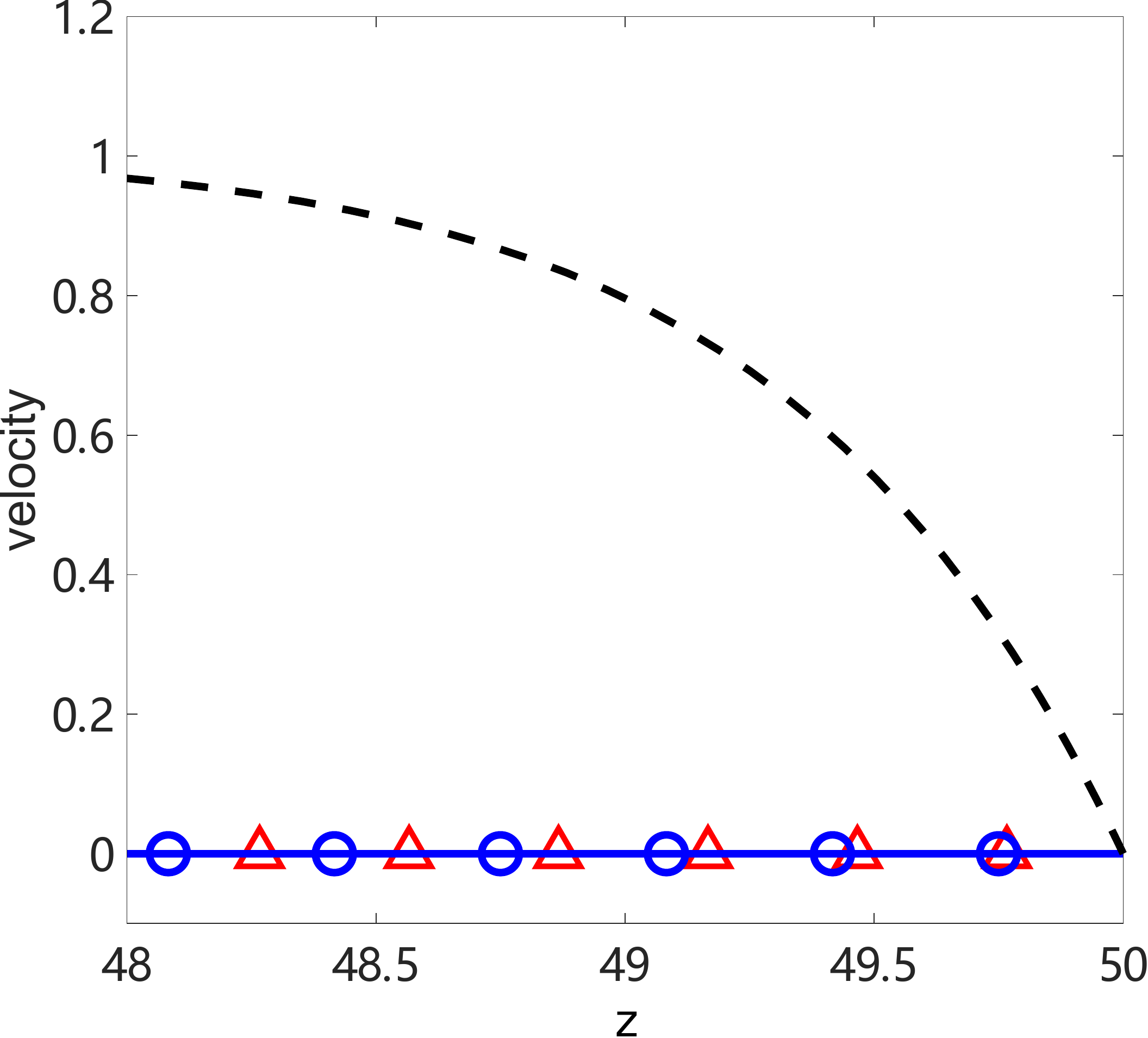}}
	\centerline {(d)}
	\end{minipage}
	\quad
	\quad
	\begin{minipage}[h]{0.45\linewidth}
	\centering
	\subfigure {\includegraphics[width=1.02\textwidth]{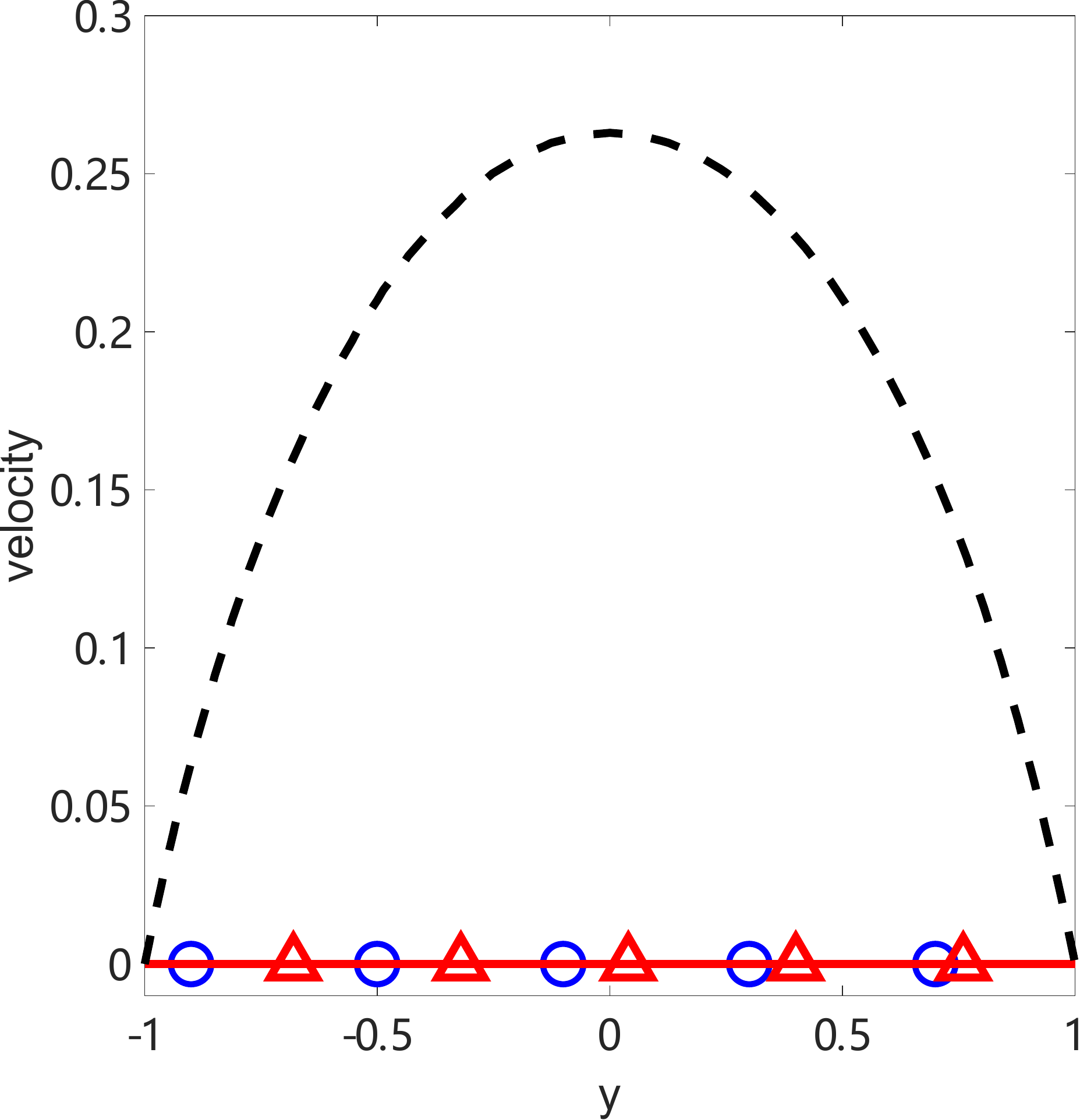}}
	\centerline {(e)}
	\end{minipage}
	\caption{Visualization of the basic flow at $Re = 1050$. (a) The contours of the streamwise velocity $u$ in the $z-y$ cross-section near one side wall. Paraview (http://www.paraview.org) is used for the visualization in this paper. (b) The distribution of the velocity along $z$ in the center plane $y=0$. (c) Velocity profiles at $z=0$. The parabolic profile of the Poiseuille flow is shown as the solid line for comparison. (d) Velocities near one side wall in the range $z\in [48,50]$. (e) Velocity profiles at $z = 49.8$.}
\label{fig:cornerplot}
\end{figure}

Figure \ref{fig:cornerplot}(a) shows the contours of the streamwise velocity $u$ in the $z-y$ plane near one of the side walls for $Re=1050$. No secondary flow structures can be observed in the contour plot. Figure \ref{fig:cornerplot}(b) shows the distribution of the streamwise velocity $u$ over $z$ in the mid-plane of $y=0$. It can be seen that the streamwise velocity $u(y=0)$ is nearly 1.0, i.e. the centerline velocity of the laminar Plane Poiseuille flow, in most of $z$ range. This is because the aspect ratio of the channel is large ($L_z/L_y=50$) so that the flow should resemble the plane Poiseuille flow in most region of the channel except for the regions near the side walls. Indeed, figure \ref{fig:cornerplot}(c) shows the velocity profile at $z=0$ (the dashed line) and the parabola of the Poiseuille flow (thin solid line), which are very close to each other. The actual profile is slightly higher than the parabola because the side walls reduce the flux in their neighborhood due to the no-slip boundary condition (see figure \ref{fig:cornerplot}(a)) so that the flux far from the side walls must be slightly larger than the parabola, given that we require the volume flux in our simulations to be the same as that of the Poiseuille flow. Figure \ref{fig:cornerplot}(d) shows the velocities along $z$ at $y=0$ near one side wall and figure \ref{fig:cornerplot}(e) shows the velocity profiles at $z=49.8$. These two panels clearly show that the transverse components $w$ and $v$ are zero, indicating that no secondary flow structures (such as corner vortices) exist at this Reynolds number. It can be inferred that the basic flow is free of secondary flows at lower Reynolds numbers. 

\subsection{Generation of a single turbulent band}
\label{sec:Re750}
When the basic flow had been obtained, we perturbed the flow by switching on the external force $\bm F$. As Ref \cite{Song2020} pointed out, $\bm F$ alone is sufficient to generate a band because even small numerical errors can be magnified by the instability induced by $\bm F$. However, it may take a long time to generate the turbulent band as it takes time for the very small disturbances to grow. To speed up the generation of a band, besides $\bm F$, we added random initial velocity disturbances of $\mathcal{O}(10^{-4})$ as suggested by \cite{Song2020}. We found in our simulations that the basic flows are nearly identical in the considered Reynolds number regime, therefore, we only generated a band with the forcing technique at $Re=750$, and the band will later be used as initial conditions for other Reynolds numbers. 

The force will be initialized close to one side wall and move towards the other side wall. Given that $L_z=100$ and the spanwise speed of the moving force is set to 0.1, we have at most 1000 time units before the force/band arrives at the other side wall. Ref \cite{Song2020} showed that it takes a few hundred time units to generate a band using the external force, and we require the band to evolve naturally for at least a few hundred time units after the forcing is switched off and before colliding with the side wall. These requirements determine the initial position, strength and the duration of the forcing. Specifically, we initialized the force $\bm F$ at $(x_0, z_0)= (0,35)$ close to the side wall at $z=50$ and moves the force towards the other side wall at $z=-50$ in order to provide the flow with enough space and time to form a turbulent band.

\begin{figure}[h]
\begin{minipage}[h]{0.29\linewidth}
		\centering
		\subfigure[t=320]{\includegraphics[width=1\textwidth]{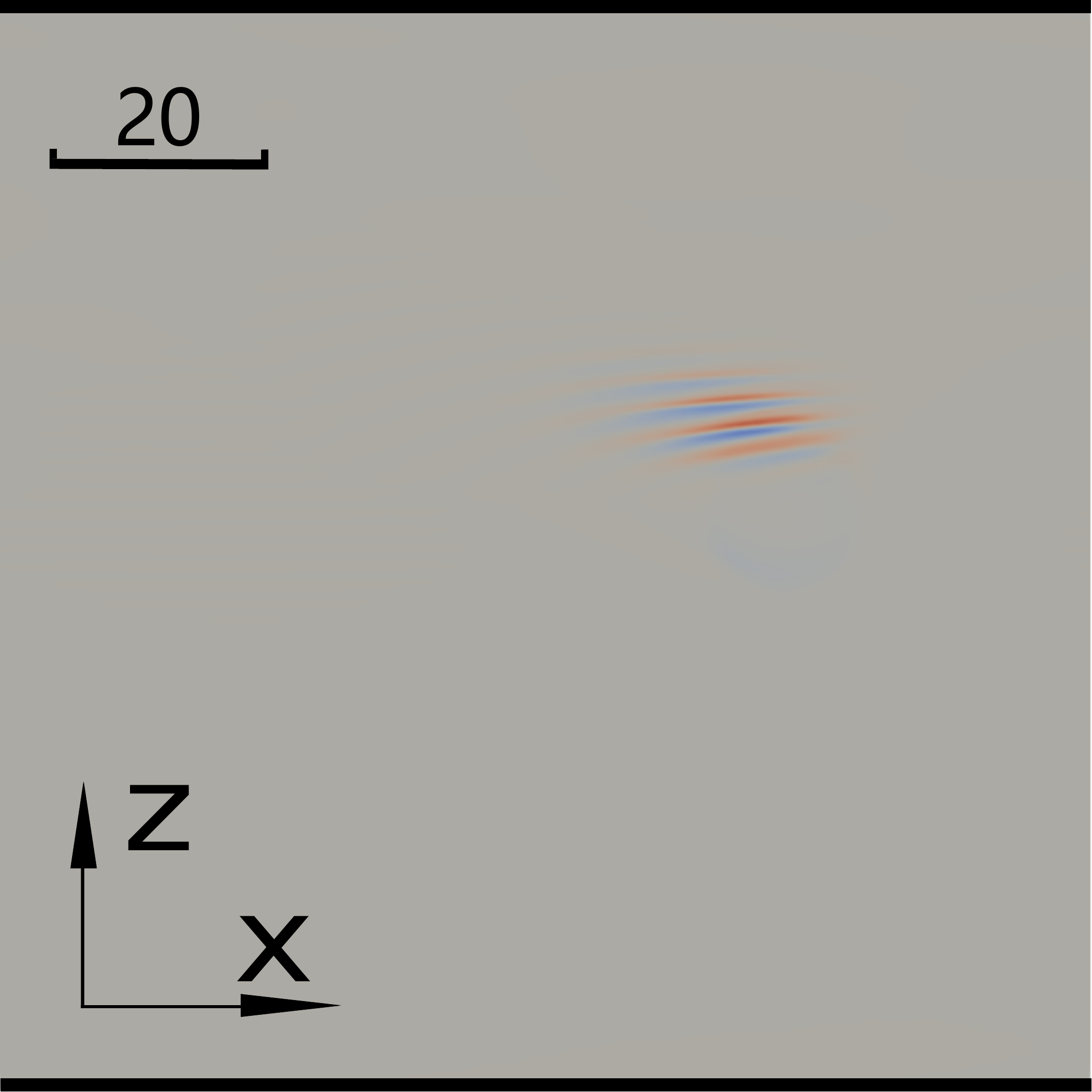}}
	\end{minipage}
	\begin{minipage}[h]{0.29\linewidth}
		\centering
		\subfigure[t=400]{\includegraphics[width=1\textwidth]{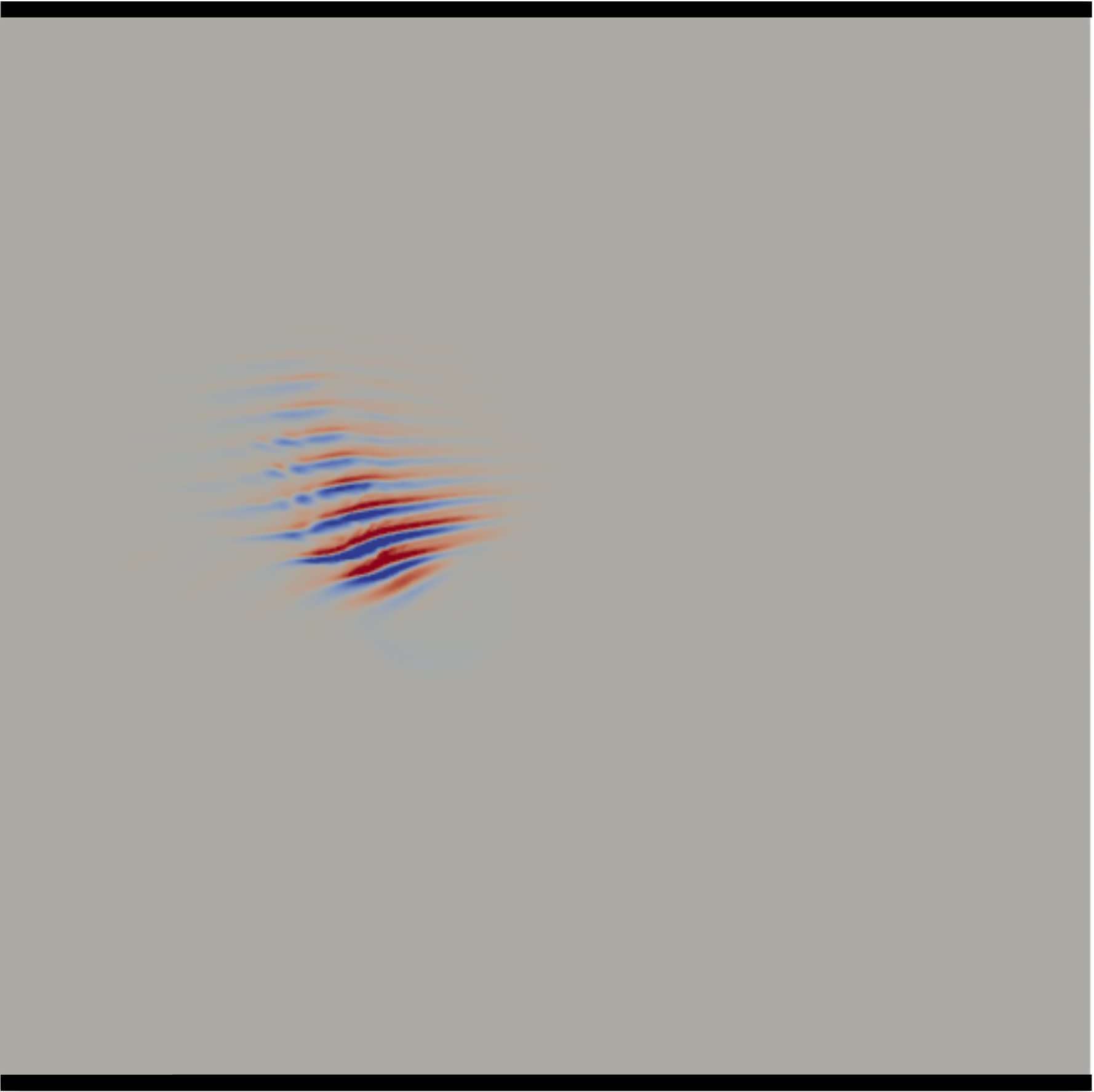}}
	\end{minipage}
	\begin{minipage}[h]{0.29\linewidth}
	\centering
	\subfigure[t=550]{\includegraphics[width=1\textwidth]{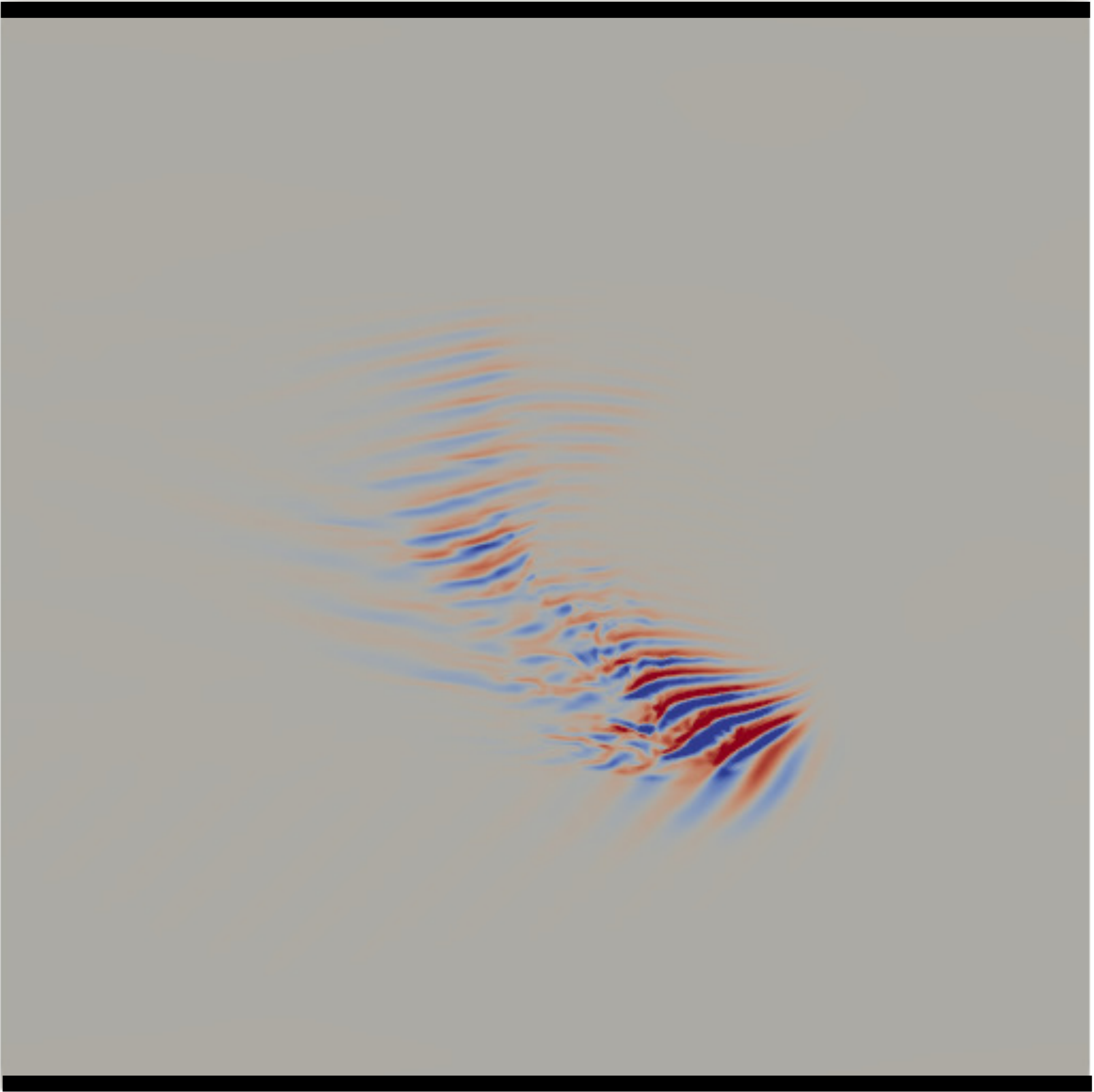}}
	\end{minipage}
	\begin{minipage}[h]{0.1\linewidth}
	\centering
	\subfigure{\includegraphics[width=1\textwidth]{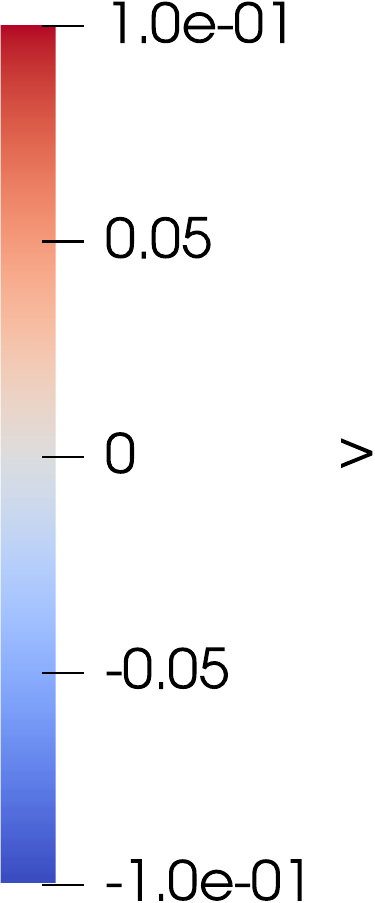}}
	\end{minipage}
\caption{The formation of a turbulent band under the external force $\bm F$ at $Re=750$. Contours of the wall-normal velocity in the $x-z$ cut plane at $y = -0.5$ at three time instances. The flow is from left to right and the basic flow has been subtracted to highlight the velocity deviations. Red color represents higher velocity and blue color represents lower velocity than the basic flow. {\color{black}{The channel side walls are marked with the bold lines at the top and bottom.}}}\label{fig:turbulentband}
\end{figure}

Figure \ref{fig:turbulentband} shows the formation process of a turbulent band under the forcing at $Re=750$. The force $\bm F$ was switched on at $t=0$, and the velocity streaks resulting from the induced instability can be clearly seen at $t=320$, and a seed for a turbulent band is formed at $t=400$. A short turbulent band is successfully generated at $t=550$. Similar formation process was shown in \cite{Song2020} in the Poiseuille flow. The generated turbulent band at $t=550$ has the typical characteristics of turbulent bands reported in the literature, such as a tilt angle about the streamwise direction, wave-like high-speed streaks (red stripes in the figure) and low-speed streaks (blue stripes) aligned alternately in the band, an active downstream end and a relatively diffusive upstream end. This validated our simulations and the implementation of the forcing technique in Nektar++.

\subsection{The collision between turbulent bands and the side wall}
The main objective of this paper is to determine whether or not turbulent bands can survive the collision with the side walls. We chose $Re=750$, 950, 975, 1000 and 1050 for this study. We selected the flow field at $t=400$ of $Re=750$ (see section \ref{sec:Re750} and figure \ref{fig:turbulentband}(b)), in which a short banded structure consisting of high- and low-speed streaks already formed, as the initial condition for other Reynolds numbers. Such a band seed is probably sufficient for generating turbulent bands at other higher Reynolds numbers. It should be noted that the flow will evolve naturally without the force $\bm F$ from this time on.

\begin{figure}[H]
	\subfigbottomskip = 10pt
	\subfigcapskip=-5pt
	\begin{minipage}[h]{0.22\linewidth}
	\centering
	\subfigure[Re=950 t=700]{\includegraphics[width = 0.995\textwidth]{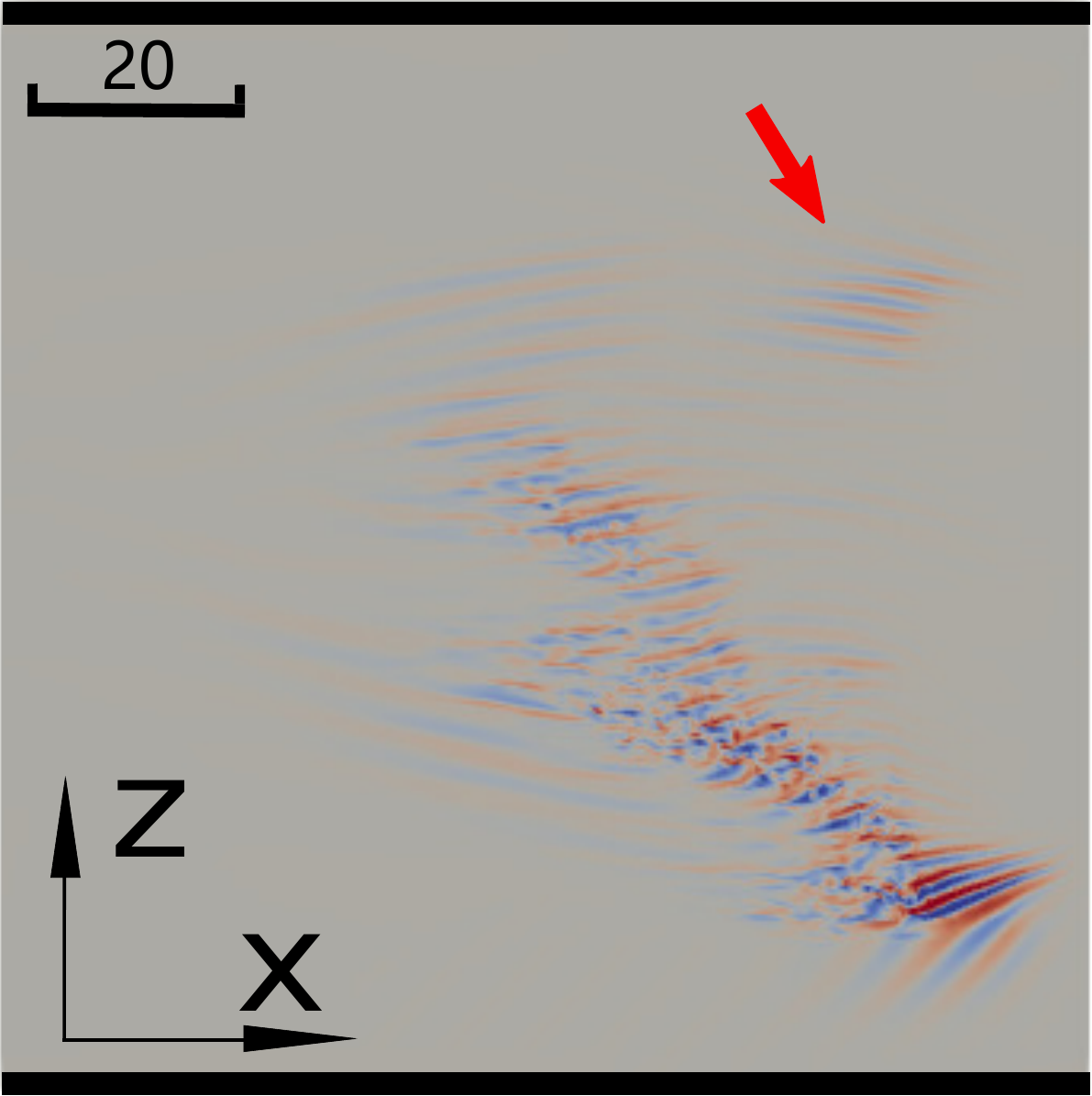}}
	\end{minipage}
	\begin{minipage}[h]{0.22\linewidth}
	\centering
	\subfigure[t=940]{\includegraphics[width = 1\textwidth]{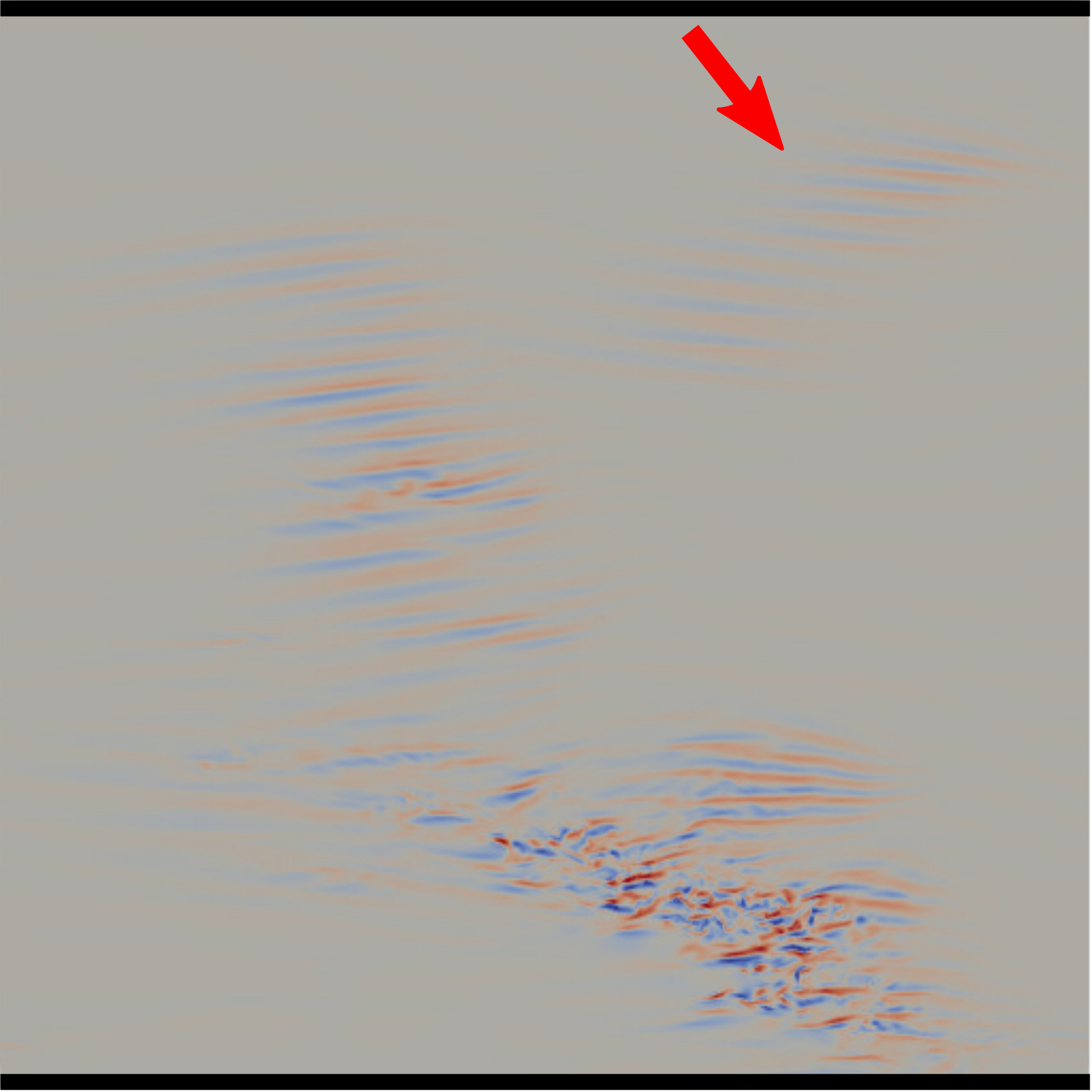}}
	\end{minipage}
	\begin{minipage}[h]{0.22\linewidth}
	\centering
	\subfigure[t=1360]{\includegraphics[width = 0.995\textwidth]{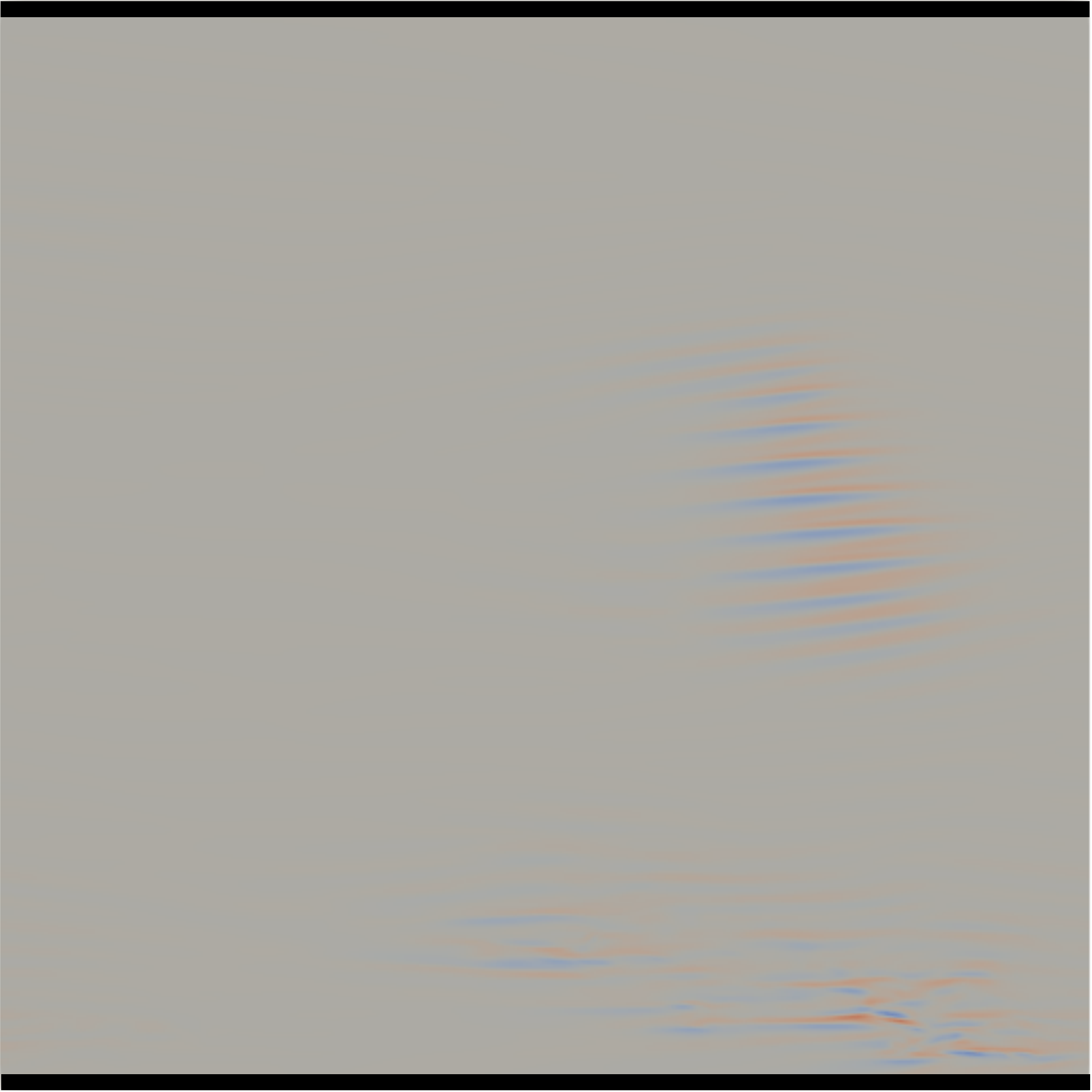}}
	\end{minipage}
	\begin{minipage}[h]{0.22\linewidth}
	\centering
	\subfigure[t=1840]{\includegraphics[width = 0.995\textwidth]{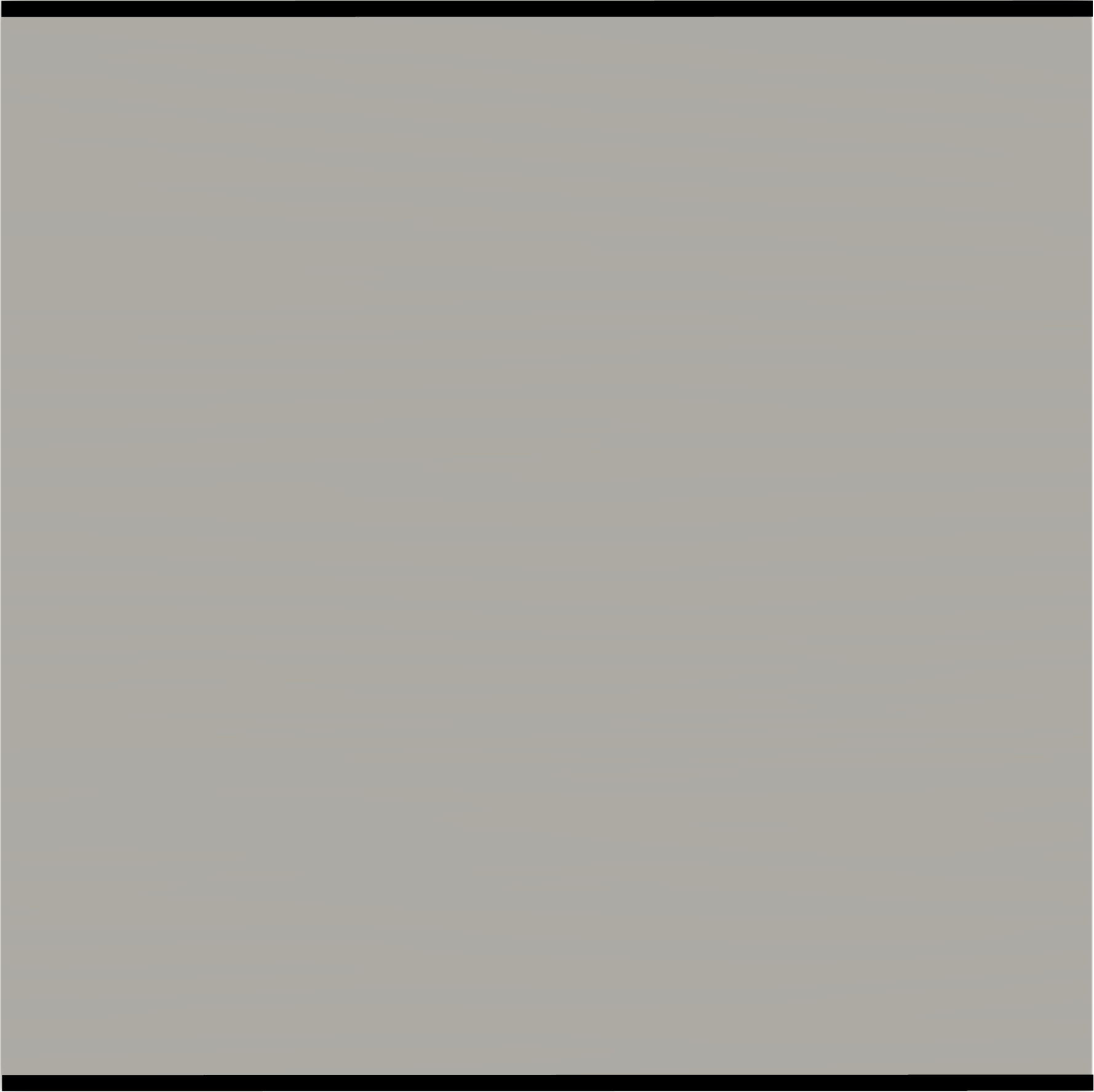}}
	\end{minipage}
	\begin{minipage}[h]{0.22\linewidth}
	\centering
	\subfigure[Re=975 t=700]{\includegraphics[width = 0.995\textwidth]{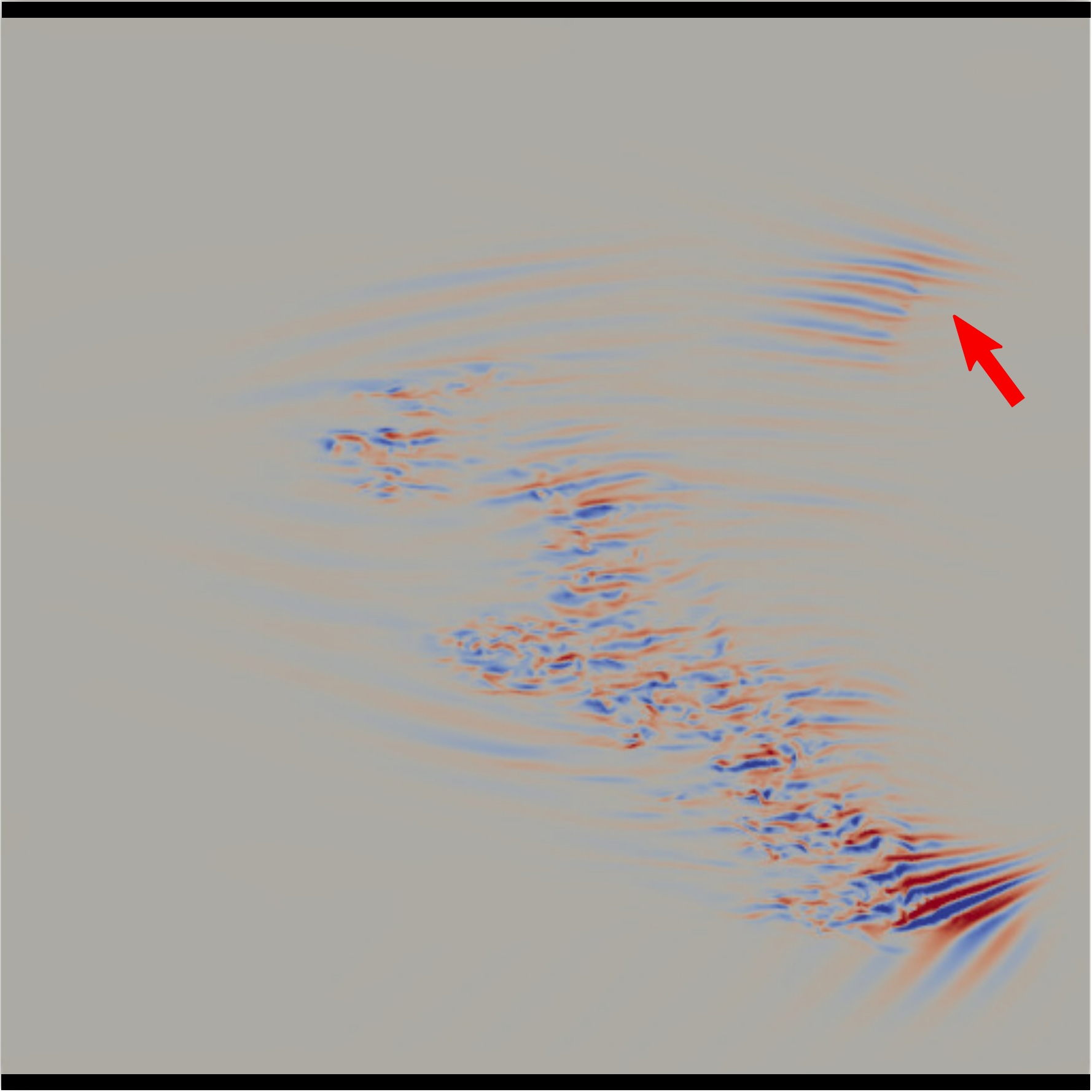}}
	\end{minipage}
	\begin{minipage}[h]{0.22\linewidth}
	\centering
	\subfigure[t=820]{\includegraphics[width = 1\textwidth]{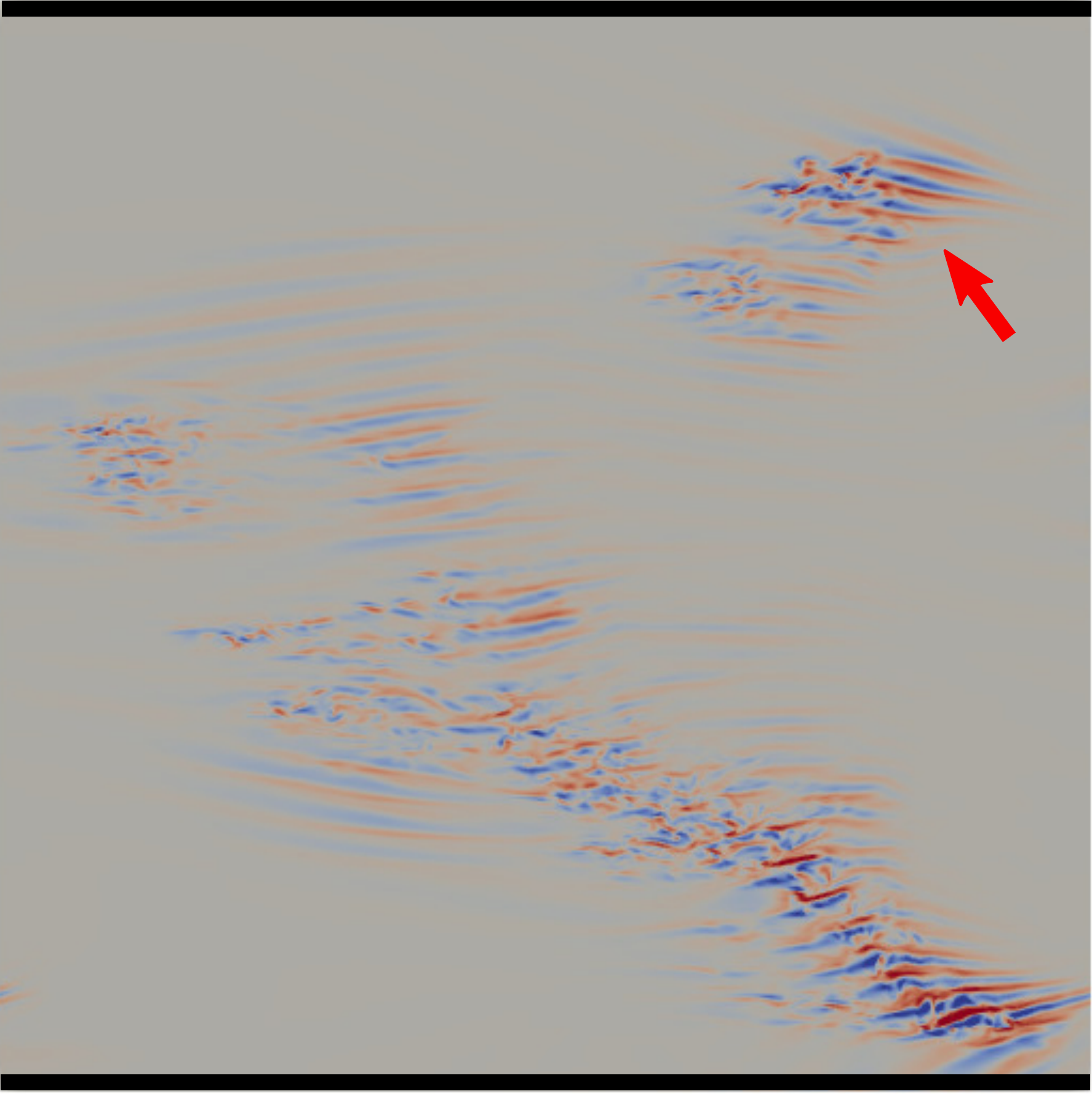}}
	\end{minipage}
	\begin{minipage}[h]{0.22\linewidth}
	\centering
	\subfigure[t=1104]{\includegraphics[width = 0.995\textwidth]{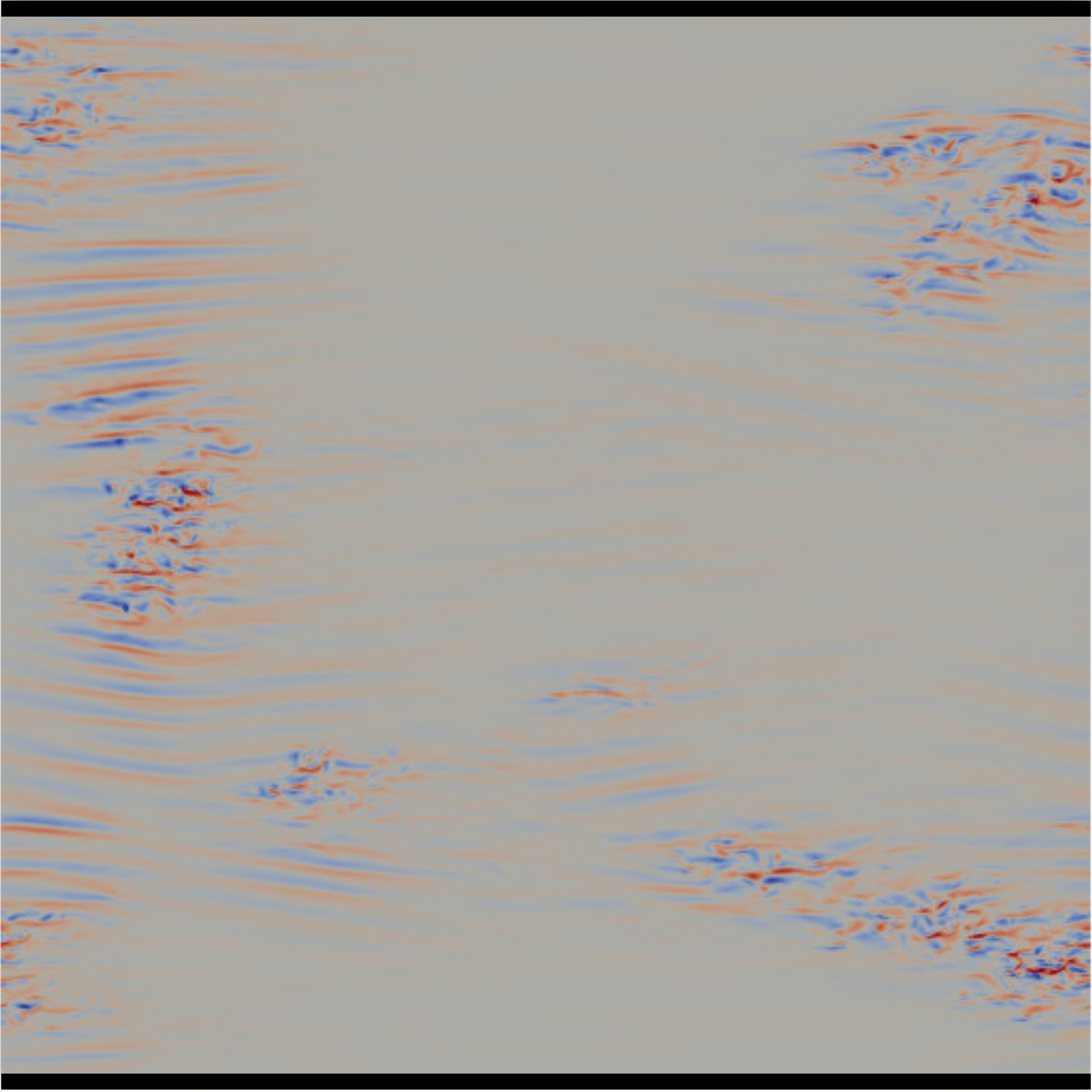}}
	\end{minipage}
	\begin{minipage}[h]{0.22\linewidth}
	\centering
	\subfigure[t=2545]{\includegraphics[width = 0.995\textwidth]{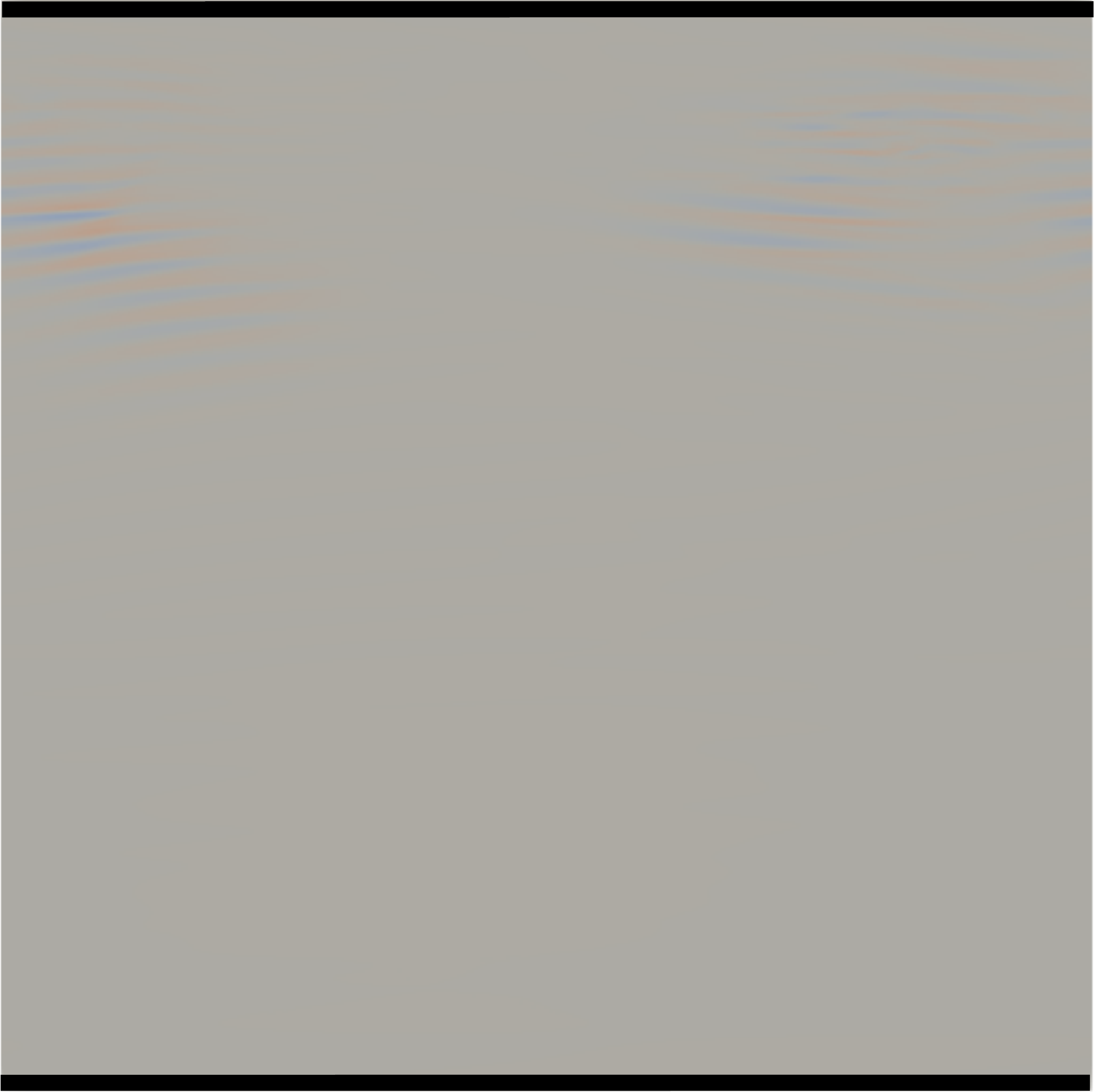}}
	\end{minipage}
	\begin{minipage}[h]{0.22\linewidth}
	\centering
	\subfigure[Re=1000 t=700]{\includegraphics[width = 0.995\textwidth]{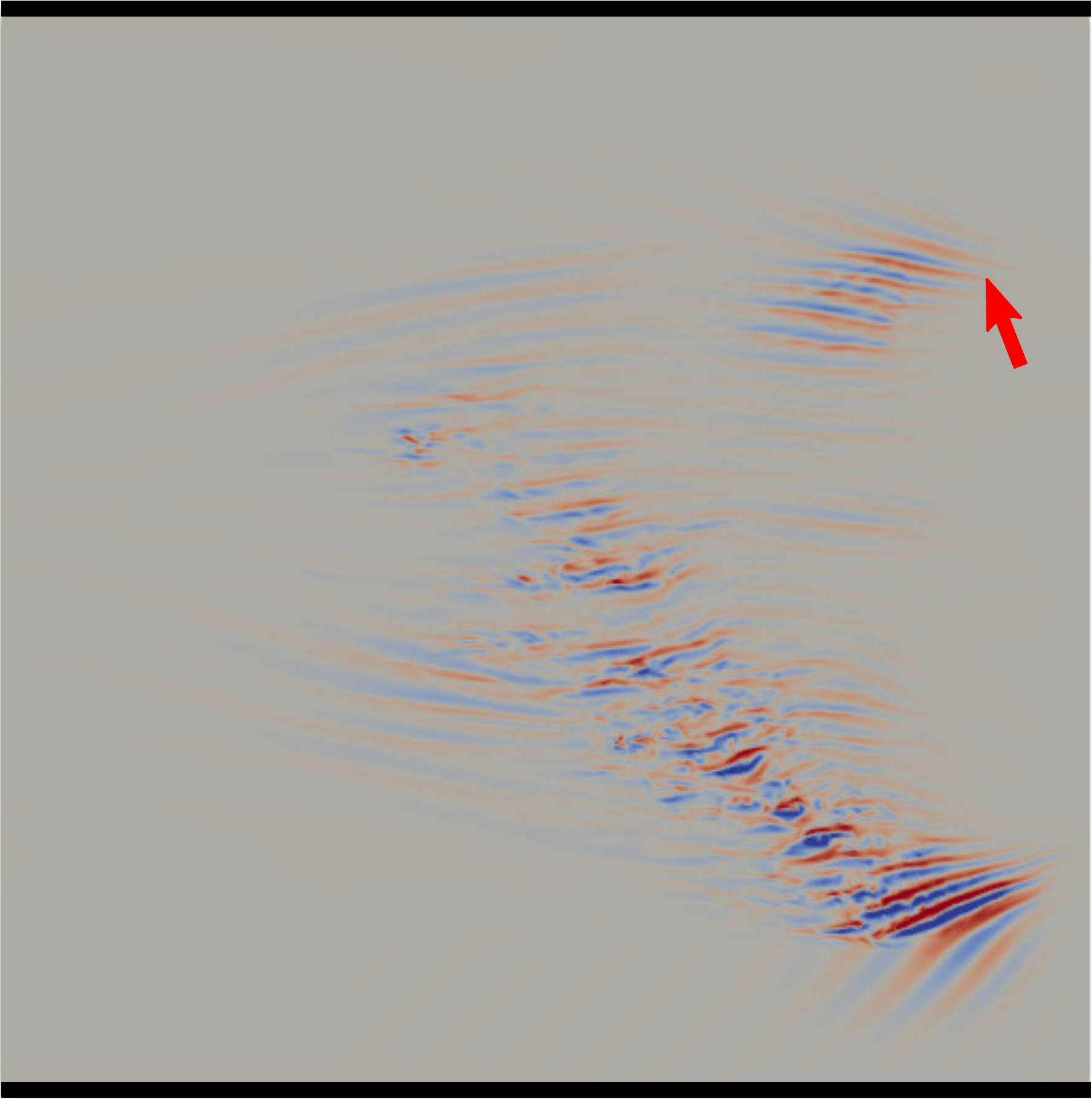}}
	\end{minipage}
	\begin{minipage}[h]{0.22\linewidth}
	\centering
	\subfigure[t=820]{\includegraphics[width = 0.995\textwidth]{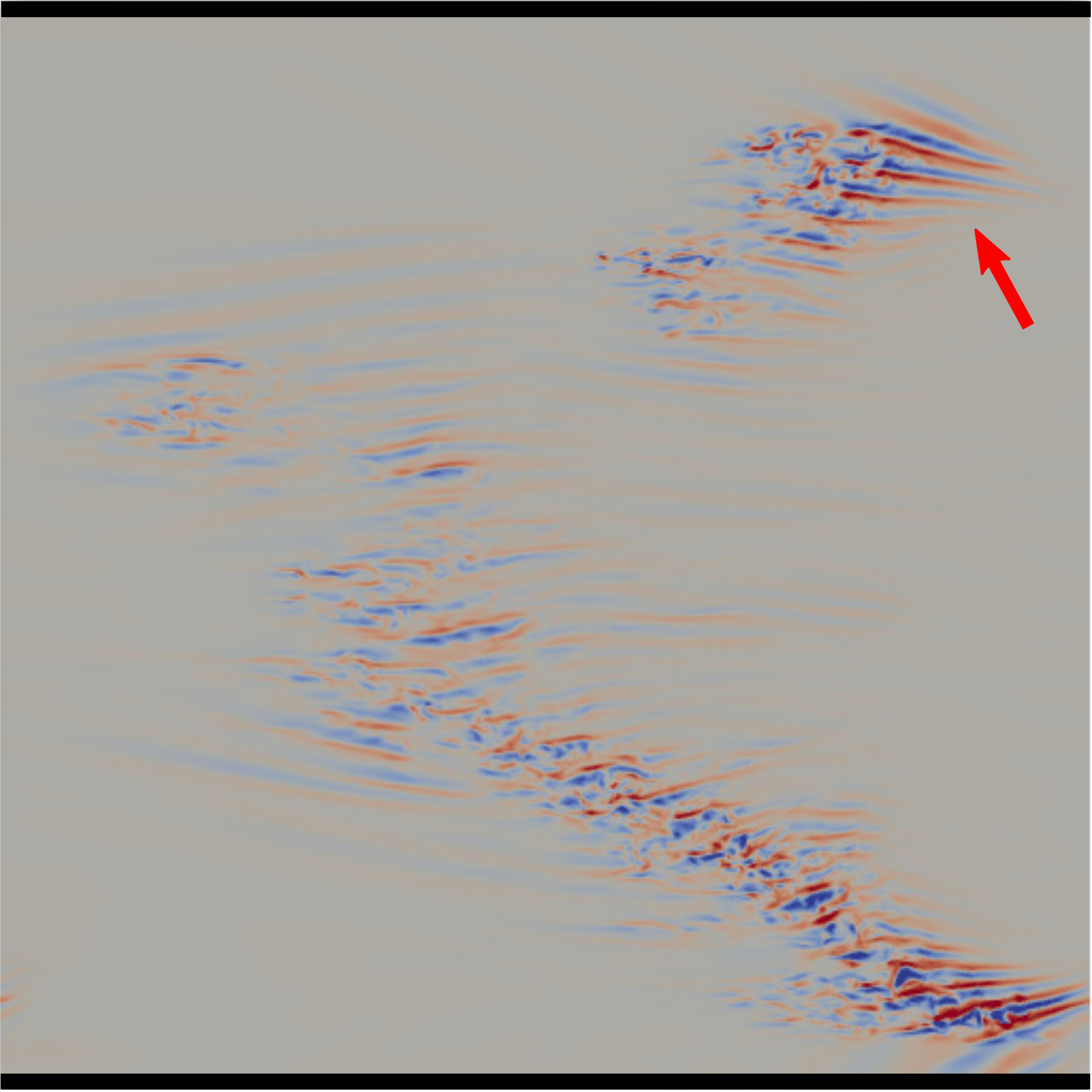}}
	\end{minipage}
	\begin{minipage}[h]{0.22\linewidth}
	\centering
	\subfigure[t=1600]{\includegraphics[width = 0.995\textwidth]{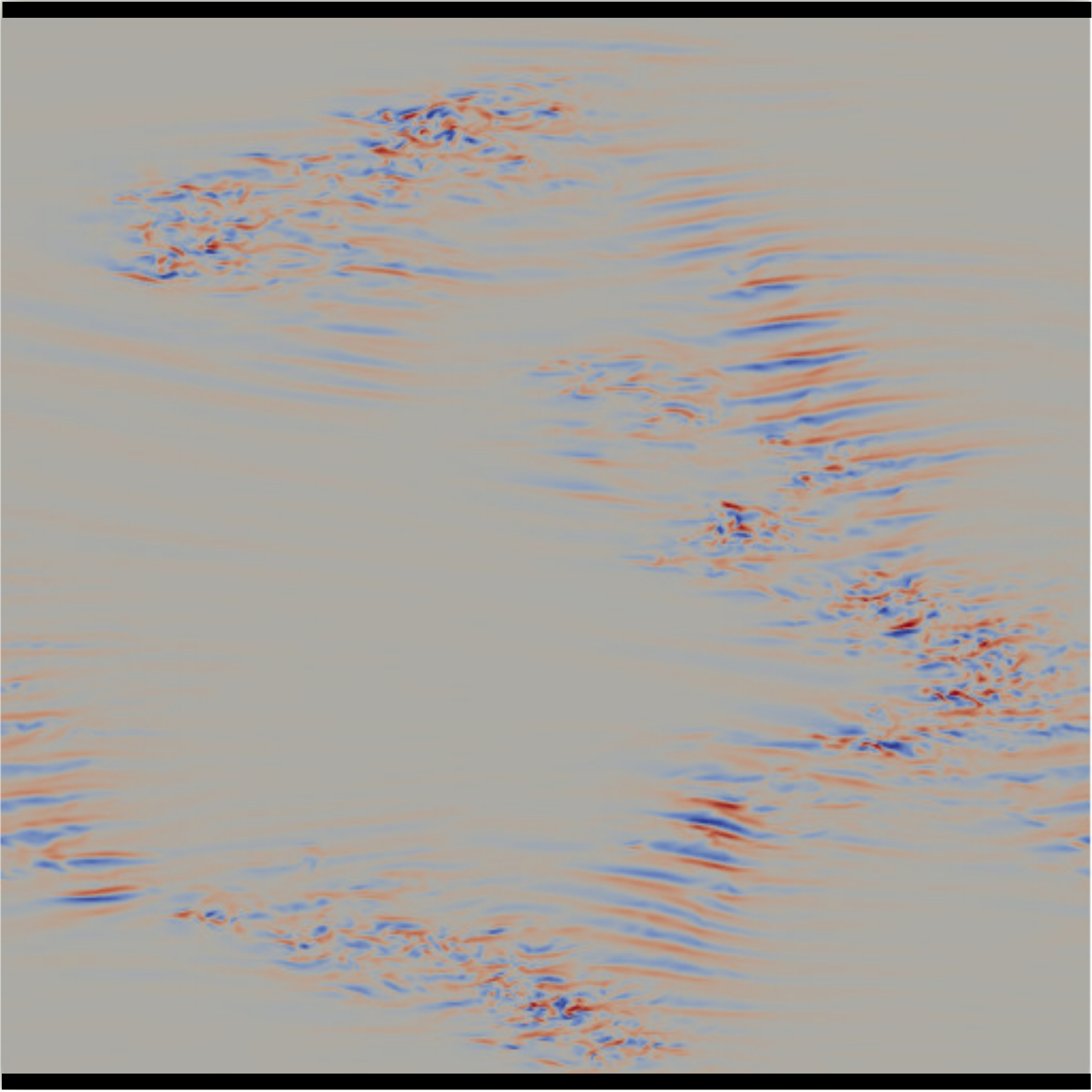}}
	\end{minipage}
	\begin{minipage}[h]{0.22\linewidth}
	\centering
	\subfigure[t=3490]{\includegraphics[width = 0.995\textwidth]{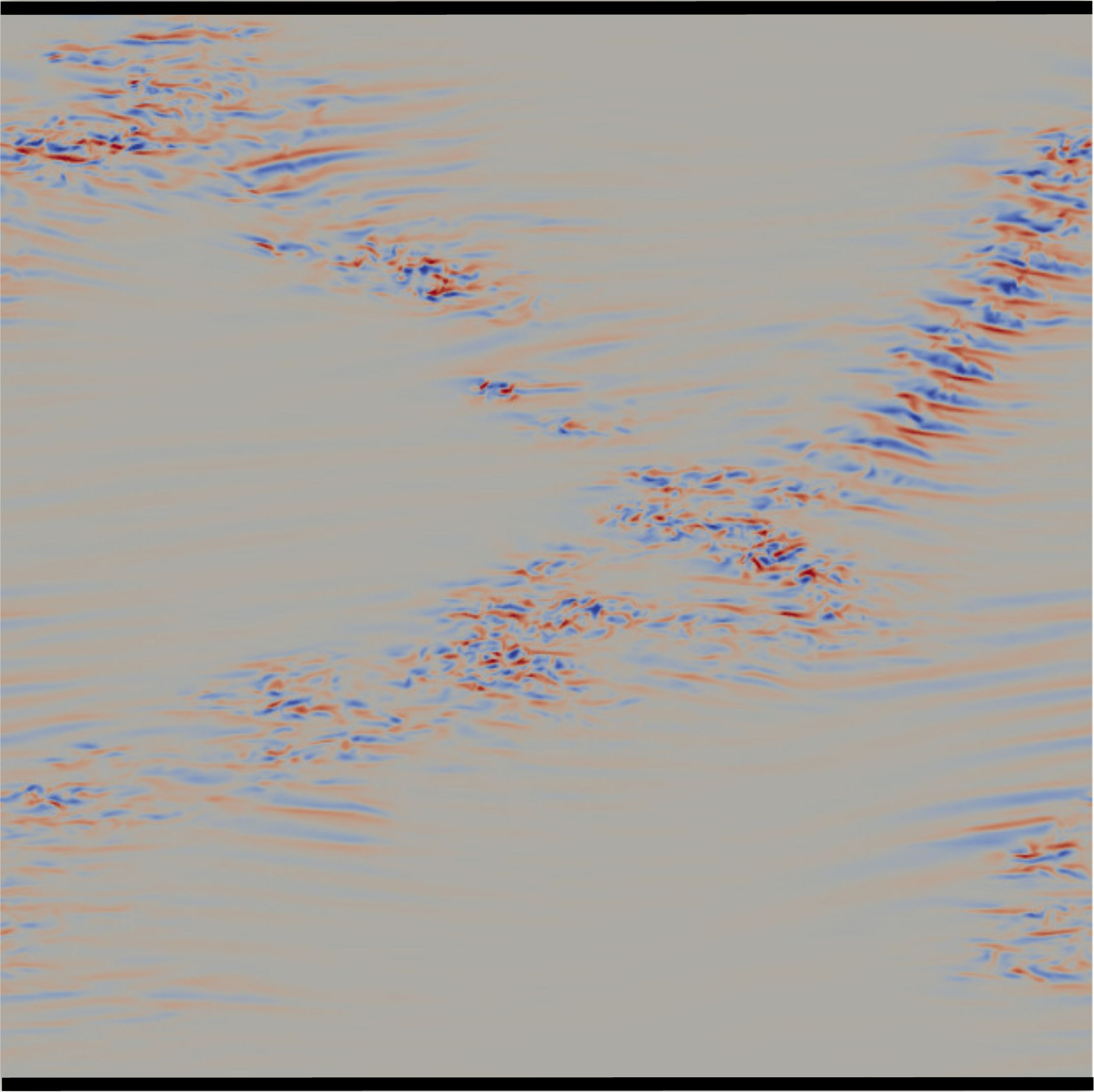}}
	\end{minipage}
	\begin{minipage}[]{0.22\linewidth}
	\subfigure[Re=1050 t=670]{\includegraphics[width = 0.995\textwidth]{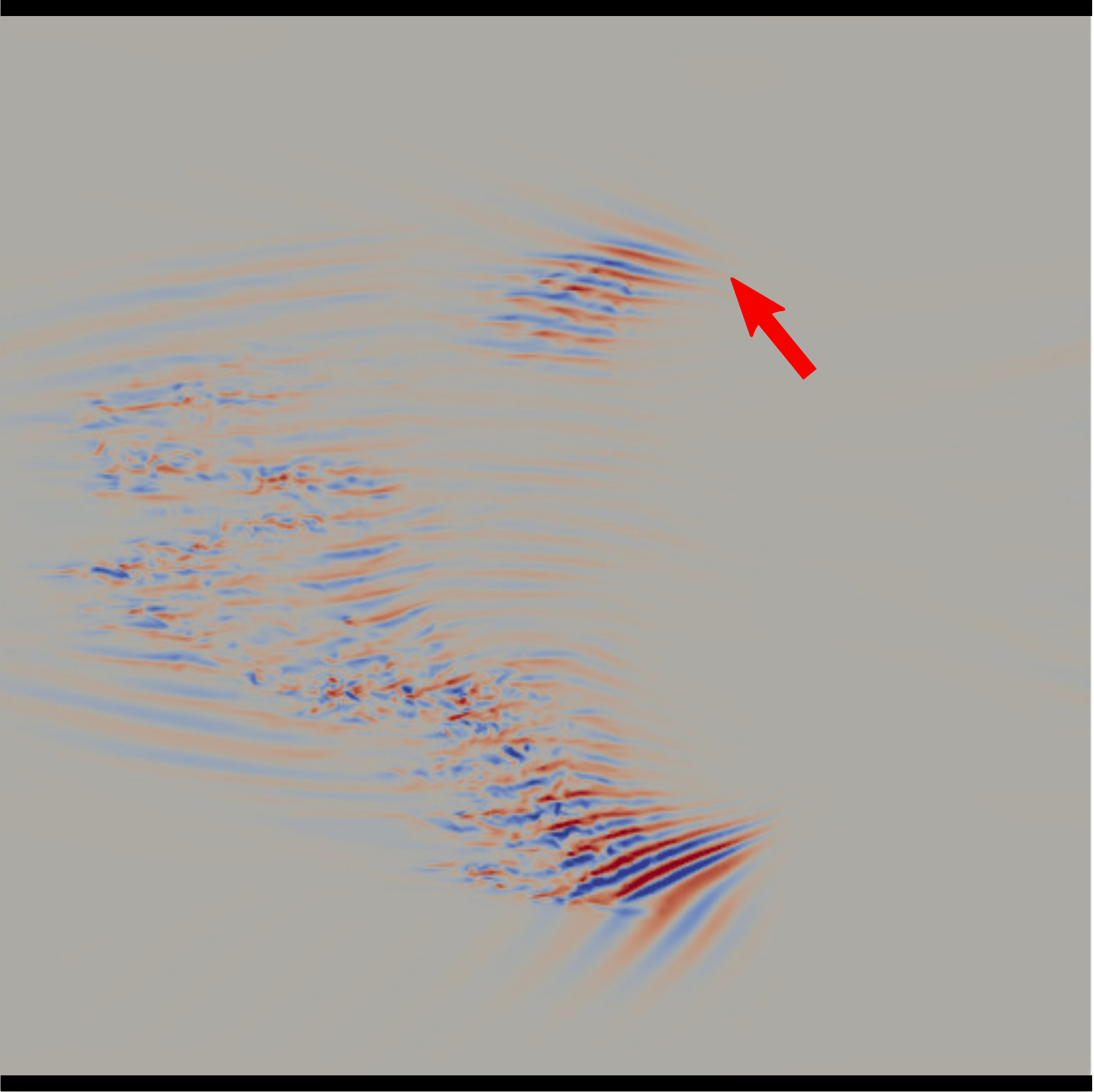}}
	\centering
	\end{minipage}
	\begin{minipage}[h]{0.22\linewidth}
	\centering
	\subfigure[t=760]{\includegraphics[width = 0.995\textwidth]{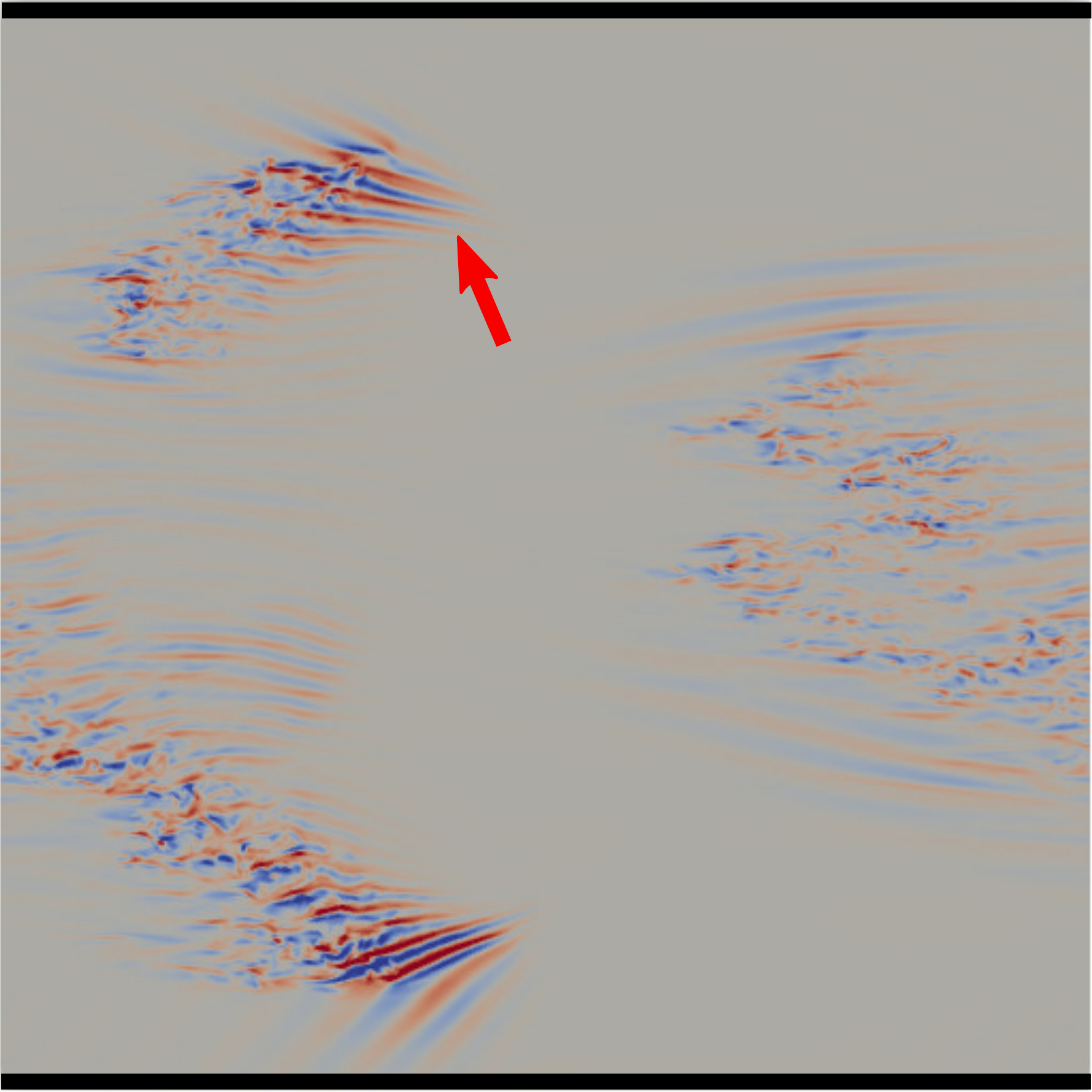}}
	\end{minipage}
	\begin{minipage}[h]{0.22\linewidth}
	\centering
	\subfigure[t=1300]{\includegraphics[width = 0.995\textwidth]{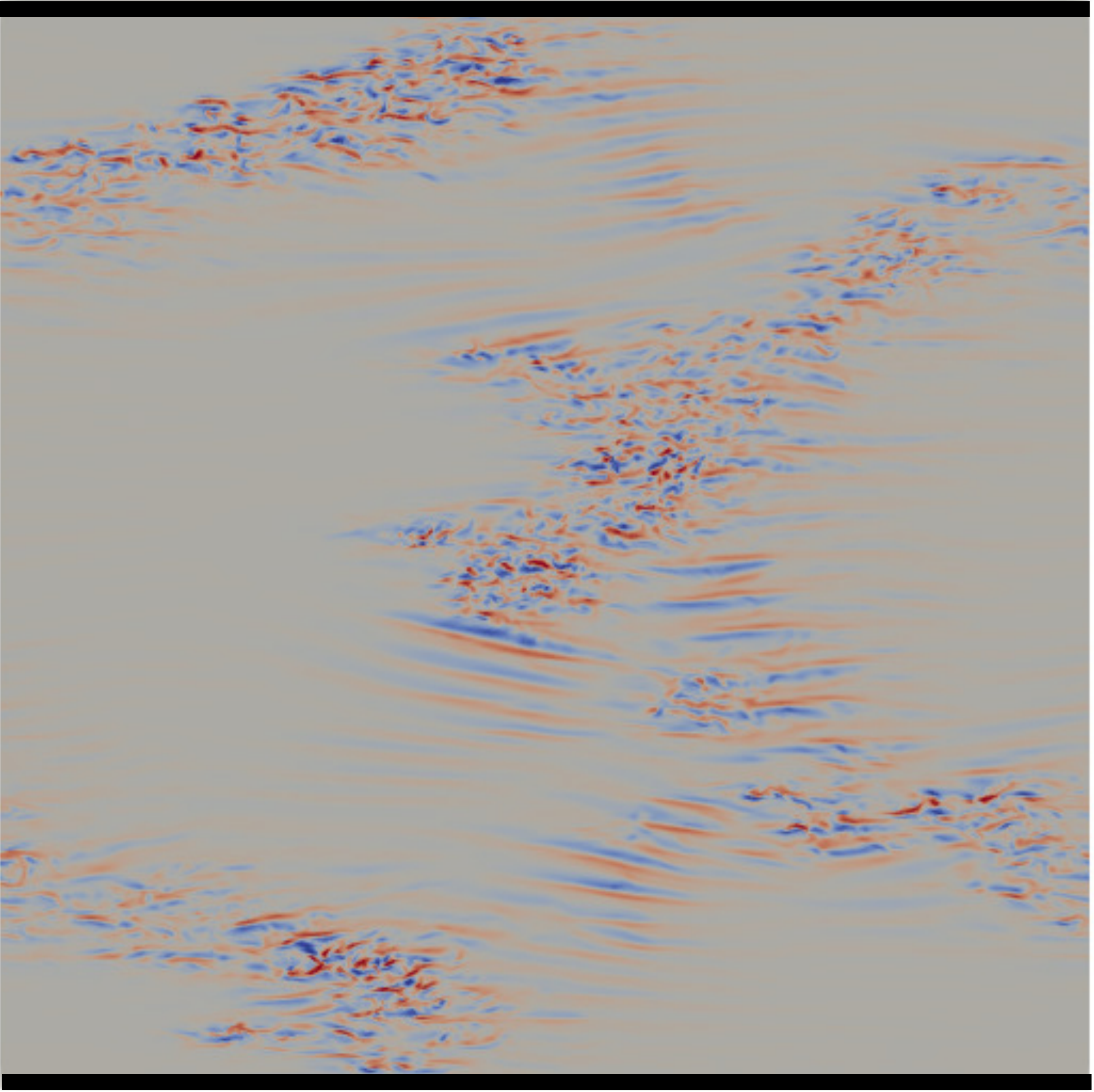}}
	\end{minipage}
	\begin{minipage}[h]{0.22\linewidth}
	\centering
	\subfigure[t=4502]{\includegraphics[width = 0.995\textwidth]{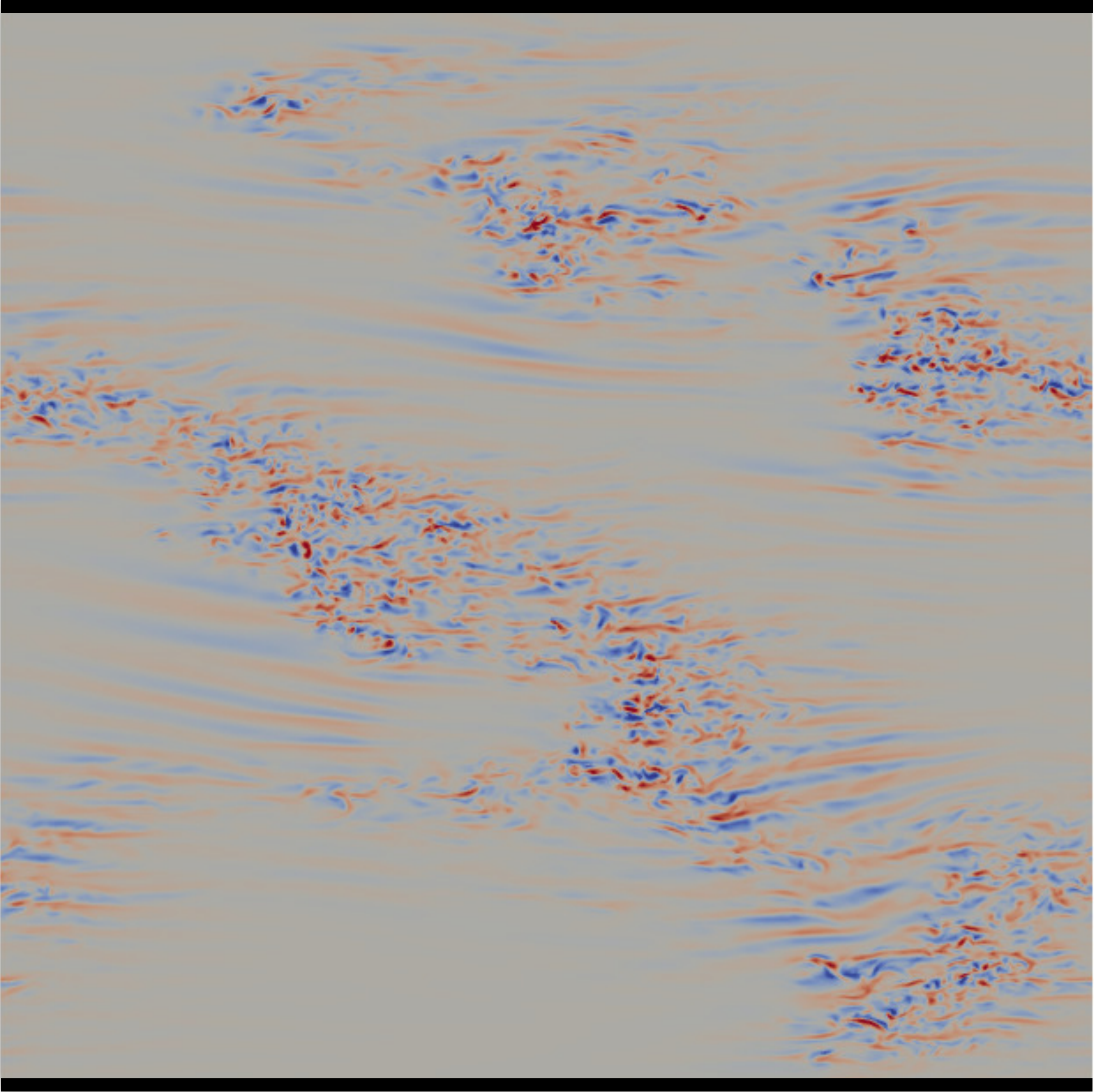}}
	\end{minipage}
	\begin{minipage}[h]{0.4\linewidth}
	\centering
	\subfigure{\includegraphics[width = 0.995\textwidth]{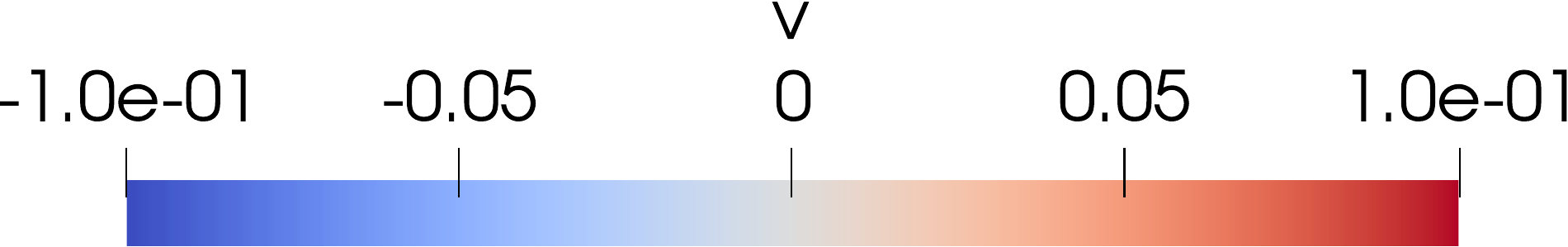}}
	\end{minipage}
	\caption{The collision between the turbulent band with channel side walls at Re = 950 (a-d), 975 (e-h), 1000 (i-l) and 1050 (m-p). Contours of the wall normal velocity is plotted. {\color{black}{The nucleated turbulent bands are pointed to by the red arrows}}.}
	\label{fig:turbulent950-1050}
\end{figure}

Figure \ref{fig:turbulent950-1050} shows the development of the flow at these Reynolds numbers. At $Re=950$ and 975, it can be seen that the original band attempts to nucleate a band (pointed to by the red arrows) with the opposite orientation at the upstream end only after a few hundred time units. This is consistent with the observations at similar Reynolds numbers in prior studies that turbulent bands frequently split and branch \cite{Paranjape2019,Shimizu2019}. The nucleated band at $Re=950$ has not well developed before hitting the wall, see panel (b) at $t=940$, but the nucleated band has well developed at $Re=975$, see panel (e) at $t=820$. Both the original bands and the nucleated bands completely decay after the collision with the side wall, see panel (d, h), and the flow cannot become turbulent anymore without introducing strong perturbations due to the sub-criticality of the transition. The results suggest that, once the downstream end is eliminated by the collision, the remaining part of the band starts to decay, indicating the importance of the turbulence-generation mechanism at the downstream end to the entire band. This phenomena is similar to the collision between turbulent bands at lower Reynolds numbers reported by \cite{Shimizu2019,Song2020}, in which the authors showed that once the downstream end of a band is destroyed by the collision, the remaining part of the band also decays. Therefore, it seems that turbulent bands below $Re=975$ share the same self-sustaining mechanism (although not shown in this figure, the band also decayed at $Re=750$ after the collision).

At $Re=1000$ and 1050, we can see that the original band quickly branch and the two nucleated bands have well-developed before they reach the side walls. After the collision with the side walls, the remaining part of the band close to the side wall continues to exist until the end of our simulations ({\color{black}{roughly 2700 time units at $Re=1000$ and 3800 time units at $Re=1050$ after the collision}}). Nevertheless, the remaining turbulence does not fill the whole domain in the channel but forms a net-like turbulence pattern interspersed with laminar flow region. This is consistent with the flow pattern reported by \cite{Paranjape2019, Shimizu2019, Duguet2020} for similar Reynolds numbers in plane Poiseuille flow. Note that the remaining turbulence away from the side walls does not create new wave-like downstream ends after the net-like pattern is formed because there is no sufficiently large laminar gap between turbulent regions for the downstream end to form, as pointed out by \cite{Shimizu2019}. This result indicates that, at sufficiently high Reynolds numbers, (parts of) a turbulent band can be locally sustained even if the downstream end is absent. 
 
\begin{figure}
\centering
\includegraphics[width = 0.8\textwidth]{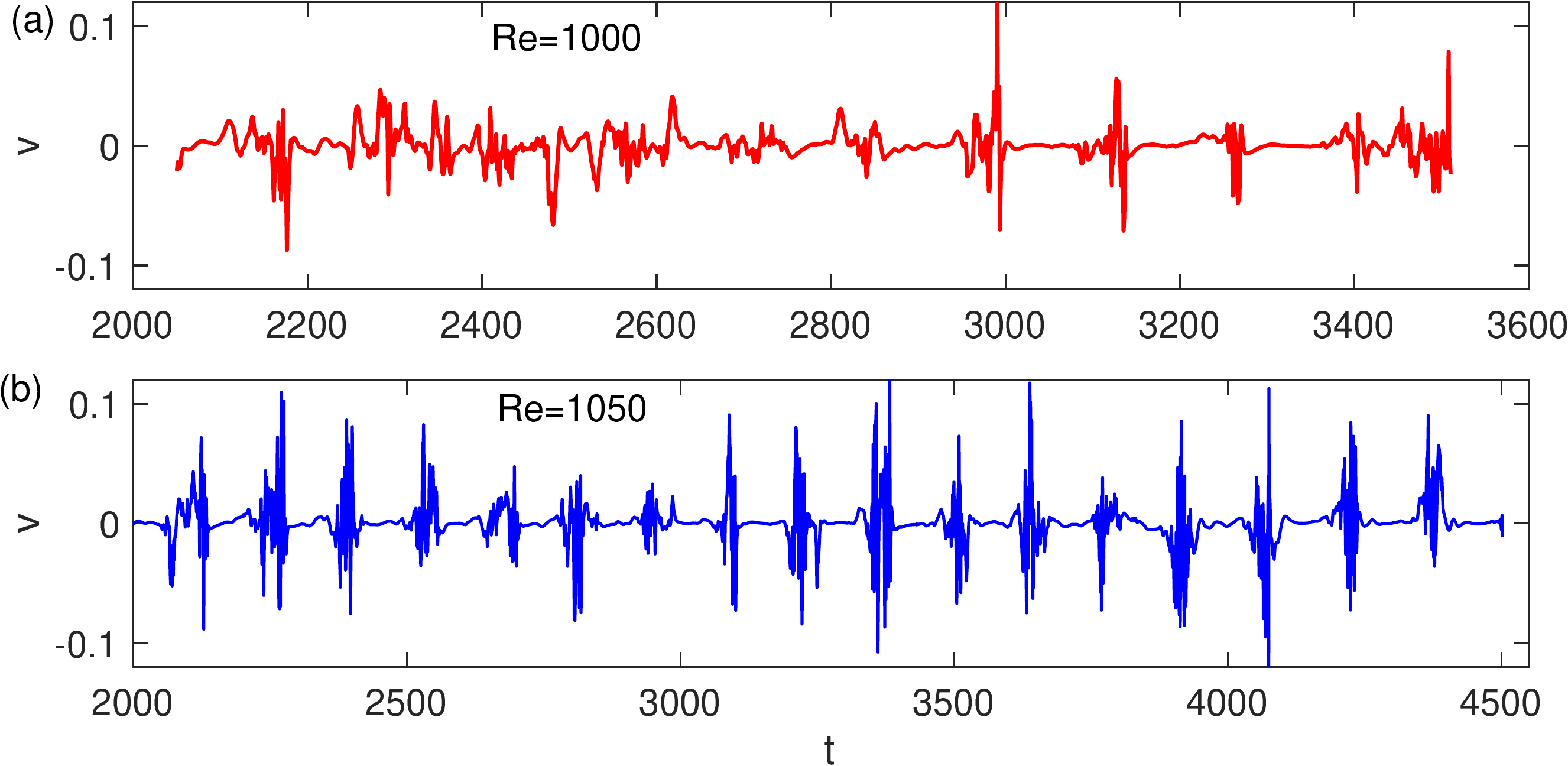}
\caption{\label{fig:monitor_point} The wall-normal velocity monitored at a point close to the side wall at $z=-50$. The streamwise location is arbitrarily set at $x=90$ and the spanwise location is at $z=-40$. This is to show the passage of turbulence close to the side wall. }
\end{figure}

{\color{black}{In order to show more details about the survival of turbulence close to the side wall after the collision at $Re=1000$ and 1050, Figure \ref{fig:monitor_point} shows the wall-normal velocity monitored at $(x,z)=(90, -40)$, a point close to the side wall at $z=-50$. The signal for $Re=1000$ shows that the turbulence may temporarily form an extended turbulent region rather than remain as a distinct banded structure, see the signal between about 2200 and 2700. Nevertheless, the turbulence didn't completely decay, otherwise there would be a large segment of the curve showing vanishing wall-normal velocity. After $t=2800$, the signal suggests that a turbulence patch started passing by nearly periodically, and the time separation between two passes is roughly 150 time units. This suggests that the turbulent patch survived and advected downstream at a streamwise speed of approximately 0.67, which is very close to the streamwise advection speed of velocity streaks in the bulk region of a turbulent band measured by \cite{Xiao2020b} (the authors reported 0.65 at $Re=950$ and 0.63 at $Re=1050$). At $Re=1050$, we observed a periodic passage of turbulence patch in the whole monitoring time window between $t=2000$ and 4500, and the period of passage is nearly the same as that between $t=2800$ and 3400 in the $Re=1000$ case. This also suggests that a turbulent patch survived near the side wall and advected downstream at a nearly constant speed. It should be noted that the specific development of the flow is initial condition dependent, and the signals shown in figure \ref{fig:monitor_point} may not be a general picture. Nevertheless, these two simulations indicate that turbulence could survive (at least for thousands of time units) close to the side wall after the collision above $Re\simeq 1000$.}} 

In summary, our results suggest that the critical Reynolds number $Re_{cr}$ should roughly sit between $Re=975$ and 1000. Below $Re_{cr}$, a turbulent band tends to decay after colliding with the side wall, as the band needs an active downstream end to sustain itself. In other words, a turbulent band can only be sustained as a whole. Whereas above $Re_{cr}$, a turbulent band can be sustained even without an active downstream end, suggesting that turbulence becomes locally sustained. {\color{black}{However, the turbulence may not be able to keep a distinct banded strucutre but form turbulent patches and even more extended turbulent regions, see figure \ref{fig:turbulent950-1050} (k, l, o, p).}}
{\color{black}{ This $Re_{cr}$ seems to be close to the minimum $Re$ for transiently sustained localized turbulence in much narrower channels (aspect ratio up to 9) \cite{Takeishi2015}. Our simulation time is much longer than that in \cite{Takeishi2015}, which is about 400 time units in our normalization. As the authors pointed out, a single localized turbulent patch is of a transient nature and is destined to decay after its finite lifetime, but our simulation times at $Re=1000$ and 1050 are one order of magnitude longer than the period of the self-sustaining cycle assessed in a minimum flow unit for Poiseuille flow at a modestly higher $Re$, which is about 140 time units \cite{Jimenez1991}. Therefore, the turbulence close to the side walls at $Re=1000$ and 1050 in our simulations can be considered as sustained, and the possible decay at longer times should be attributed to the intrinsic transient nature of localized turbulence rather than the loss of the downstream end of the band.}} Crossing $Re_{cr}$, the self-sustaining mechanism of a turbulent band seems to change qualitatively.

\subsection{Collision details}
\begin{figure}[h]
	\subfigbottomskip = 2pt
	\subfigcapskip=-5pt
	\begin{minipage}[h]{0.22\linewidth}
	\centering
	\subfigure[Re=750 t=780.5]{\includegraphics[width = 1\textwidth]{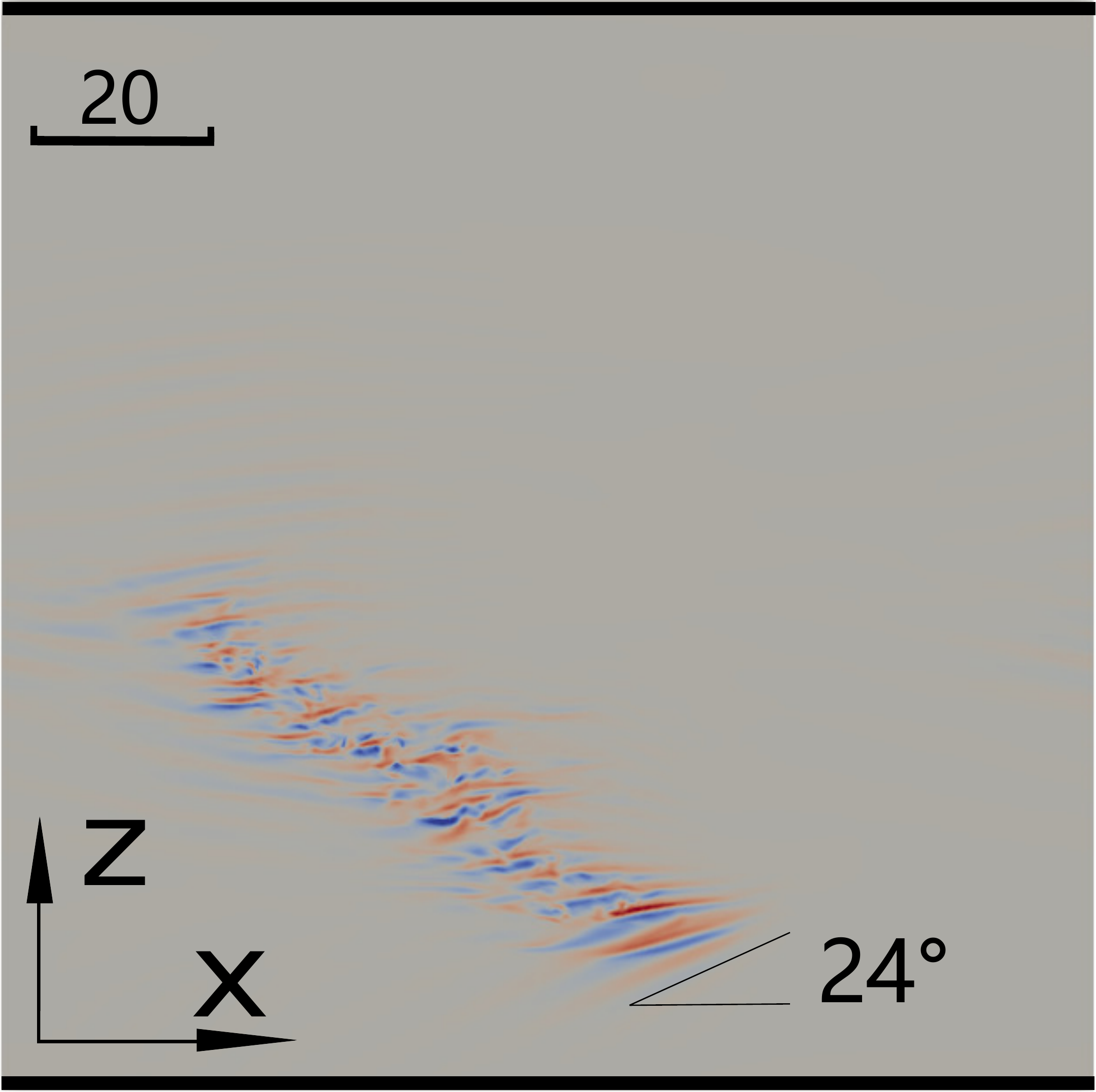}}
	\end{minipage}
	\quad
	\begin{minipage}[h]{0.22\linewidth}
	\centering
	\subfigure[t=825.5] {\includegraphics[width = 1\textwidth]{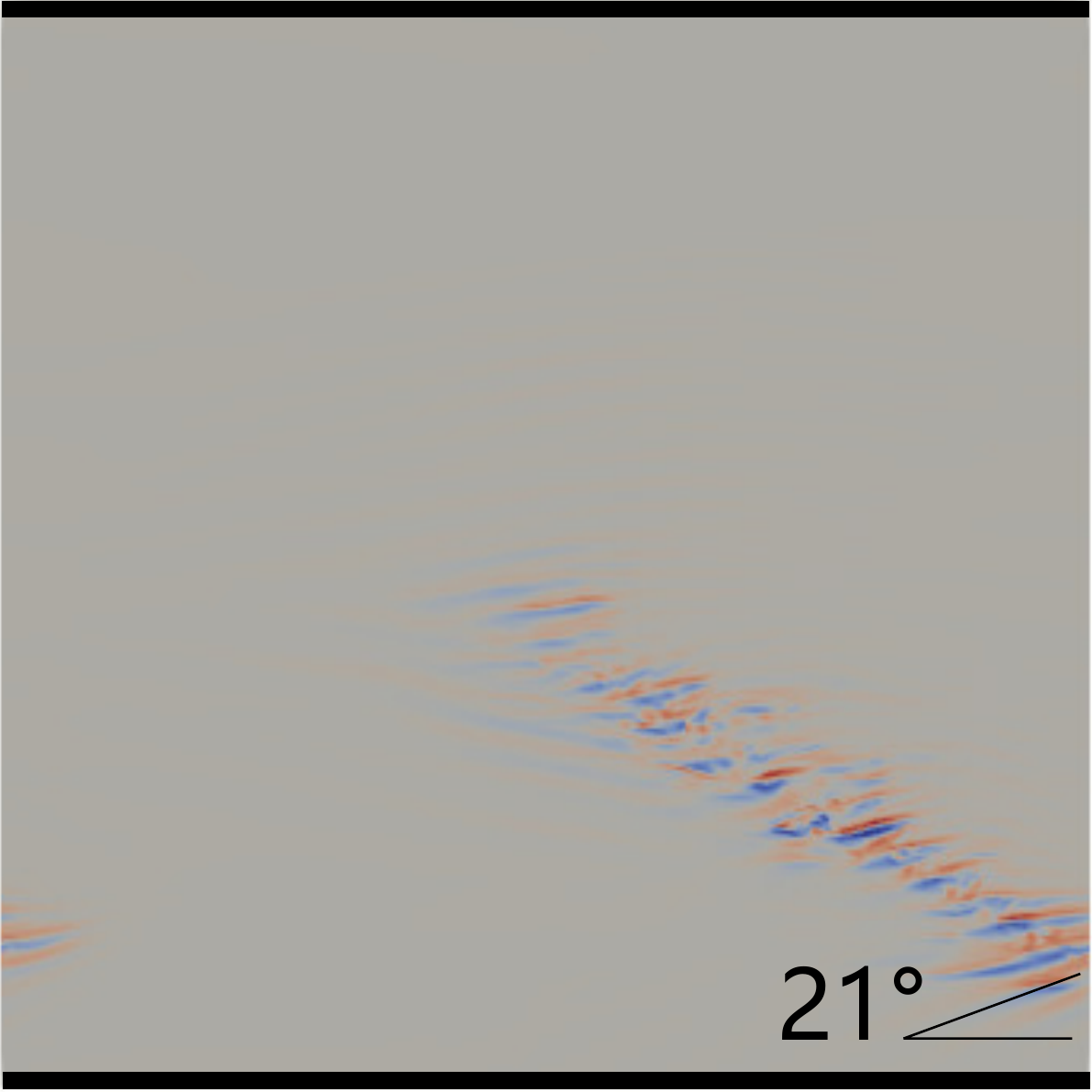}}
	\end{minipage}
	\quad
	\begin{minipage}[h]{0.22\linewidth}
	\centering
	\subfigure[t=870.5] {\includegraphics[width = 1\textwidth]{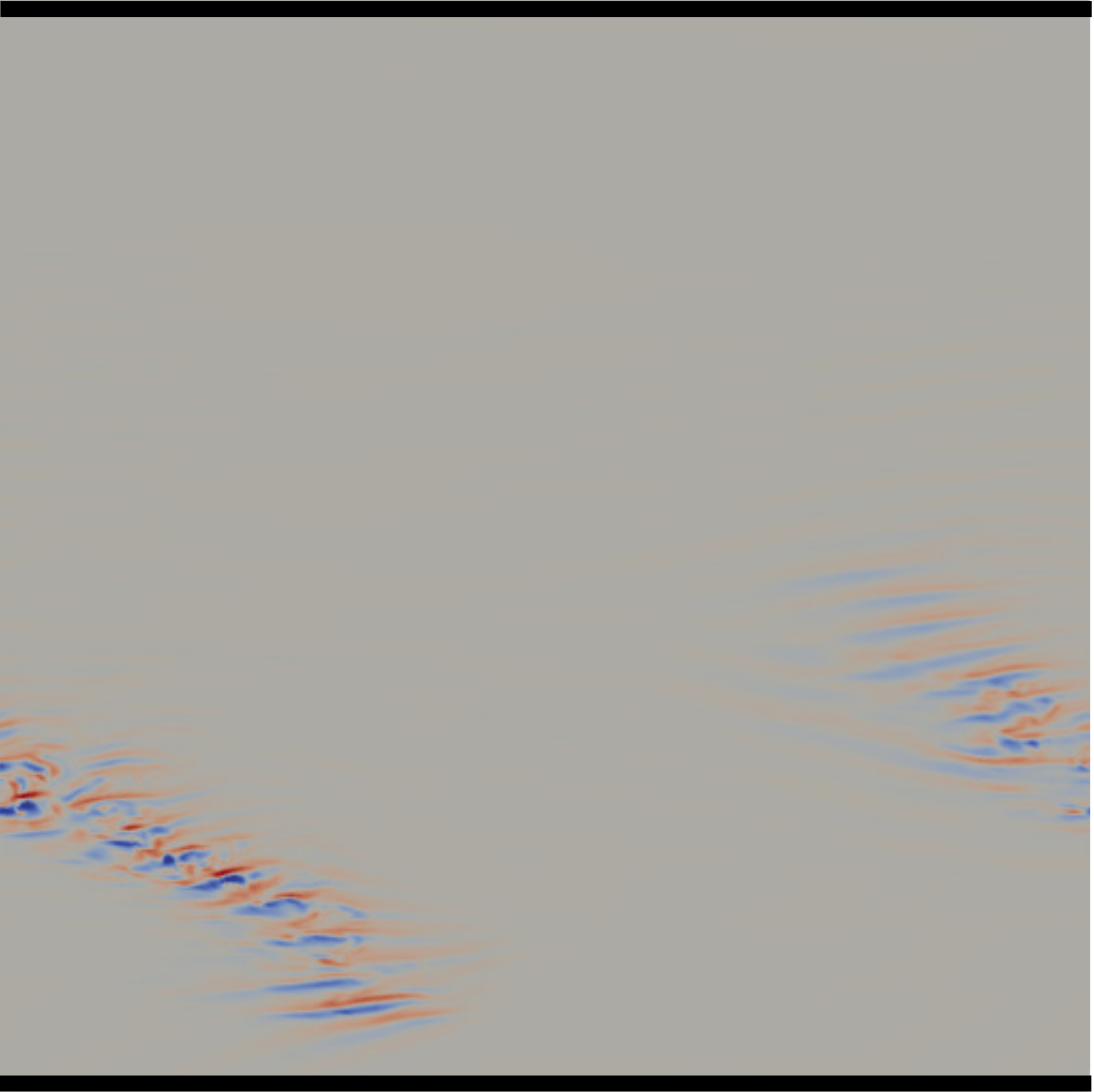}}
	\end{minipage}
	\quad
	\begin{minipage}[h]{0.22\linewidth}	
	\centering
	\subfigure[t=900.5] {\includegraphics[width = 1\textwidth]{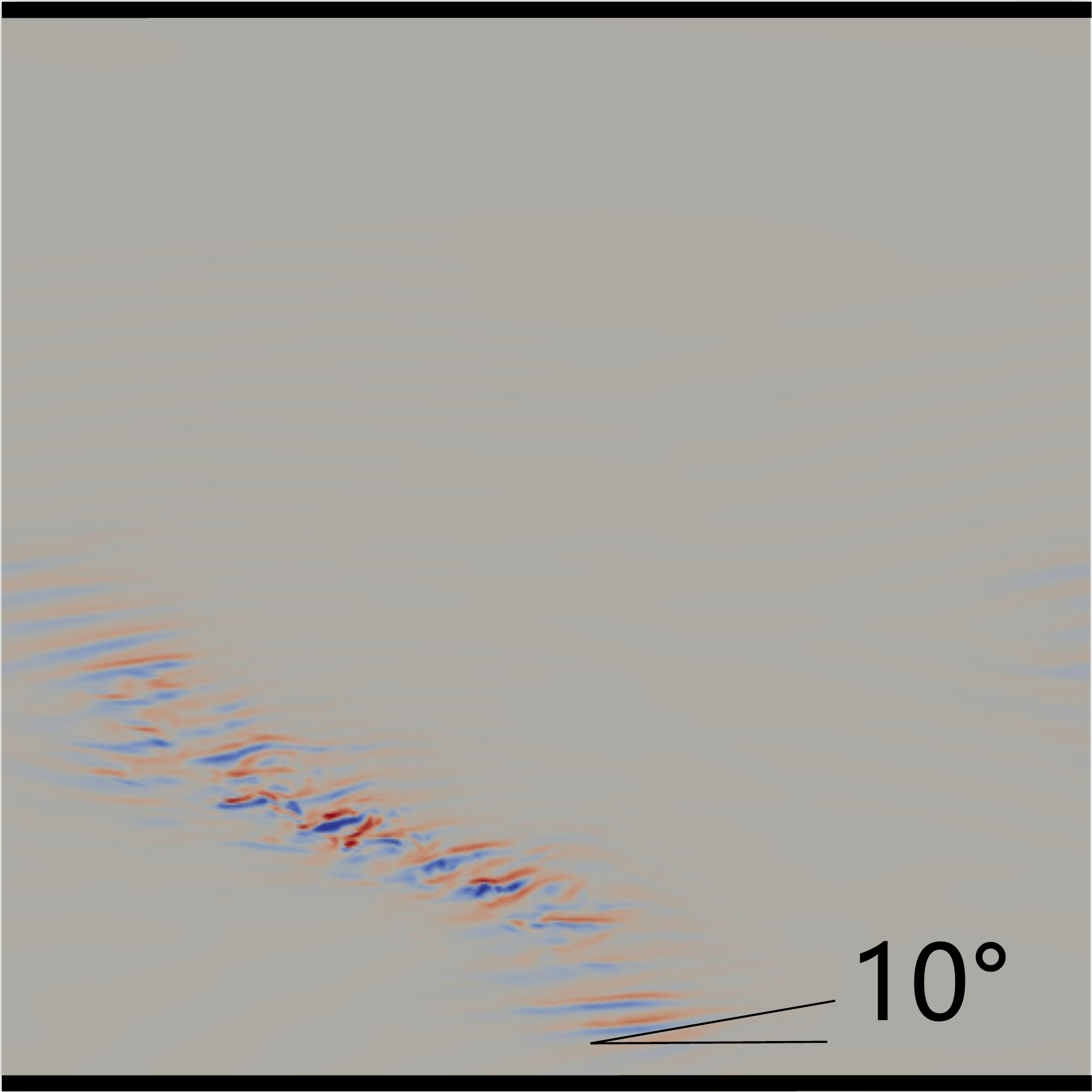}}
	\end{minipage}
	\begin{minipage}[h]{0.22\linewidth}
	\centering
	\subfigure[Re=1050 t=700]{\includegraphics[width = 1\textwidth]{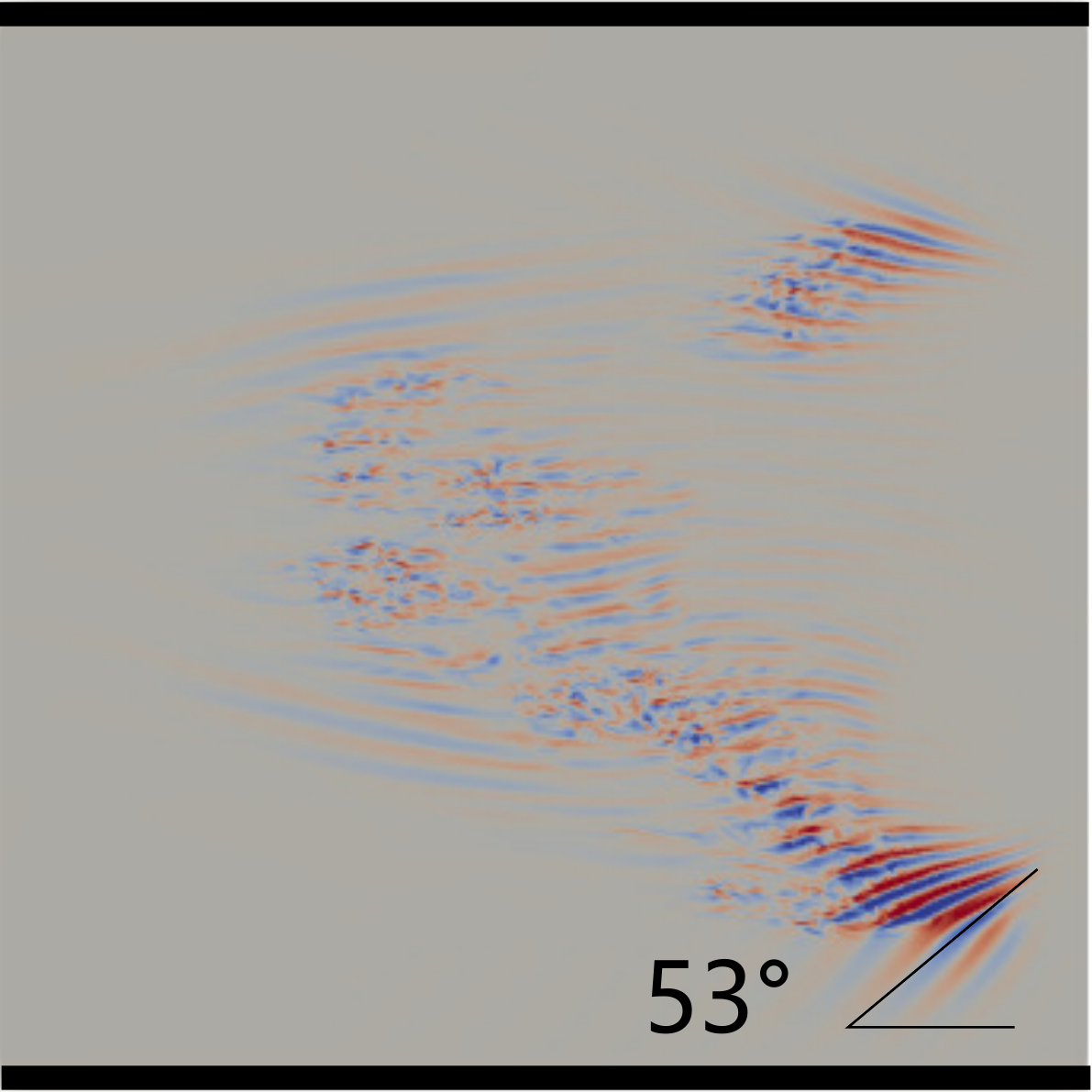}}
	\end{minipage}
	\quad
	\begin{minipage}[h]{0.22\linewidth}
	\centering
	\subfigure[t=745] {\includegraphics[width = 1\textwidth]{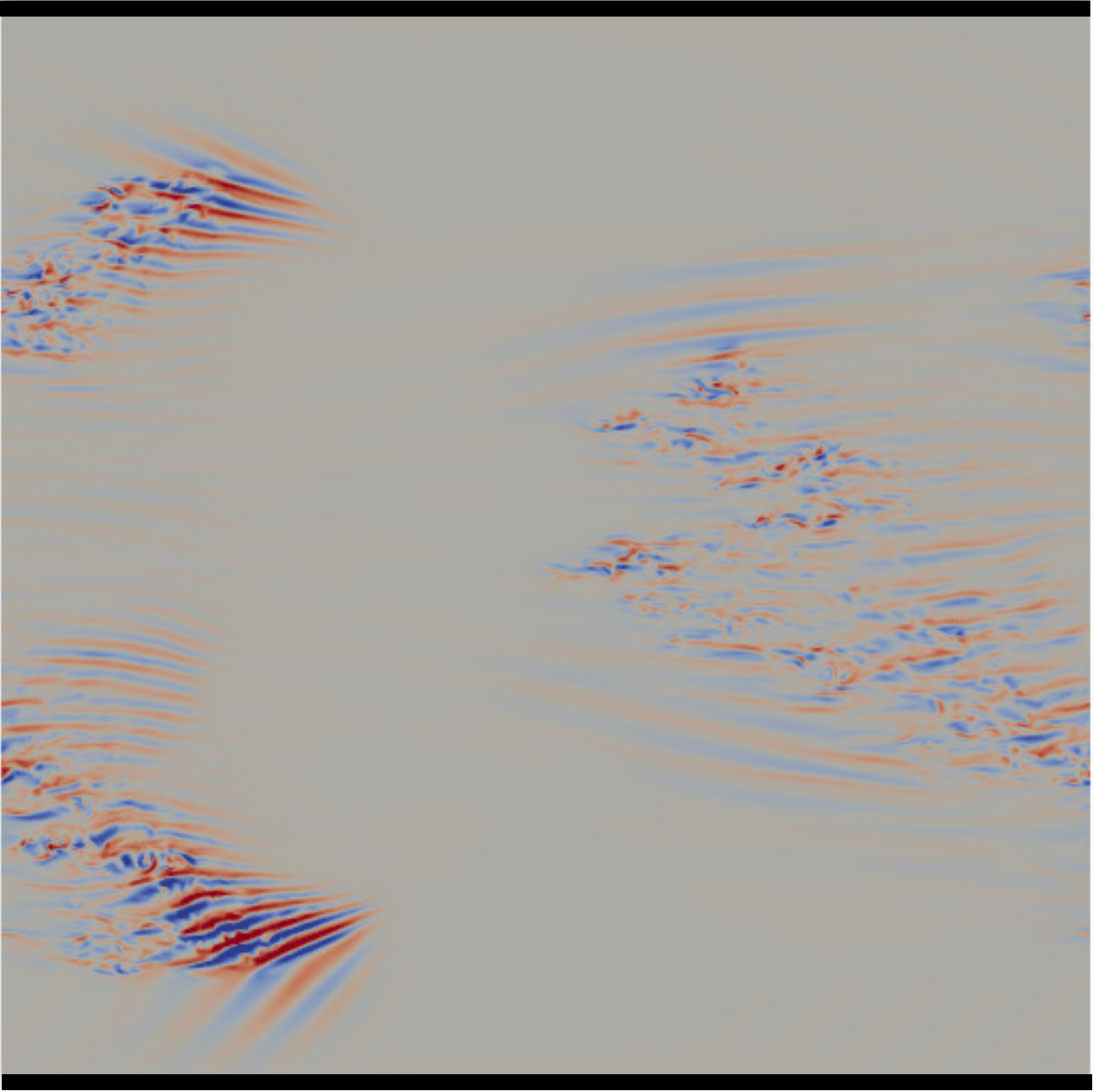}}
	\end{minipage}
	\quad
	\begin{minipage}[h]{0.22\linewidth}
	\centering
	\subfigure[t=790] {\includegraphics[width = 1\textwidth]{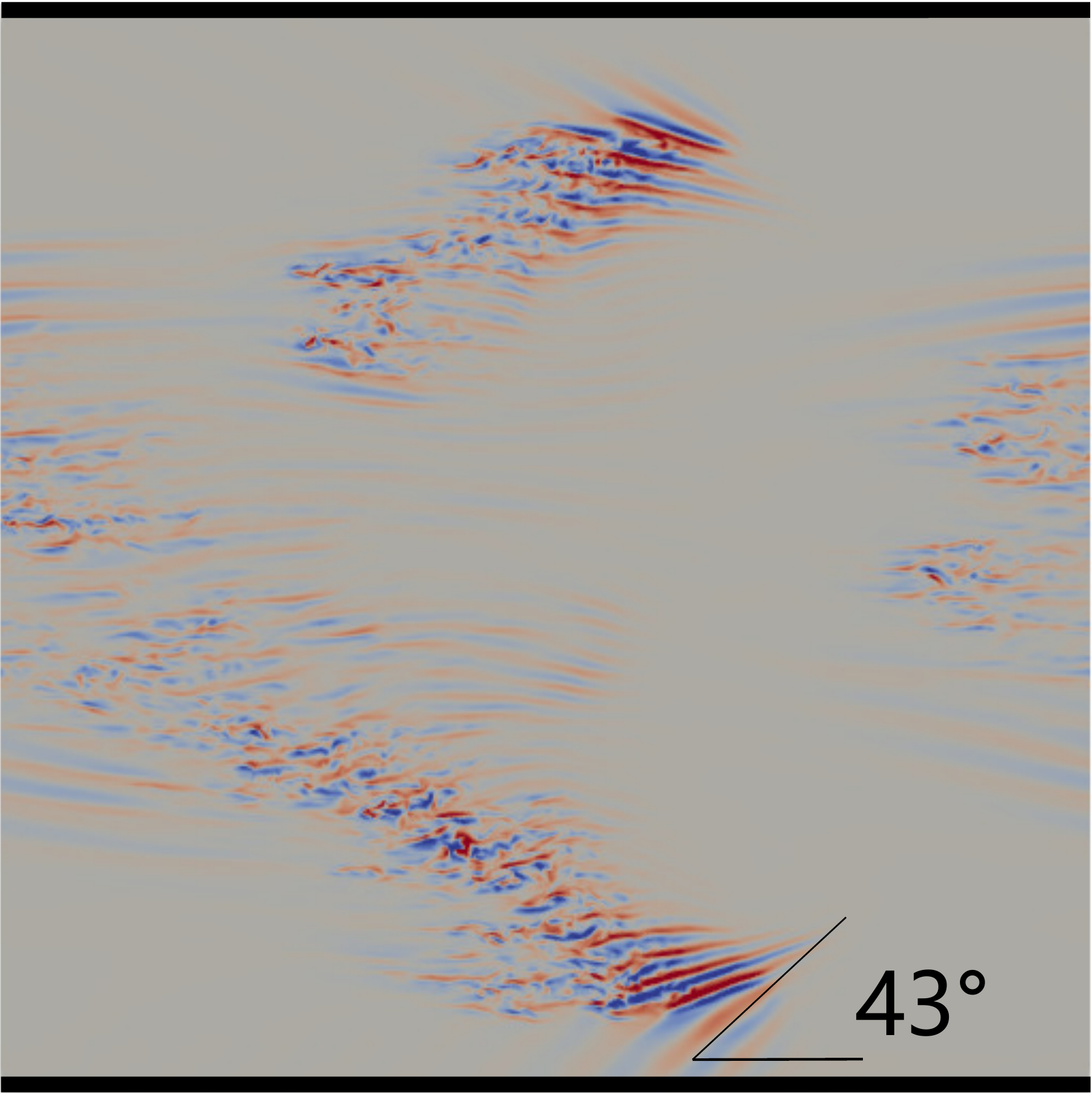}}
	\end{minipage}
	\quad
	\begin{minipage}[h]{0.22\linewidth}	
	\centering
	\subfigure[t=835] {\includegraphics[width = 1\textwidth]{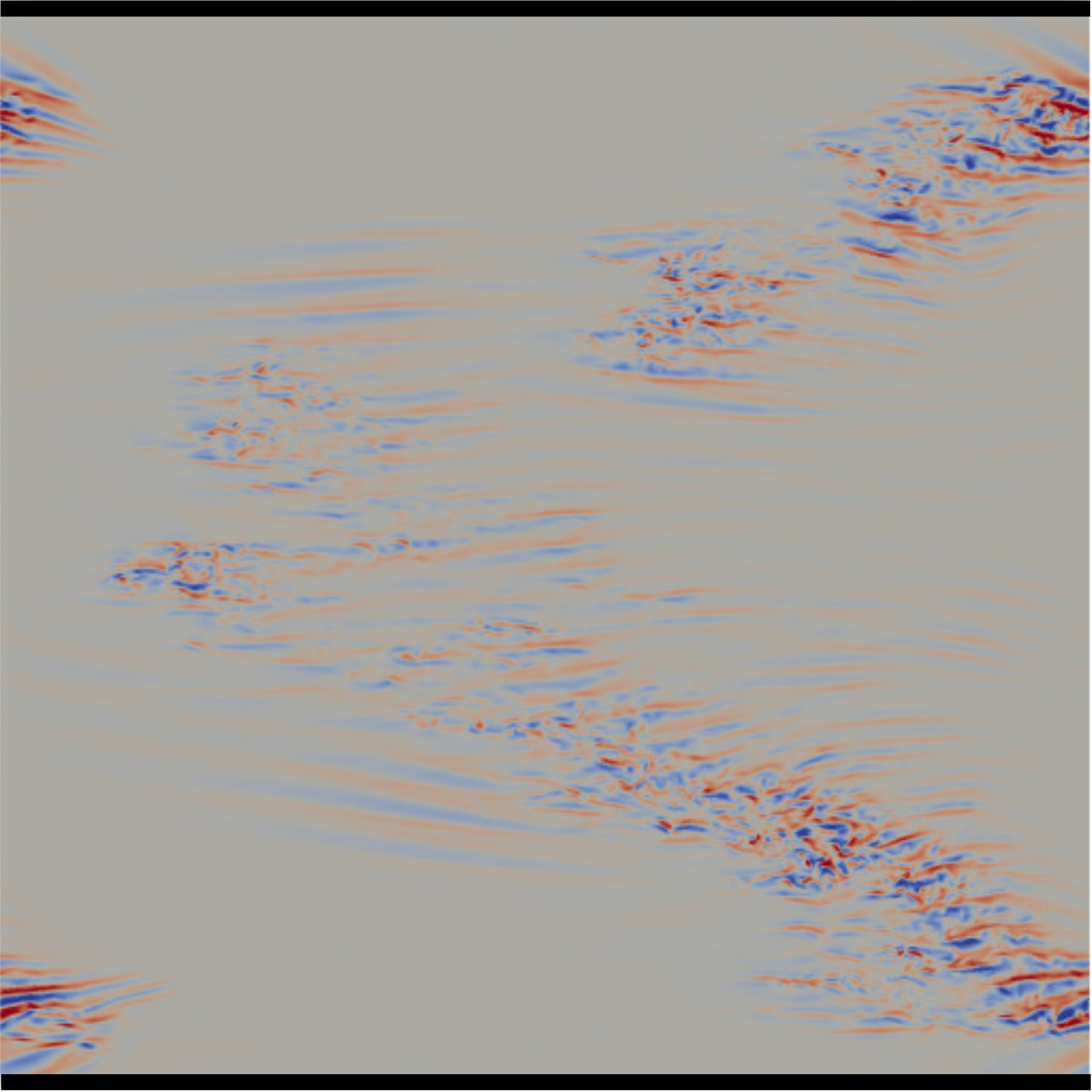}}
	\end{minipage}
	\begin{minipage}[h]{0.4\linewidth}
	\centering
	\subfigure{\includegraphics[width = 0.995\textwidth]{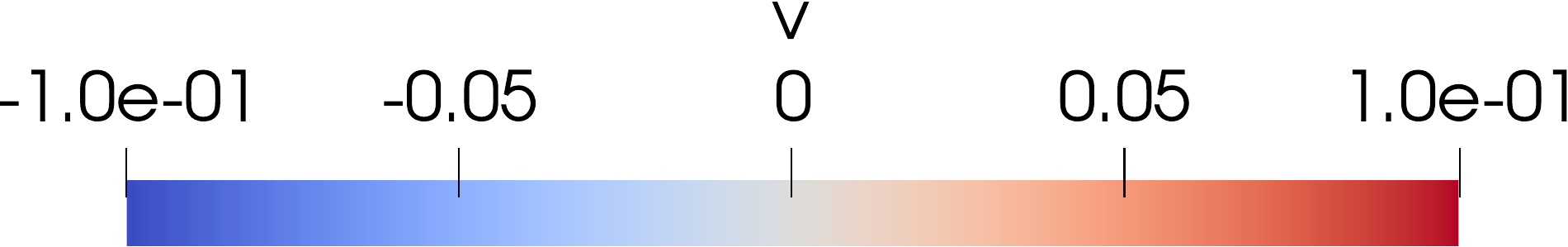}}
	\end{minipage}
	\caption{Detailed collision process at Re = 750 (a-d) and Re = 1050 (e-h). The angles shown in the figure are approximate tilt angles of the high-speed streaks at the downstream end of the turbulent bands.}
\label{fig:750-1050-wall}
\end{figure}

We selected $Re = 750$ and 1050 to show the details of the collision process, see figure \ref{fig:750-1050-wall}. At $Re=750$, the most obvious change when the downstream end hits the side wall is the reduction in the tilt angle of the wave-like structures (vortices and streaks) about the streamwise direction. 
Ref. \cite{Xiao2020} reported that the tilt angle at $Re=750$ is approximately $38^\circ$ in plane Poiseuille flow. 
Here, while the band approaches the side wall, it can be seen that the angle is approximately $24^\circ$ in panel (a) and decreases to about $21^\circ$ in panel (b). These angles are already significantly lower than that in the absence of side walls. The angle further decreases to below $10^\circ$ after the collision, see panel (c,d), i.e. the flow structures nearly align with the streamwise direction. At $Re=1050$, the tilt angle of the streaks at the downstream end is much larger than at $Re=750$ (a hint of a stronger instability), and although the tilt angle also decreases while colliding with the side wall, the value of the angle remains relatively large (from $53^\circ$ to $43^\circ$).

Another noticeable difference between the two Reynolds numbers can be observed. At $Re=750$, after the downstream end collides with the wall, the reduction of the tilt angle also spreads away from the side wall into the bulk region of the turbulent band. With the reduction of the tilt angle come reduced turbulent activities (see figure \ref{fig:750-1050-wall}(c, d)). This suggests that after losing the active downstream end, the flow structures inside the bands cannot maintain the tilt angle and the turbulence intensity, which eventually leads to the gradual decay of the whole band. On the contrary, at $Re=1050$, although the wave-like downstream end disappears after the collision, no significant change in the flow pattern and turbulence intensity can be observed in the remaining parts of the band (even close to the side wall). These are further evidences for that a turbulent band can only sustain itself as a whole at low Reynolds numbers, requiring an active turbulence-generating downstream end, whereas could be locally sustained inside the turbulent band at higher Reynolds numbers.
\section{Discussion}\label{sec:discussion}
\subsection{Mechanism of the decay at low Reynolds numbers}
At relatively low Reynolds numbers, e.g. at $Re=750$, the downstream end starts to feel the side wall at a certain distance, exhibiting decreasing tilt angles and lowered turbulent activity of the flow structures. Such structures can only slowly decay due to viscosity. These changes gradually spread towards the bulk region of the band, leading to the decay of the entire band. The tilt angle about the streamwise direction of the wave-like structures generated at the downstream end is a characteristic of the local instability as proposed by \citep{Xiao2020,Song2020}. 
The authors provided the local mean flow $\bm U$ (i.e. spatially and temporally averaged flow) at the downstream end of a turbulent band, as shown in (\ref{equ:target_profile}$-$\ref{equ:target_profile_2}). They showed that the linear instability of this profile indeed can capture some characteristics of the wave-like structures observed at the downstream end of turbulent bands, and proposed that this instability is responsible for the generation of turbulence. The generation of turbulent bands by forcing such an instability as shown in section \ref{sec:Re750} and in \cite{Song2020} also support this proposed mechanism.

Here we try to more quantitatively illustrate the relationship between the instability and tilt angle using the profile (\ref{equ:target_profile_2}$-$\ref{equ:target_profile}). We modify one of the components to show the change in the instability and the tilt angle of the most unstable wave. The details of the modal instability analysis was given by \cite{Xiao2020}. Table \ref{tab:stability}  shows the change in the tilt angle and growth rate of the most unstable wave resulting from the change in the streamwise velocity $U_x$ and spanwise velocity $U_z$. It can be seen that a stronger instability (a larger growth rate) is associated with a larger tilt angle. Besides, increasing $U_x$ by a factor of 1.3 {\color{black}{or decreasing by a factor of 0.8}} only very slightly changes the growth rate and tilt angle of the most unstable wave, but increasing or decreasing $U_z$ by the same factors results in considerable changes in the growth rate and tilt angle. {\color{black}{It should be noted that the factors 1.3 and 0.8 are chosen arbitrarily, and the purpose is only to illustrate the importance of $U_x$ and $U_z$ in the linear instability.}} Therefore, the results suggest that the spanwise velocity component of the local mean flow at the downstream end of a turbulent band dominates the local instability. Although the actual flow at the downstream end is more complex than the oversimplified profile $\bm U$, our analysis at least qualitatively illustrates how the velocity components affect the local instability and the flow pattern.
\begin{table}
\centering
\begin{tabular}{c|c|c|c|c}
\hline
profiles & $\alpha$ & $\beta$ & $\gamma$ & tilt angle \\
\hline
$(U_x+1-y^2, U_y, U_z)$  & 0.31 & -1.96 & 0.0188 & $9.0^\circ$ \\
$(1.3U_x+1-y^2, U_y, U_z)$  & 0.31 & -1.94 & 0.0192 & $9.1^\circ$ \\
$(U_x+1-y^2, U_y, 1.3U_z)$  & 0.37 & -1.96 & 0.0298 & $10.7^\circ$ \\
$(0.8U_x+1-y^2, U_y, U_z)$  & 0.31 & -1.94 & 0.0185 & $9.1^\circ$ \\
$(U_x+1-y^2, U_y, 0.8U_z)$  & 0.27 & -1.94 & 0.0114 & $7.9^\circ$ \\
\hline
\end{tabular}
\caption{\label{tab:stability} Influence of the streamwise velocity component $U_x$ and spanwise velocity component  $U_z$ on the local linear instability and tilt angle of the most unstable wave for the basic flow ($U_x+1-y^2, U_y, U_z$), where $U_x$, $U_y$ and $U_z$ are given by  (\ref{equ:target_profile}$-$\ref{equ:target_profile_2}). Note that $U_x$ has to be added by the parabola $1-y^2$ for the linear stability analysis because $U_x$ shown in (\ref{equ:target_profile}) is only the deviation from the parabola \cite{Song2020}. $\alpha$ and $\beta$ are the streamwise and spanwise wavenumbers, respectively, and $\gamma$ is the exponential growth rate of the most unstable wave, {\color{black}{which is obtained by an eigenvalue analysis, see \cite{Xiao2020}. The tilt angle about the streamwise direction of the most unstable wave is calculated as $\arctan{|\alpha/\beta|}$.}}}
\end{table}

Therefore, it naturally follows that the decrease in the tilt angle of flow structures is likely a sign of a reduced instability. Besides, the lowered turbulent activity is a sign of reduced nonlinearity, because one can expect that only sufficiently strong nonlinearity can bring in activities at various length and time scales. Reasonably, this can also be related to the reduced instability. These changes can be understood from the fact that the side wall can certainly affect the flow close to it due to the no-slip boundary condition. 
Figure \ref{fig:cornerplot}(d) shows that the streamwise velocity component is only affected by the side wall in a narrow region very close to the wall ($z\in [48, 50]$), which is even smaller than the spanwise wavelength of the unstable waves at the downstream end as shown by \cite{Xiao2020,Song2020} (see also table \ref{tab:stability}). Besides, the streamwise velocity component at the downstream end was shown to affect little the instability. Therefore, the only possibility left is that the side wall affects (presumably suppresses) the spanwise velocity component of the local mean flow, which has been proposed to play a central role for the local instability and turbulence generation \cite{Xiao2020,Song2020}. Therefore, a reduced instability can be expected when the downstream end gets sufficiently close to the side wall so that the spanwise velocity is affected by the wall. Above $Re_{cr}$, similar effects (e.g. decreasing tilt angles of the wave-like structures) are also observed, but the effects seem not to affect the remaining part of the turbulent band and the band is still sustained after losing the downstream end upon colliding with the wall. In other words, turbulent bands at higher Reynolds numbers do not rely on the local instability at the downstream end.

\subsection{Domain size effect}
We only considered a single channel width $L_z=100$ in this work. One may question that the final state of the flow may be different if the width of the channel is larger. We are aware of this possibility. First of all, the objective of this work is to study the wall-effects on the flow relatively close to the side wall. In a very wide channel with side walls, of course one can expect that the flow state will resemble that in plane Poiseuille flow in the region far from the side walls, see \cite{Shimizu2019,Paranjape2019}.

Second of all, even in much wider channels, our conclusion may still apply at $Re\le 924$, in which regime turbulence can only form one-sided band pattern (parallel band pattern) according to \cite{Shimizu2019}. This is because neighboring bands with opposite orientations would collide with each other and the interaction between the bands would result in the one-sided band pattern. Ref \cite{Shimizu2019} in a very large computational domain determined that the transition from the one-sided pattern to a two-sided pattern (i.e. a band pattern with both orientations) occurs at $Re\simeq 924$. In the one-sided pattern regime, all bands would collide with one of the two side walls because they all propagate in the same direction along the spanwise direction. All turbulent bands colliding with the side wall would not survive the collision according to our results, given a sufficiently long time for the flow to evolve. However, a turbulent band may split before it decays and whether the flow would completely relaminarize depends on the competition between the decay due to the collision with the side wall and the proliferation through splitting. While at which Reynolds number the two processes balance each other is still unknown. Nevertheless, the splitting is rare at relatively low Reynolds number in this regime \cite{Paranjape2019}, and therefore the whole flow would relaminarize.

Between $Re\simeq 924$ and $Re_{cr}$, indeed, our conclusion would only apply to the turbulence sufficiently close to the wall, because the flow far from the side walls may form a two-sided band pattern which can sustain itself by frequently nucleating new bands through branching and splitting \cite{Paranjape2019,Shimizu2019}, as long as the nucleating rate is higher than the decay rate of turbulence caused by the collision with the side wall.

\section{Conclusion}
Using the spectral element code Nektar++, we implemented the forcing technique of \cite{Song2020} with the ability to precisely control turbulent bands in channel flow and studied the side-wall effects on turbulent bands at transitional Reynolds numbers. We observed the decay of turbulent bands at relatively low Reynolds numbers, agreeing with the observation of \cite{Paranjape2019}. At higher Reynolds numbers, however, turbulent bands can survive the collision with the side wall. We narrowed down the critical Reynolds number $Re_{cr}$ to be in the range between 975 and 1000. The flow would completely relaminarize below $Re_{cr}$ if the width of the channel is insufficiently large as shown in our simulations with $L_z=100$. The decay of turbulent bands below $Re_{cr}$ is likely due to the effects of the side wall on the spanwise velocity component of the local mean flow at the downsteam end, which dominates the local instability that the turbulent band rely on for its self-sustainment. Above $Re_{cr}$, a turbulent band does not rely on the instability at the downstream end, therefore, the remaining turbulence is still sustained after the collision with the side wall, {\color{black}{though the remaining turbulence may not be able to keep a distinct banded structure}}. In experimental studies of long-time characteristics of the flow, such as the pattern formation and turbulence fraction measurements, our results suggest that the side wall effects have to be carefully taken care of below $Re_{cr}$.

\section*{Acknowledgements}
\indent We acknowledge financial support from the National Natural Science Foundation of China under grant number 91852105 and from Tianjin University under grant number 2018XRX-0027. We acknowledge the computing resources from TianHe-2 at the National Supercomputer Centre in Guangzhou and TianHe-1(A) at the National Supercomputer Centre in Tianjin. We thank the reviewers for their comments and suggestions.

\section*{Conflict of interests}
The authors declare no conflict of interests.

\bibliographystyle{elsarticle-num-names}
\biboptions{sort&compress}
\bibliography{mybibfile}

\begin{thebibliography}{10}
\expandafter\ifx\csname url\endcsname\relax
  \def\url#1{\texttt{#1}}\fi
\expandafter\ifx\csname urlprefix\endcsname\relax\def\urlprefix{URL }\fi
\expandafter\ifx\csname href\endcsname\relax
  \def\href#1#2{#2} \def\path#1{#1}\fi

\bibitem{Tsukahara2005}
T.~Tsukahara, Y.~Seki, H.~Kawamura, D.~Tochio, {DNS} of turbulent channel flow
  at very low {R}eynolds numbers, in: Proceedings of Fourth International
  Symposium on Turbulence and Shear Flow Phenomena, Williamsburg, USA, 2005,
  pp. 935--940.

\bibitem{TsukaharaKawaguchi2014}
T.~Tsukahara, Y.~Kawaguchi, H.~Kawamura, An experimental study on
  turbulent-stripe structure in transitional channel flow, arXiv:1406.1378
  [physics.flu-dyn].

\bibitem{Tuckerman2014}
L.~S. Tuckerman, T.~Kreilos, H.~Shrobsdorff, T.~M. Schneider, J.~F. Gibson,
  Turbulent{-}laminar patterns in plane {P}oiseuille flow, Phys.\ Fluids 26
  (2014) 114103.

\bibitem{Xiong2015}
X.~M. Xiong, J.~Tao, S.~Chen, L.~Brandt, Turbulent bands in plane-{P}oiseuille
  flow at moderate {R}eynolds numbers, Phys.\ Fluids 27 (2015) 041702.

\bibitem{Tao2018}
J.~J. Tao, B.~Eckhardt, X.~M. Xiong, Extended localized structures and the
  onset of turbulence in channel flow, Phys. Rev. Fluids 3 (2018) 011902.

\bibitem{Kanazawa2018}
T.~Kanazawa, Lifetime and growing process of localized turbulence in plane
  channel flow, Ph.D. thesis, Osaka University (2018).

\bibitem{Paranjape2019}
C.~S. Paranjape, Onset of turbulence in plane poiseuille flow, Ph.D. thesis,
  IST Austria (2019).

\bibitem{Shimizu2019}
M.~Shimizu, P.~Manneville, Bifurcations to turbulence in transitional channel
  flow, Phys. Rev. Fluids 4 (2019) 113903.

\bibitem{Paranjape2020}
C.~S. Paranjape, Y.~Duguet, B.~Hof, Oblique stripe solutions of channel flow,
  J.\,Fluid Mech. 897 (2020) A7.

\bibitem{Xiao2020}
X.~Xiao, B.~Song, The growth mechanism of turbulent bands in channel flow at
  low {R}eynolds numbers, J.\,Fluid Mech. 883 (2020) R1.

\bibitem{Tuckerman2020}
L.~S. Tuckerman, M.~Chantry, D.~Barkley, Patterns in {W}all-{B}ounded {S}hear
  {F}lows, Ann.\ Rev.\ Fluid Mech. 52 (2020) 343--67.

\bibitem{Liu2020}
J.~Liu, Y.~Xiao, L.~Zhang, M.~Li, J.~Tao, S.~Xu, Extension at the downstream
  end of turbulent band in channel flow, Phys.\ Fluids 32 (2020) 121703.

\bibitem{Duguet2020}
P.~V. Kashyap, Y.~Duguet, O.~Dauchot, Flow statistics in the transitional
  regime of plane channel flow, Entropy 22 (2020) 1001.

\bibitem{Song2020}
B.~Song, X.~Xiao, Trigger turbulent bands directly at low {R}eynolds numbers in
  channel flow using a moving-force technique, J.\,Fluid Mech. 903 (2020) A43.

\bibitem{Xiao2020b}
X.~Xiao, B.~Song, Kinematics and dynamics of turbulent bands at low {R}eynolds
  numbers in channel flow, Entropy 22 (2020) 1167.

\bibitem{Takeishi2015}
K.~Takeishi, G.~Kawahara, H.~Wakabayashi, M.~Uhlmann, A.~Pinelli, Localized
  turbulence structures in transitional rectangular-duct flow, J.\,Fluid Mech.
  782 (2015) 368--379.

\bibitem{Sano2016}
M.~Sano, K.~Tamai, A universal transition to turbulence in channel flow, Nat.
  Phys. 12 (2016) 249--253.

\bibitem{Yimprasert2021}
S.~Yimprasert, M.~Kvick, P.~H. Alfredsson, M.~Matsubara, Flow visualization and
  skin friction determination in transitional channel flow, Experiments in
  Fluids 62 (2021) 31.

\bibitem{2008Gmsh}
C.~Geuzaine, J.~F. Remacle, Gmsh: A three-dimensional finite element mesh
  generator with built-in pre-and post-processing facilities, Intnl J.\,Num.\
  Meth.\ Engng 79 (2008) 1309--1331.

\bibitem{nektar}
C.~Cantwell, D.~Moxey, A.~Comerford, A.~Bolis, G.~Rocco, G.~Mengaldo, D.~{De
  Grazia}, S.~Yakovlev, J.-E. Lombard, D.~Ekelschot, B.~Jordi, H.~Xu,
  Y.~Mohamied, C.~Eskilsson, B.~Nelson, P.~Vos, C.~Biotto, R.~Kirby,
  S.~Sherwin, Nektar++: An open-source spectral/hp element framework, Computer
  Physics Communications 192 (2015) 205--219.

\bibitem{KARNIADAKIS1991414}
G.~E. Karniadakis, M.~Israeli, S.~A. Orszag, High-order splitting methods for
  the incompressible navier-stokes equations, Journal of Computational Physics
  97~(2) (1991) 414--443.

\bibitem{Xiao2020c}
Y.~Xiao, J.~Tao, L.~Zhang, Self-sustaining and propagation mechanism of
  localized wave packet in plane-poiseuille flow, Phys.\ Fluids 33 (2020)
  031706.

\bibitem{Jimenez1991}
J.~Jimenez, P.~Moin, The minimal flow unit in near{-}wall turbulence, J.\,Fluid
  Mech. 225 (1991) 213--240.

\end{thebibliography}

{\color{black} {

\appendix

\section{Grid resolution}\label{sec:appendix}
The results shown in the main text were obtained using 80 elements in the spanwise direction. Using a $9^{th}$-order polynomial and GLL point distribution in the element, this resolution gives a sub-element grid spacing ranging from 0.085 ($h/\Delta z=11.8$) to 0.3 ($h/\Delta z=3.3$) in the largest element at the channel center ($z=0$ and $y=0$). Although the largest grid spacing at the element center is a bit larger than that used in the literature for Poiseuille flow at similar $Re$ ($h/\Delta z\approx 6$, see \cite{Tao2018,Kanazawa2018,Shimizu2019,Xiao2020b}), where uniform grid was used, the average grid spacing in the largest element is comparable to the latter. As the element size decreases towards the side walls, see figure \ref{fig:wallmesh} and equation \ref{grid}, the grid resolution would be much higher in the near wall region than at the channel center and should be sufficient.

Nevertheless, in order to show that the results are not qualitatively affected by the selection of the grid resolution, we increased the number of element to 120 in the spanwise direction, which results in a sub-element grid spacing ranging from 0.057 ($h/\Delta z=17.5$) to 0.2 ($h/\Delta z=5$) in the largest element at the channel center. The average grid spacing is smaller than that used in the literature for Poiseuille flow \cite{Tao2018,Kanazawa2018,Shimizu2019,Xiao2020b}, and the grid spacing close to the side walls is further smaller. With this resolution, we repeated the simulation at $Re=975$. The comparison between the low- and high-resolution results is shown in figure \ref{fig:comparison-between-precision}.

\begin{figure}[h]
	\subfigbottomskip = 2pt
	\subfigcapskip=-5pt
	\begin{minipage}[h]{0.22\linewidth}
	\centering
	\subfigure[low t=820]{\includegraphics[width = 1\textwidth]{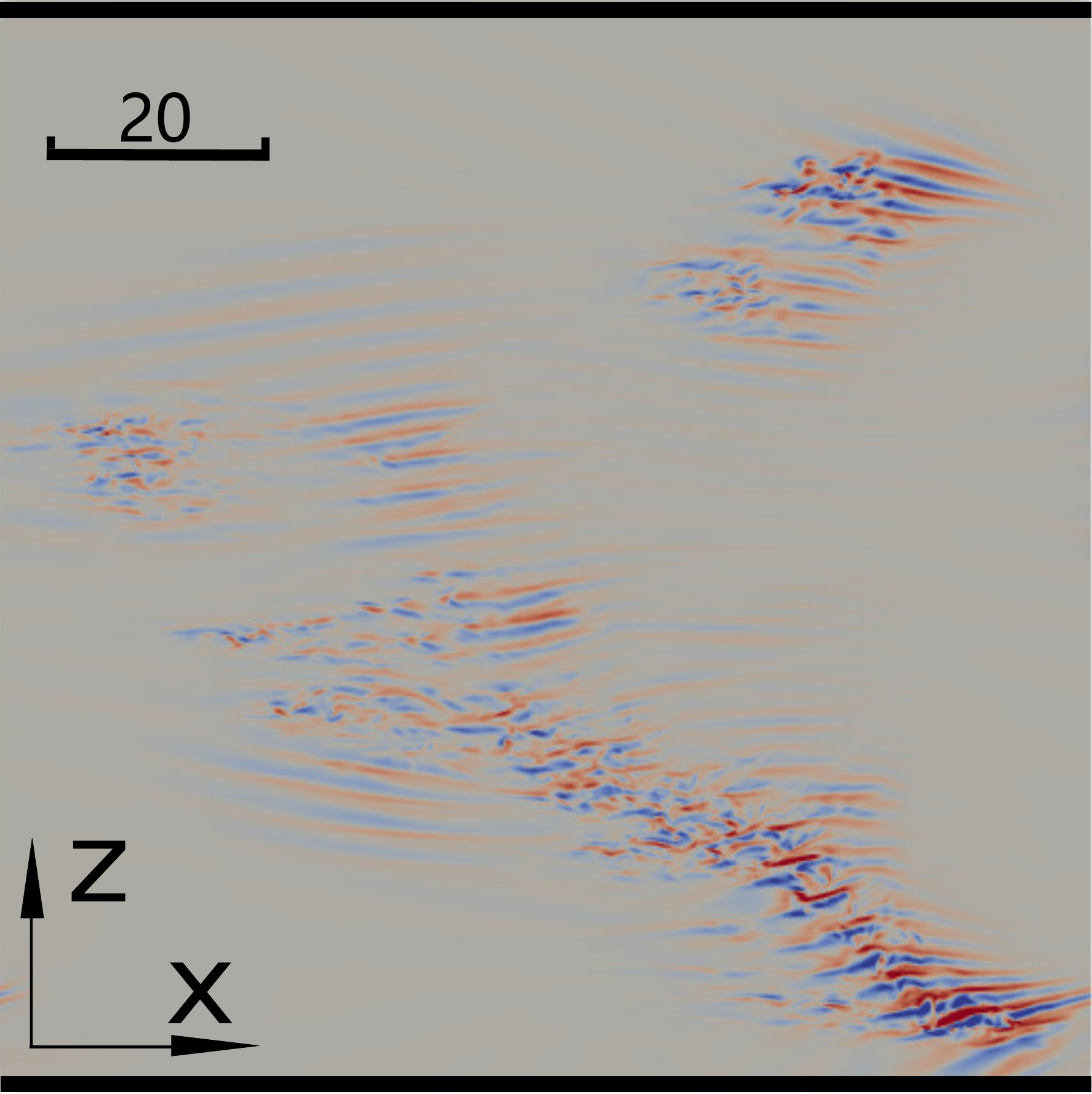}}
	\end{minipage}
	\quad
	\begin{minipage}[h]{0.22\linewidth}
	\centering
	\subfigure[t=1195]{\includegraphics[width = 1\textwidth]{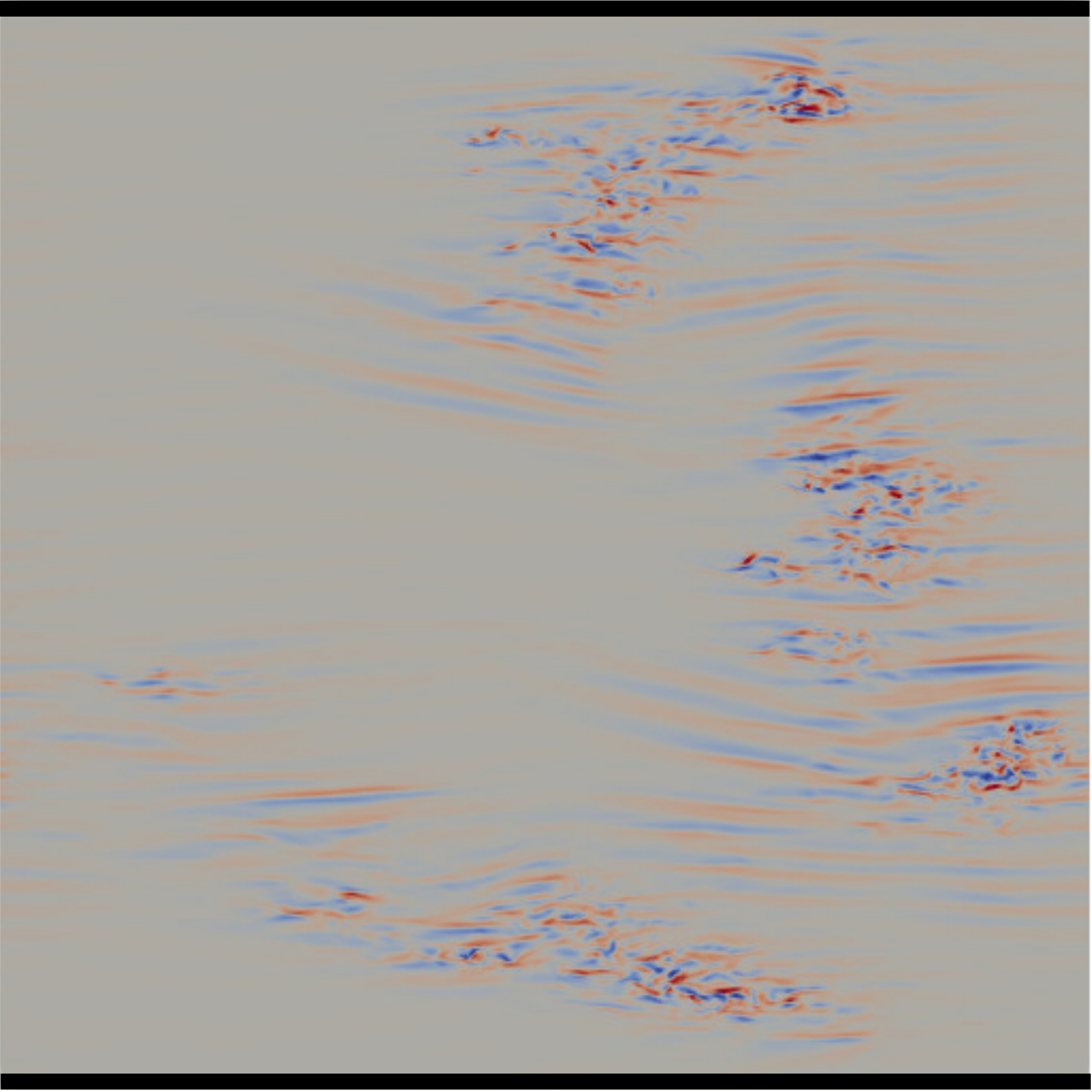}}
	\end{minipage}
	\quad
	\begin{minipage}[h]{0.22\linewidth}
	\centering
	\subfigure[t=1555]{\includegraphics[width = 1\textwidth]{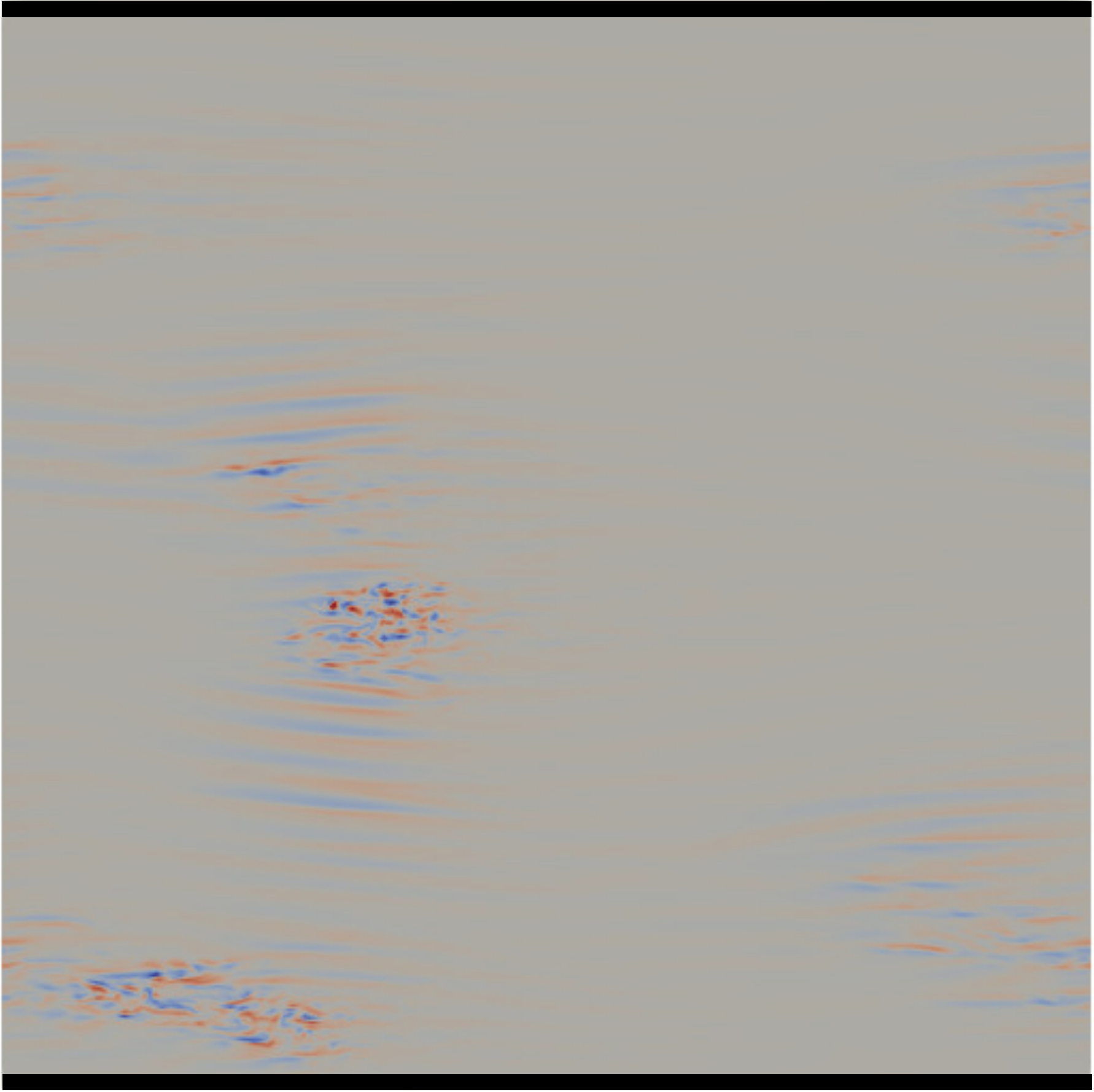}}
	\end{minipage}
	\quad
	\begin{minipage}[h]{0.22\linewidth}
	\centering
	\subfigure[t=2545]{\includegraphics[width = 1\textwidth]{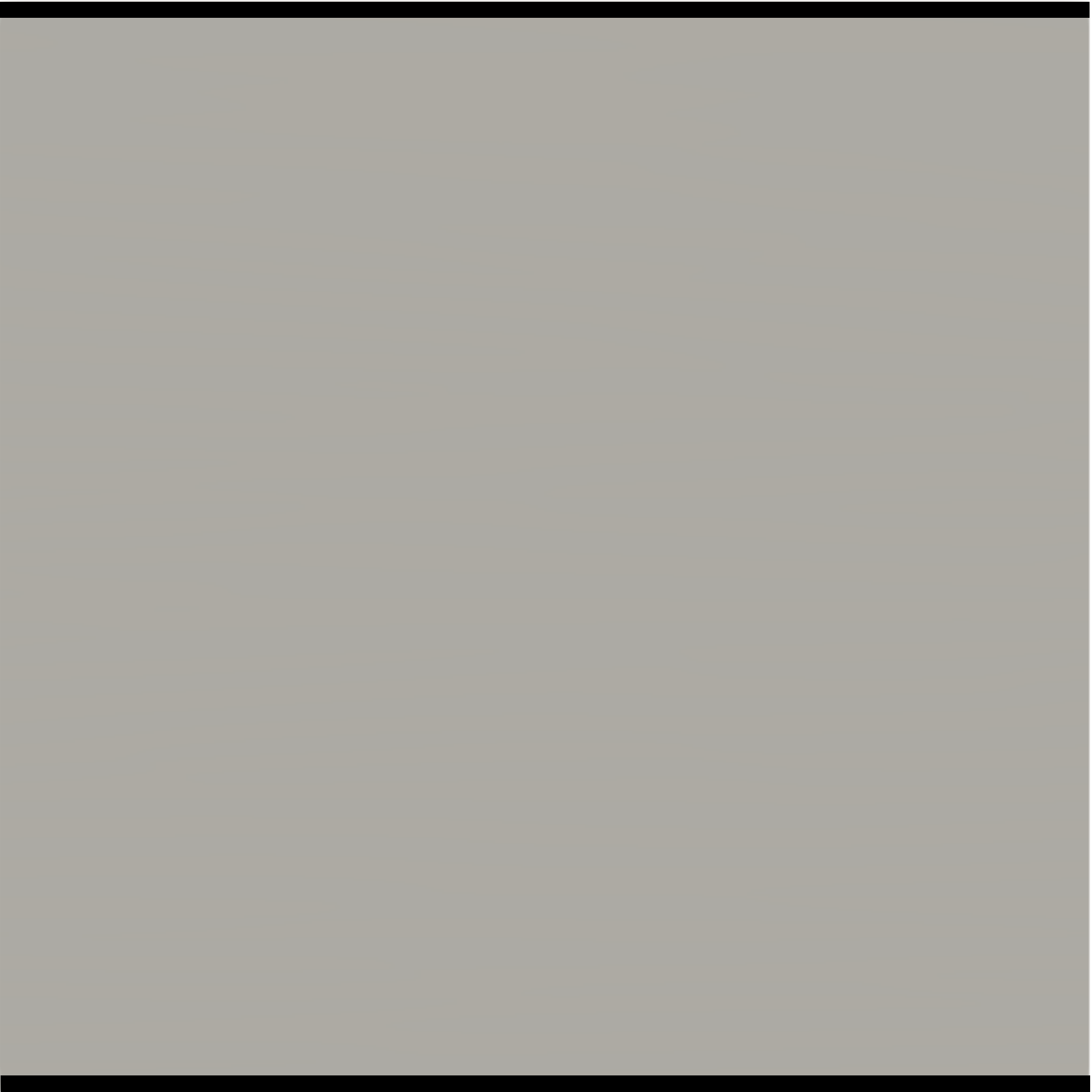}}
	\end{minipage}

	\begin{minipage}[h]{0.22\linewidth}
	\centering
	\subfigure[Re=975 t=850]{\includegraphics[width = 1\textwidth]{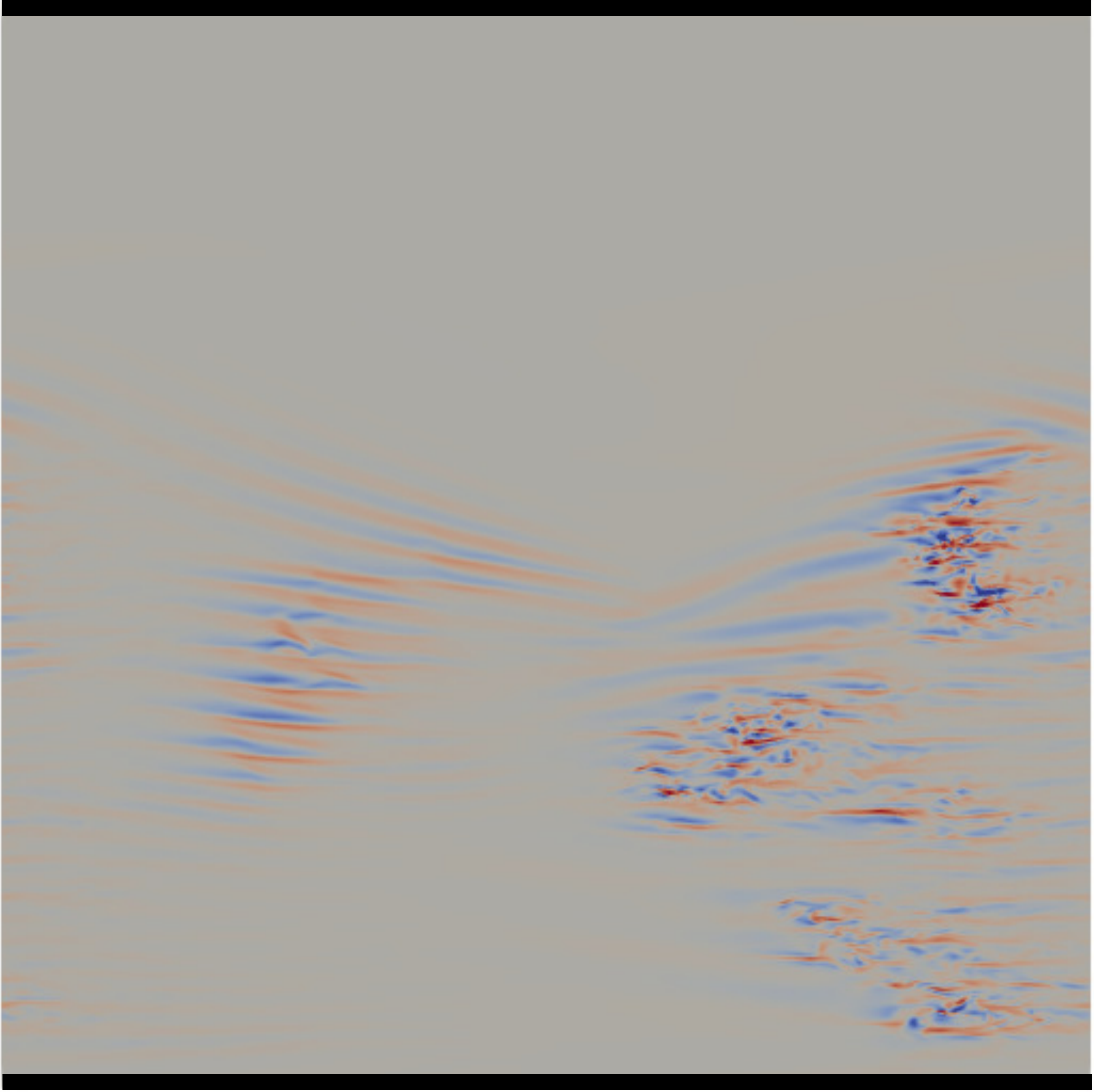}}
	\end{minipage}
	\quad
	\begin{minipage}[h]{0.22\linewidth}
	\centering
	\subfigure[t=1150]{\includegraphics[width = 1\textwidth]{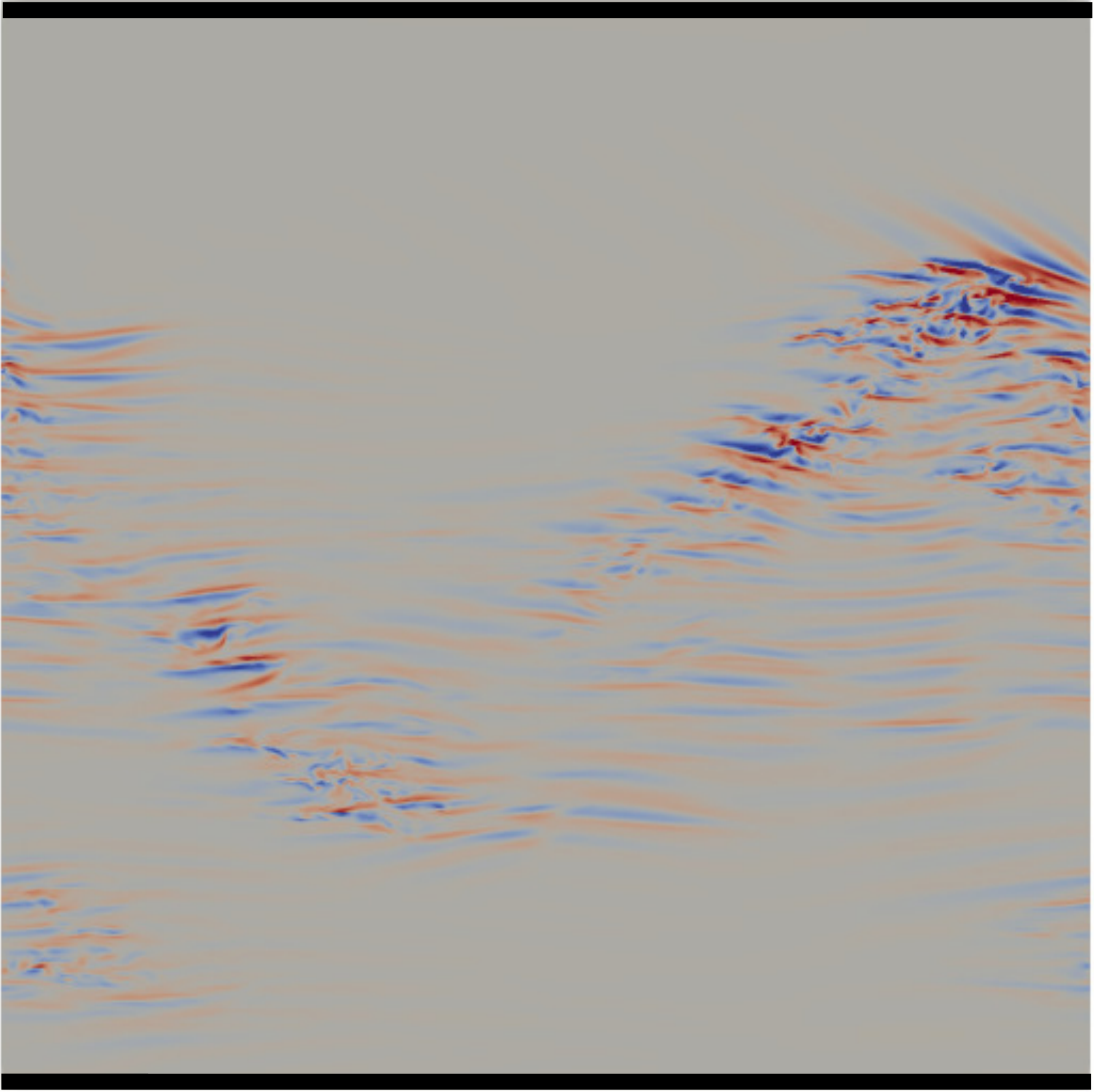}}
	\end{minipage}
	\quad
	\begin{minipage}[h]{0.22\linewidth}
	\centering
	\subfigure[t=1540]{\includegraphics[width = 1\textwidth]{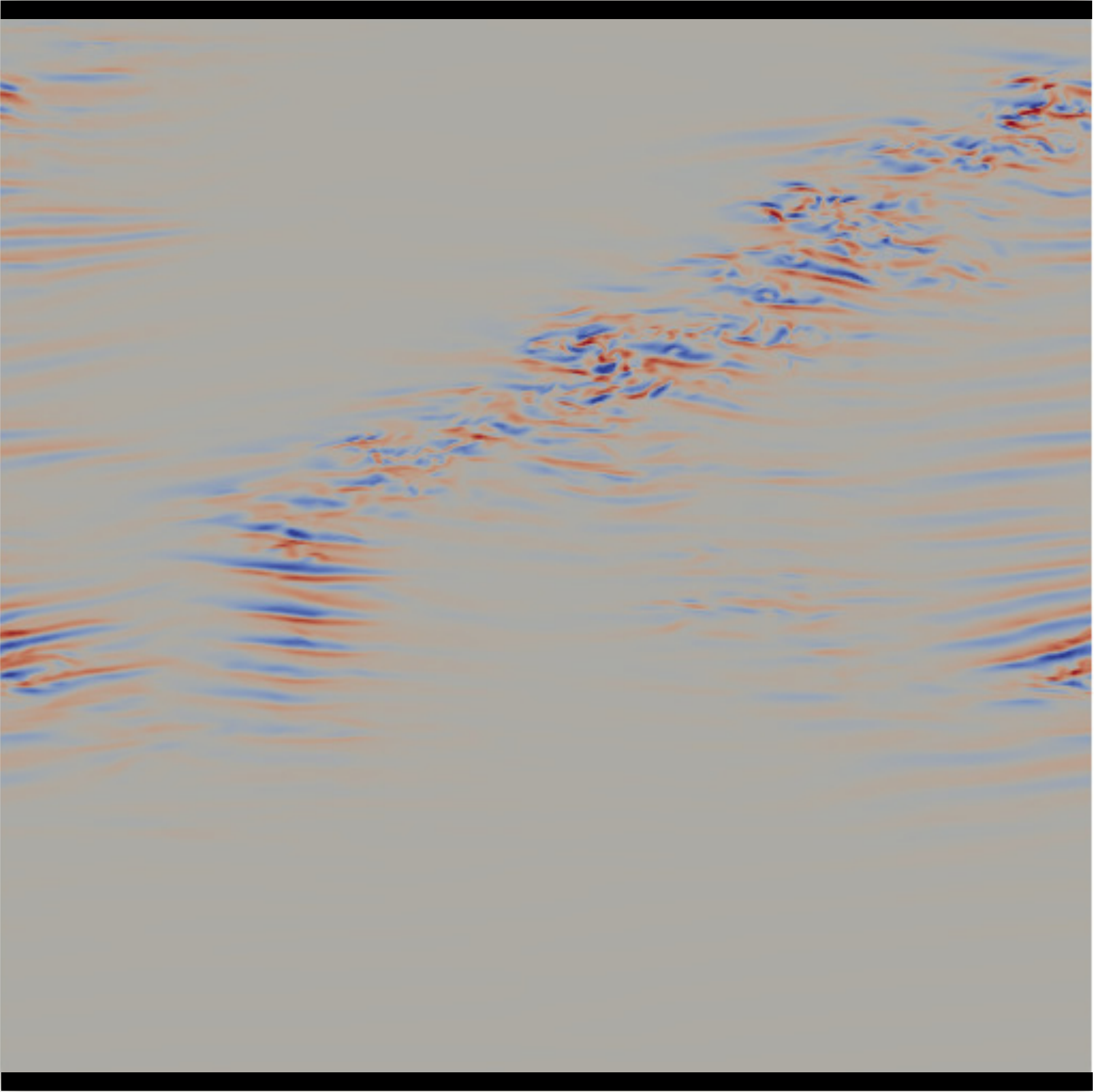}}
	\end{minipage}
	\quad
	\begin{minipage}[h]{0.22\linewidth}
	\centering
	\subfigure[t=2545]{\includegraphics[width = 1\textwidth]{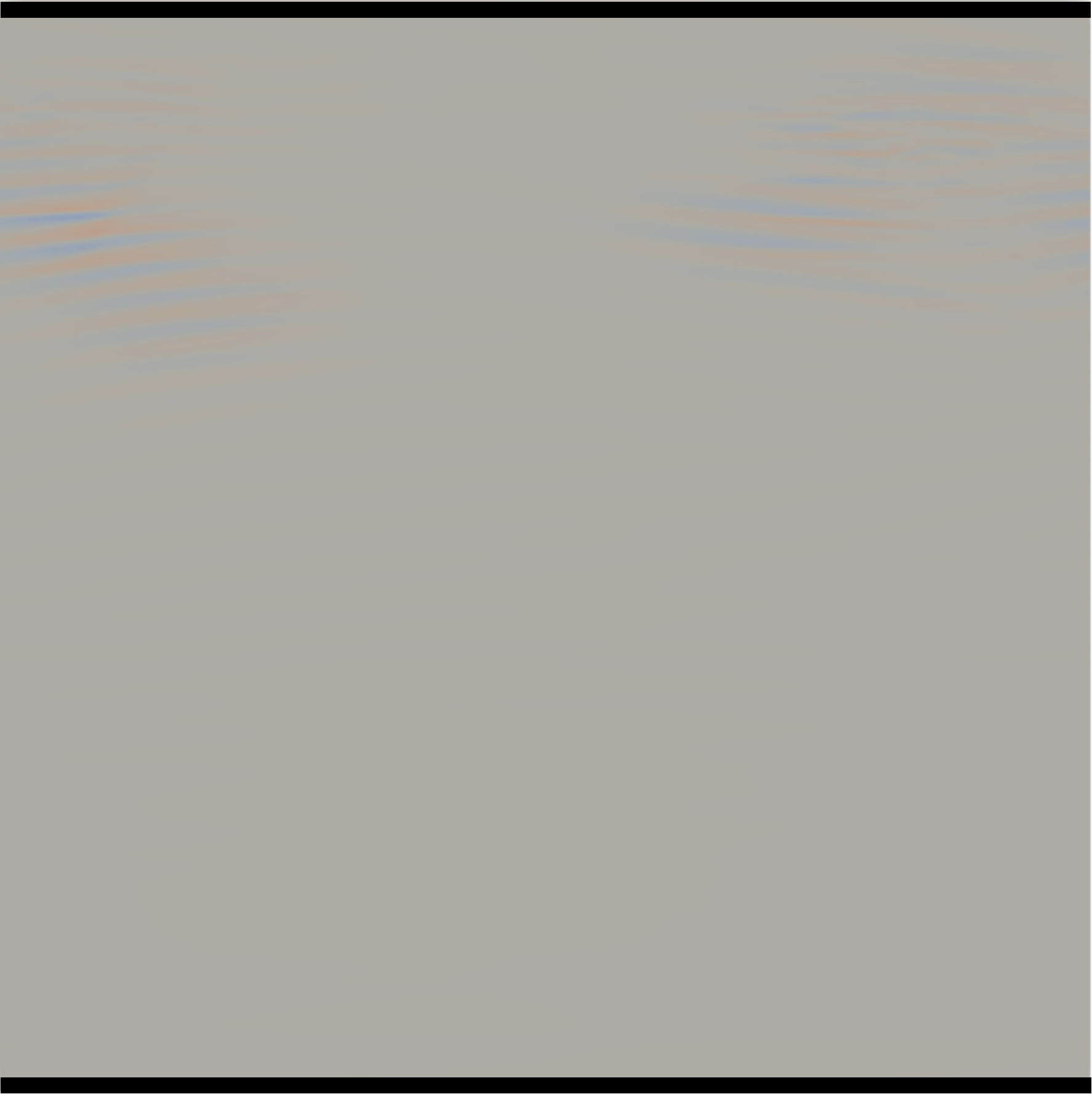}}
	\end{minipage}
	\quad
	\begin{minipage}[h]{0.4\linewidth}
	\centering
	\subfigure{\includegraphics[width = 0.995\textwidth]{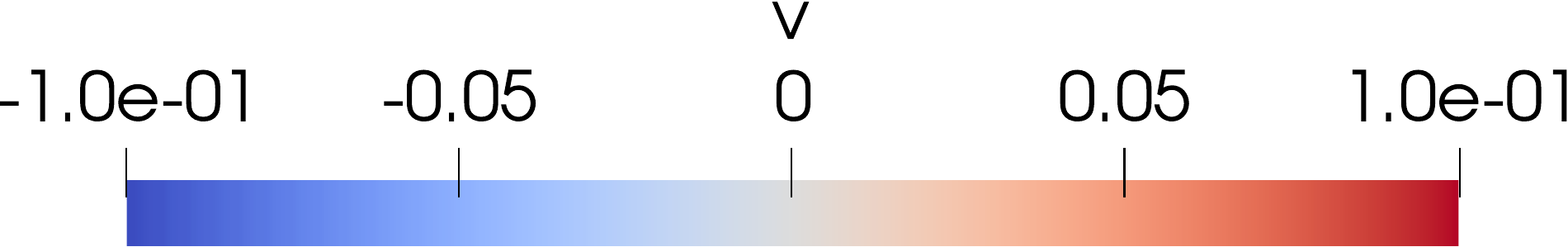}}
	\end{minipage}
	\caption{Comparison between simulations using high and low resolution grid at Re=975. The upper row uses 80 elements in the spanwise direction and the lower row uses 120 elements. The time step size is the same for both cases.}
\label{fig:comparison-between-precision}
\end{figure}

The initial condition for the low resolution case was interpolated onto the high-resolution grid and used as the initial condition. Some differences in the detailed development of the flow were observed (e.g. compare panel (a) and (e) in figure \ref{fig:comparison-between-precision}), which is not too surprising given the chaotic nature of transitional turbulence. Nevertheless, the final flow state seems not be qualitatively affected by the resolution because turbulent bands completely decayed and within similar time spans in both cases.

}}

\end{document}